\documentclass[aps,prb,preprint,superscriptaddress,amsmath,amssymb]{revtex4-2}
\usepackage{makecell}
\usepackage[table]{xcolor}
\usepackage{booktabs}
\usepackage{multirow}
\usepackage[utf8]{inputenc} 
\usepackage[T1]{fontenc}    
\usepackage{url}            
\usepackage{booktabs}       
\usepackage{amsfonts}       
\usepackage{nicefrac}       
\usepackage{microtype}      
\usepackage{fancyhdr}       
\usepackage{graphicx}       
\usepackage{amsmath}
\usepackage{svg}
\usepackage{pifont}
\usepackage{ulem}
\usepackage{comment}
\usepackage{textcomp}
\usepackage{subcaption}
\usepackage{threeparttable}

\usepackage[ruled,vlined]{algorithm2e}

\graphicspath{{Fig/}}

\newcommand{\bfR}{ {\bf R}} 
\newcommand{\bfG}{ {\bf G}} 
\newcommand{\bfGp}{ {\bf G'}} 
\newcommand{\bfq}{ {\bf q}} 
\newcommand{\bfk}{ {\bf k}} 

\newcommand{\bracket}[3]{\ensuremath{\langle #1 | #2 | #3 \rangle}}

\definecolor{ghjnote}{RGB}{45,110,185}

\usepackage{hyperref}       
\usepackage{cleveref}

\begin{document}

\title{$G^0W^0$ implementation based on the pseudopotential and numerical-atomic-orbital basis-set framework: Algorithms and benchmarks
}

\author{Huanjing Gong}
\affiliation{Institute of Physics, Chinese Academy of Sciences, Beijing 100190, China}
\affiliation{School of Physical Sciences, University of Chinese Academy of Sciences, Beijing 100049, China}
\author{Min-Ye Zhang}
\email{minyez@iphy.ac.cn}
\affiliation{Institute of Physics, Chinese Academy of Sciences, Beijing 100190, China}
\affiliation{The NOMAD Laboratory at the FHI of the Max Planck Society, Faradayweg 4-6, D-14195 Berlin-Dahlem, Germany}
\author{Peize Lin}
\affiliation{Institute of Artificial Intelligence, Hefei Comprehensive National Science Center, Hefei 230026, Anhui, China.}
\author{Bohan Jia}
\affiliation{Institute of Physics, Chinese Academy of Sciences, Beijing 100190, China}
\affiliation{School of Physical Sciences, University of Chinese Academy of Sciences, Beijing 100049, China}
\author{Ziqing Guan}
\affiliation{Institute of Physics, Chinese Academy of Sciences, Beijing 100190, China}
\affiliation{School of Physical Sciences, University of Chinese Academy of Sciences, Beijing 100049, China}
\author{Lixin He}
\email{helx@iphy.ac.cn}
\affiliation{CAS Key Laboratory of Quantum Information, University of Science and Technology of
China, Hefei 230026, Anhui, China}
\affiliation{Institute of Artificial Intelligence, Hefei Comprehensive National Science Center, Hefei 230026, Anhui, China.}
\author{Xinguo Ren}
\email{renxg@iphy.ac.cn}
\affiliation{Institute of Physics, Chinese Academy of Sciences, Beijing 100190, China}


\begin{abstract}
The $GW$ method delivers substantially improved accuracy in electronic band structure calculations over conventional Kohn-Sham density functional theory (KS-DFT) by explicitly incorporating the electron self-energy effect beyond mean-field approximations.
Despite many existing implementations, a periodic $GW$ implementation within the framework of numerical atomic orbitals (NAO) combined with the pseudopotential (PP) scheme has not been reported.
This is urgently needed given the increasing popularity of the NAO-PP framework in KS-DFT calculations and its importance for the development of machine-learning electronic-structure
approaches.
In this work, we present an efficient NAO-PP-based $G^0W^0$ computational framework by interfacing the first-principles software package ABACUS with LibRPA -- a library for 
performing low-scaling random-phase approximation and $GW$ calculations based on NAOs.
Our approach employs the localized resolution of identity (LRI) technique with a novel compression scheme, significantly improving both computational efficiency and numerical stability. In addition, an analytic treatment of the small-\(\bfq\) limit of the microscopic dielectric function reduces the need for dense \(\bfq\)-point sampling. Furthermore, we propose a practical strategy to select a suitable KS-DFT pseudopotential prior to $G^0W^0$ calculations by examining the frequency-dependent macroscopic dielectric function.
Systematic benchmarks validate the effectiveness of our compression scheme and real-space tensor filtering strategies, demonstrating both high accuracy and significant computational efficiency gains. Comparisons with established $G^0W^0$ implementations show excellent agreement in band structures and band gaps, confirming ABACUS+LibRPA as a reliable and efficient platform for large-scale $G^0W^0$ simulations.
\end{abstract}

\maketitle

\section{INTRODUCTION}
The $GW$ method, derived from Green’s function-based many-body perturbation theory (MBPT) \cite{hedin_gw_1965}, provides a physically rigorous framework for electronic band structure calculations, unlike the mean-field-type approximation within Kohn-Sham density functional theory (KS-DFT) \cite{hohenberg_inhomogeneous_1964,kohn_self-consistent_1965}. This approach has been widely applied to calculate quasiparticle (QP) energies and band gaps with high reliability across a variety of systems, including bulk semiconductors and insulators, two-dimensional (2D) materials, and clusters. \cite{gw100_2015, gloze_compendium_2019, 2d_Rasmussen_2016,gw_Reining_2018}
However, the $GW$ method is computationally demanding, often limiting its applicability to relatively simple systems. Therefore, developing efficient and robust $GW$ implementation frameworks remains an active research topic in contemporary computational chemistry and materials science\cite{vasp_gw_2014, abinit_Gonze_2005, yambo_Sangalli_2019, BerkeleyGW_Deslippe_2012, GPAW_Huser_2013,Ren_gw_2021, exciting_Gulans_2014, cp2k_Graml_2024, Rangel/etal:2020, ZhuT21}. 

An efficient implementation of any first-principles electronic-structure methods entails choosing a suitable basis set representation and a proper treatment of the nuclear-electron interactions. In this regard, the combination of numerical atomic orbitals (NAO) and the pseudopotential (PP) scheme has been highly successful in KS-DFT electronic-structure calculations, as exemplified by established NAO-based codes such as SIESTA \cite{soler_siesta_2002}, OpenMX \cite{openmx_2003,openmx_2004}, and ABACUS\cite{li_large-scale_2016,lin_ab_2024,abacus_2025_jcp}. 
In recent years, owing to its compact and localized representation, the NAO+PP framework has also become a convenient foundation for machine-learning electronic-structure methods and Hamiltonian-learning models\cite{li2022deep,wang2024deeph,wang2024universal,zhong2023transferable,cheng2025efficient,zhang2025advancing,su2023efficient}.
The inherent locality of NAOs leads to sparse Hamiltonian and overlap matrices, while the PP formalism avoids the explicit treatment of rapidly oscillating core states, thereby reducing the basis size. These features make the NAO+PP framework particularly attractive for large-scale calculations. However, despite its success in KS-DFT and machine-learning applications, a systematic implementation of $G^0W^0$ within the NAO+PP framework has not yet been reported or thoroughly benchmarked, and its efficacy in many-body perturbation theory therefore remains unclear.

For advanced electronic structure calculations based on numerical atomic orbitals (NAOs) that go beyond conventional density functional approximations, the localized resolution of identity (LRI) approximation serves as a powerful numerical technique, making high-level methods applicable to large-scale systems. By decomposing four-center integrals into two-center terms using a set of auxiliary basis functions (ABFs), LRI achieves substantial computational acceleration and memory savings while maintaining high accuracy. Furthermore, the spatial locality of NAOs naturally introduces sparsity in the relevant matrices, allowing for real-space tensor filtering techniques --- where small tensor blocks are selectively neglected according to a threshold --- to further reduce computational costs. The combination of NAOs and LRI has facilitated the development of low-scaling algorithms and codes for exact exchange (EXX) \cite{lin_force_2025, lin_accuracy_2020, lin_efficient_2021, lin_ab_2024, cao_sym_2025, exx_Levchenko_2015,exx_kokott_2024} and random-phase approximation (RPA) calculations \cite{rpa_shi_2024,librpa_shi_2025} in both all-electron and pseudopotential (PP) frameworks. For $GW$ calculations, LRI has been shown to deliver adequate numerical precision within the all-electron NAO scheme \cite{Ren_gw_2021}. Given these developments, we anticipate that LRI will also be effective for NAO-PP-based $GW$ calculations, although a robust computational infrastructure and systematic benchmarks are still lacking.

In this work, we develop an efficient NAO+PP-based $G^0W^0$ computational framework by integrating the LibRPA library \cite{rpa_shi_2024,librpa_shi_2025} with the ABACUS code \cite{li_large-scale_2016,lin_ab_2024,abacus_2025_jcp}.
LibRPA is a standalone library package designed for low-scaling NAO-based RPA and $GW$ calculations \cite{rpa_shi_2024,librpa_shi_2025,Zhang2026arXiv}, whereas ABACUS is a DFT software package that employs NAO and plane-wave dual basis sets 
\cite{chen_systematically_2010,chen_electronic_2011,lin_strategy_2021,abacus_2025_jcp} within
the norm-conserving PP framework. 
Novel features in our $G^0W^0$ implementation in the ABACUS+LibRPA infrastructure include
the incorporation of relativistic effects (spin-orbit coupling, SOC) via the two-component spinor formalism, utilizing the PseudoDojo pseudopotentials.\cite{van_setten_pseudodojo_2018} 
In addition, several key optimizations are introduced to enhance both efficiency and accuracy.
First, we develop an analytic treatment of the small-\(q\) limit of the microscopic dielectric function, substantially accelerating \(q\)-point convergence.
This reduces the required k-point sampling density and thus lowers computational cost.
Second, a compression scheme is introduced to the LRI-based $G^0W^0$ workflow, significantly reducing the memory usage and computational cost while preserving numerical precision.
Third, we propose a practical guideline for selecting a pseudopotential suitable for $GW$ calculations, based on an analysis of the frequency dependence of the macroscopic dielectric function.
Furthermore, we systematically benchmark real-space tensor-filtering thresholds for the RI coefficients, Green's function, and screened Coulomb interaction, demonstrating a significant reduction in wall time with minimal accuracy loss. 
We validate our implementation through benchmark studies on representative semiconductors, demonstrating its accuracy and reliability through comparison with established $G^0W^0$ implementations such as those in FHI-aims \cite{blum_ab_2009} and VASP \cite{hafner_ab-initio_2008,vasp_gw_2014}.

The paper is organized as follows. In \cref{sec:theoretical framework}, we present the theoretical framework of our $G^0W^0$ implementation, starting from the formulation based on the pseudopotential NAO framework and then introducing the representation in an auxiliary basis set, including the treatment of SOC.
In \cref{sec:implement}, we discuss key implementation details designed to enhance efficiency, including the $\Gamma$-point singularity treatment to accelerate the ${\bf k}$-point convergence, a compression scheme for the LRI-based $G^0W^0$ workflow to reduce the memory consumption, and a practical strategy for selecting a suitable KS-DFT pseudopotential for $G^0W^0$ calculations. In \cref{sec:result}, we assess the impact of these optimizations and present systematic benchmarks validating the computational efficiency gains from the compression scheme and real-space tensor filtering, 
as well as the accuracy through comparisons with established $G^0W^0$ implementations, demonstrating the reliability and efficiency of our implementation. Finally, we conclude this work in \cref{sec:conclusion}.

\section{THEORETICAL FRAMEWORK}
\label{sec:theoretical framework}

In this section, we present the theoretical framework of $G^0W^0$ within the pseudopotential and NAO basis set framework. We begin by reviewing the workflow of the $G^0W^0$ implementation in LibRPA, based on a modified space-time formalism. Next, we derive the $G^0W^0$ equations in the NAO-PP framework, building on the preceding KS-DFT calculations in ABACUS. Finally, we introduce the auxiliary-basis formulation of $G^0W^0$ using the LRI technique, which is crucial for reducing the computational cost of the calculations.

Starting from a mean-field Green's function, the $G^0W^0$ approach corresponds to the first iteration of Hedin's equations\cite{hedin_gw_1965}, neglecting vertex corrections. From Dyson's equation for the one-particle Green's function, one can derive an effective single-particle eigenvalue problem as
\begin{equation}
    h^{\mathrm{MF}}(\mathbf{r})\Psi_{n\mathbf{k}}(\mathbf{r})-\int d\mathbf{r}^{\prime}v^{\mathrm{xc}}(\mathbf{r},\mathbf{r}^{\prime})\Psi_{n\mathbf{k}}(\mathbf{r}^{\prime})+\int d\mathbf{r}^{\prime}\Sigma^{\mathrm{xc}}(\mathbf{r},\mathbf{r}^{\prime},\epsilon_{n\mathbf{k}})\Psi_{n\mathbf{k}}(\mathbf{r}^{\prime})=\epsilon_{n\mathbf{k}}\Psi_{n\mathbf{k}}(\mathbf{r}),
    \label{eq:Dyson}
\end{equation}
where $h^{\mathrm{MF}}(\mathbf{r})$ is a mean-field Hamiltonian, $v^{\mathrm{xc}}(\mathbf{r},\mathbf{r}^{\prime})$ is the mean-field exchange-correlation (XC) 
potential contained in $\hat{h}^{\mathrm{MF}}(\mathbf{r})$, $\Sigma^{\mathrm{xc}}(\mathbf{r},\mathbf{r}^{\prime},\epsilon_{n\mathbf{k}})$ is the XC self-energy, and $\{\Psi_{n\mathbf{k}}(\mathbf{r})\}$, $\{\epsilon_{n\mathbf{k}}\}$ are the QP wavefunctions and energies with band index $n$ and Bloch wavevector $\mathbf{k}$ in the first Brillouin zone (BZ). In practice, the QP wavefunctions are often approximated by the mean-field eigenfunctions $\{\psi_{n\mathbf{k}}(\mathbf{r})\}$, which can be chosen as the Kohn-Sham (KS) or Hartree-Fock (HF) orbitals. Projecting these wavefunctions onto Eq.~\eqref{eq:Dyson} yields the QP equation as
\begin{equation}
    \epsilon_{n\mathbf{k}}=\epsilon_{n\mathbf{k}}^0+\langle\psi_{n\mathbf{k}}|\Sigma^x+\Sigma^c(\epsilon_{n\mathbf{k}})-v^\mathrm{xc}|\psi_{n\mathbf{k}}\rangle,
    \label{eq:eqp}
\end{equation}
where $\{\epsilon_{n\mathbf{k}}^0\}$ are the mean-field eigenvalues of $h^{\mathrm{MF}}(\mathbf{r})$, and the XC self-energy is divided into the frequency-independent exact-exchange energy $\Sigma^x$ in HF and the correlation part $\Sigma^c$. To solve Eq.~\eqref{eq:eqp}, the central task is the evaluation of the dynamical $G^0W^0$ correlation self-energy $\Sigma^{c}_{n\mathbf{k}}(\omega)$ in the KS representation. In a periodic system using the space-time formalism, the spatially non-local self-energy matrix element in the imaginary-time domain is given by
\begin{equation}
    \begin{gathered}
        \Sigma^c_{n\mathbf{k}}(\mathrm{i}\tau)\begin{aligned}=\langle n\mathbf{k}|\hat{\Sigma}^c(\mathrm{i}\tau)|n\mathbf{k}\rangle=\int\mathrm{d}\mathbf{r}\mathrm{d}\mathbf{r^{\prime}}\psi_{n\mathbf{k}}^*(\mathbf{r})\Sigma^c(\mathbf{r},\mathbf{r^{\prime}},\mathrm{i}\tau)\psi_{n\mathbf{k}}(\mathbf{r^{\prime}})\end{aligned}\\=-\int\mathrm{d}\mathbf{r}\mathrm{d}\mathbf{r^{\prime}}\psi_{n\mathbf{k}}^*(\mathbf{r})G^0(\mathbf{r},\mathbf{r^{\prime}},\mathrm{i}\tau)W^{0(c)}(\mathbf{r},\mathbf{r^{\prime}},\mathrm{i}\tau)\psi_{n\mathbf{k}}(\mathbf{r^{\prime}}).
    \end{gathered}
    \label{eq:self-energy}
\end{equation}
Finally, the correlation self-energy on the real-frequency axis is obtained via Fourier transform from the imaginary-time domain to the imaginary-frequency domain, followed by an analytic continuation. The QP energies ${\epsilon_{n\mathbf{k}}}$ are then determined iteratively by solving Eq.~\eqref{eq:eqp}.

\subsection{Space-time \texorpdfstring{$G^0W^0$}{GW} formalism}
\label{ssec:space-time GW}

The main idea of the space-time $G^0W^0$ formalism is to switch between the real-space/reciprocal-space representations and between the time/frequency domains via Fourier transformations, so that the computational steps involving convolutions become pointwise multiplications\cite{spacetime_rojas_1995,spacetime_rieger_1999}. 
To this end, the time/frequency Fourier transformations are carried out on the imaginary axis, and this can be efficiently performed by the recently introduced minimax integration grids\cite{kaltak_minimax_2014,kaltak_minimax2_2014,Liu_minimax_2016}. 
A key feature of this algorithm is that it reduces the formal scaling of the computational cost from $O(N^4)$ to $O(N^3)$. In the conventional plane-wave-based space-time algorithm, the real-space representation of the two-point quantities (such as the polarization function) requires tabulation on uniform real-space grids. This causes a substantial increase of memory requirement. The situation will be quite different in atomic-orbital based formalism, as will be discussed later.

In the $G^0W^0$ approach for a periodic system, the non-interacting Green's function $G^0$ is typically constructed from the KS eigenvalues $\epsilon_{n\mathbf{k}}$ and eigenfunctions $\psi_{n\mathbf{k}}(\mathbf{r})$. In its imaginary-time representation, $G^0$ reads
\begin{equation}
    G^0(\mathbf{r},\mathbf{r}';i\tau)=\frac{1}{N_\mathbf{k}}\sum_{n\mathbf{k}}\psi_{n\mathbf{k}}(\mathbf{r})\psi_{n\mathbf{k}}^*(\mathbf{r}')e^{-\epsilon_{n\mathbf{k}}\tau}\left[\Theta(\tau)(1-f_{n\mathbf{k}})-\Theta(-\tau)f_{n\mathbf{k}}\right]\, ,
    \label{eq:G0}
\end{equation}
where $N_\mathbf{k}$ denotes the total number of $\mathbf{k}$ points, $\Theta(\tau)$ is the Heaviside step function, and $f_{n\mathbf{k}}$ is the occupation number of the corresponding KS state. 

The required dynamically screened Coulomb interaction $W^{0(c)}(\mathbf{r},\mathbf{r^{\prime}},\mathrm{i}\tau)$ is transformed from its counterpart in the reciprocal space and imaginary frequency domain, where it is computed as
\begin{equation}
    W^{0(c)}_{\mu\nu}(\mathbf{q},i\omega)=\sum_{\nu'}\left(\varepsilon^{-1}_{\mu\nu'}(\mathbf{q},i\omega)-\delta_{\mu\nu'}\right)V_{\nu'\nu}(\mathbf{q})\, ,
    \label{eq:W}
\end{equation}
where $\varepsilon^{-1}(\mathbf{q},i\omega)$ is the inverse dielectric function, and $V(\mathbf{q})$ is the bare Coulomb matrix. 
In the standard space-time algorithm, the $\mu,\nu$ indices in Eq.~\eqref{eq:W} denote the plane waves $\bfG$, and the Coulomb matrix
$V_{\nu'\nu}(\bfq)=V_{\bfG,\bfGp}(\bfq)=\frac{4\pi}{|\bfq+\bfG|^2}\delta_{\bfG\bfGp}$. However, in the NAO-based space-time algorithm to be
detailed later,  $\mu,\nu$ corresponds to the Bloch-summed atom-centered auxiliary basis functions (ABFs).

Within the random phase approximation (RPA), the dielectric function is given by
\begin{equation}
    \varepsilon_{\mu\nu}(\mathbf{q},i\omega)=\delta_{\mu\nu}-\sum_{\nu'}\chi^0_{\mu\nu'}(\mathbf{q},i\omega)V_{\nu'\nu}(\mathbf{q})\, ,
    \label{eq:epsinv}
\end{equation}
with $\chi^0_{\mu\nu}(\mathbf{q},i\omega)$ being the independent-particle polarizability (density response function) in the representation of
plane waves or ABFs. It is obtained via a transformation of its counterpart in the real space and imaginary time domain
\begin{equation}
    \chi^0(\mathbf{r},\mathbf{r}';i\tau)=G^0(\mathbf{r},\mathbf{r}';i\tau)G^0(\mathbf{r}',\mathbf{r};-i\tau)\, .
    \label{eq:chi0}
\end{equation}

In our NAO-based space-time implementation, the real-space/reciprocal-space transformation is carried out for lattice-vector-dependent ABF matrices $X_{\mu\nu}(\mathbf{R})$, which are transformed to their reciprocal-space counterparts $X_{\mu\nu}(\mathbf{q})$ by discrete Fourier transforms over the  Born-von K\'{a}rm\'{a}n (BvK) supercell with $\mathbf{q}$ sampled in the first BZ. The transformation between the frequency and time domains is evaluated on a discrete set of nonuniform imaginary-time points with associated weights. In this work, we employ the minimax time-frequency grid \cite{kaltak_minimax_2014,kaltak_minimax2_2014,Liu_minimax_2016}, as implemented in the GreenX library\cite{Azizi_greenx_2023,Azizi_minimax_2024}, which provides optimized imaginary-time nodes and weights to accurately perform these transformations.

\subsection{\texorpdfstring{$G^0W^0$}{GW} equations in the NAO-PP representation}
\label{ssec:GW in pp and naos}

In the $G^0W^0$ framework, full consistency between pseudopotential and all-electron treatments is inherently difficult to achieve. In principle, one would need a $G^0W^0$-level pseudopotential, but such a potential does not exist due to the complexity of pseudoizing the $G^0W^0$ self-energy. Even if available, it would break the consistency when used in the underlying KS-DFT calculation \cite{gloze_compendium_2019}. As a result, current $G^0W^0$ workflows rely on well-tested KS-DFT pseudopotentials, with particular care taken to ensure that their unoccupied (scattering) states are accurately described and free of ghost states.

In $G^0W^0$ calculations based on pseudopotentials, the choice of the KS-DFT pseudopotential (PP) is crucial for obtaining accurate quasiparticle energies. A KS-DFT PP that is adequate for ground-state calculations may still lead to significant errors in $G^0W^0$ \cite{Li_gwpp_2012, Gomez_gwpp_2008}. In our implementation, the non-interacting Green's function in Eq.~\eqref{eq:G0} relies on the frozen-core approximation, core-valence partitioning, and pseudo-wavefunctions~\cite{Li_gwpp_2012}. As a result, selecting a reliable KS-DFT PP is essential for achieving consistent and trustworthy $G^0W^0$ results.

Despite its importance, no systematic protocol currently exists for assessing the suitability of a given KS-DFT PP for $G^0W^0$ calculations. To address this issue, we introduce a practical strategy for evaluating and selecting an appropriate KS-DFT PP prior to performing $G^0W^0$ calculations, as discussed in Sec.~\ref{imp:select pp}.

On the other hand, the KS wavefunctions $\psi_{n\mathbf{k}}(\mathbf{r})$ are expanded in terms of a set of localized numerical atom-centered orbitals as
\begin{equation}
    \psi_{n\mathbf{k}}(\mathbf{r})=\sum_{i}c_{i,n}(\mathbf{k})\sum_{\mathbf{R}}\varphi_{i}(\mathbf{r}-\mathbf{R}-\boldsymbol{\tau}_I)e^{i\mathbf{k\cdot R}}\, ,
    \label{eq:Bloch_expansion}
\end{equation}
where $\varphi_{i}(\mathbf{r}-\mathbf{R}-\boldsymbol{\tau}_I)$ denotes the NAO centered on atom $I$ in the cell at lattice vector $\mathbf{R}$, and $c_{i,n}(\mathbf{k})$ are the corresponding expansion coefficients. The composite index $i$ includes the atom index $I$, angular momentum quantum numbers $(l,m)$, and the radial function index $\zeta$. In ABACUS, the NAO basis functions are generated using the CGH scheme \cite{chen_systematically_2010,lin_strategy_2021}, where the radial part is constructed from linear combinations of spherical Bessel functions, enabling systematic convergence by increasing the number of radial functions $\zeta$ for each angular momentum channel. Within the NAO basis framework, the non-interacting Green's function $G_0$ in the imaginary-time domain can be written as
\begin{equation}
    G^0(\mathbf{r},\mathbf{r}^{\prime};\mathrm{i}\tau)=\sum_{i,j}\sum_{\mathbf{R}_1,\mathbf{R}_2}\varphi_i(\mathbf{r}-\mathbf{R}_1-\boldsymbol{\tau}_I)G^{0}_{ij}(\mathbf{R}_2-\mathbf{R}_1,\mathrm{i}\tau)\varphi_j(\mathbf{r}^{\prime}-\mathbf{R}_2-\boldsymbol{\tau}_J).
    \label{eq:G0 in naos}
\end{equation}

Substituting Eq.~\eqref{eq:G0 in naos} into Eq.~\eqref{eq:chi0}, the independent-particle polarizability $\chi^0$ in the real-space and imaginary-time domain can be expressed as
\begin{equation}
    \begin{aligned}
    \chi^0(\mathbf{r},\mathbf{r^{\prime}},\mathrm{i}\tau)=&\sum_{i,j,k,l}\sum_{\mathbf{R}_1,\mathbf{R}_2,\mathbf{R}_3,\mathbf{R}_4}\varphi_i(\mathbf{r}-\mathbf{R}_1-\boldsymbol{\tau}_I)\varphi_k(\mathbf{r}-\mathbf{R}_3-\boldsymbol{\tau}_K)G^{0}_{ij}(\mathbf{R}_2-\mathbf{R}_1,\mathrm{i}\tau)\\
    &\times G^{0}_{lk}(\mathbf{R}_3-\mathbf{R}_4,-\mathrm{i}\tau)\varphi_j(\mathbf{r}^{\prime}-\mathbf{R}_2-\boldsymbol{\tau}_J)\varphi_l(\mathbf{r}^{\prime}-\mathbf{R}_4-\boldsymbol{\tau}_L) \, .
    \label{eq:chi0 in naos}
    \end{aligned}
\end{equation}

The localization of NAOs in real space enables efficient exploitation of sparsity in the $G^0W^0$ formalism. Since the basis functions decay to zero beyond a cutoff radius, the NAO representations of Green's function, polarizability, and self-energy matrices become sparse, with only nearby orbital pairs contributing significantly. This sparsity can be exploited through real-space tensor filtering techniques~\cite{lin_efficient_2021,rpa_shi_2024,Zhang2026arXiv}
to reduce computational cost, as detailed in Sec.~\ref{res:filtering}.

The expressions above show how the polarizability can be represented in the NAO basis, but direct evaluation in real space is computationally demanding due to the four-orbital integrals involved. To overcome this bottleneck and enable an efficient $G^0W^0$ implementation, we employ the LRI technique, which significantly reduces the computational scaling. In the next section, we introduce the auxiliary basis framework for expanding the polarization function, screened Coulomb interaction, and self-energy, forming the foundation of our efficient $G^0W^0$ methodology.

\subsection{Auxiliary basis representation of \texorpdfstring{$G^0W^0$}{GW} equations}
\label{ssec:GW in abfs}
To reduce the computational cost of $G^0W^0$ calculations, we employ the LRI technique \cite{ihrig_accurate_2015,lin_accuracy_2020,Ren_gw_2021,rpa_shi_2024} to expand products of NAO basis functions in terms of a set of ABFs $\{P_{\mu}(\mathbf{r})\}$ as
\begin{equation}\begin{aligned}
    &\varphi_i(\mathbf{r}-\mathbf{R}_1-\boldsymbol{\tau}_I)\varphi_k(\mathbf{r}-\mathbf{R}_3-\boldsymbol{\tau}_K)\\&\boldsymbol{\approx}\sum_{\mu\in I}C_{i(\mathbf{R}_1),k(\mathbf{R}_3)}^{\mu(\mathbf{R}_1)}P_\mu(\mathbf{r}-\mathbf{R}_1-\boldsymbol{\tau}_I)+\sum_{\mu\in K}C_{i(\mathbf{R}_1),k(\mathbf{R}_3)}^{\mu(\mathbf{R}_3)}P_\mu(\mathbf{r}-\mathbf{R}_3-\boldsymbol{\tau}_K)\\&=\sum_{\mu\in I}C_{i(\mathbf{0}),k(\mathbf{R}_{3}-\mathbf{R}_{1})}^{\mu(\mathbf{0})}P_{\mu}(\mathbf{r}-\mathbf{R}_{1}-\boldsymbol{\tau}_{I})+\sum_{\mu\in K}C_{i(\mathbf{R}_{1}-\mathbf{R}_{3}),k(\mathbf{0})}^{\mu(\mathbf{0})}P_{\mu}(\mathbf{r}-\mathbf{R}_{3}-\boldsymbol{\tau}_{K})\mathrm{~.}
    \label{eq:LRI_expansion}
\end{aligned}\end{equation}
The central idea is to build a more compact and linearly independent representation of $\chi^0$ using the ABFs, replacing the highly linearly dependent orbital products $\{\varphi_i(\mathbf{r}-\mathbf{R}_1-\boldsymbol{\tau}_I)\varphi_k(\mathbf{r}-\mathbf{R}_3-\boldsymbol{\tau}_K)\}$. Here, $\mu\in I$ means that the ABF $P_\mu$ is centered on atom $I$, and $\mu\in K$ is defined analogously. This locality is the defining feature of the local RI approximation, in which each orbital product is expanded only in ABFs centered on the two atoms involved.
In this representation, the independent-particle polarizability $\chi^0$ in Eq.~\eqref{eq:chi0 in naos} can be expressed entirely in terms of the ABFs as
\begin{equation}
    \chi^0(\mathbf{r},\mathbf{r}^{\prime},\mathrm{i}\tau)=\sum_{\mu,\nu}\sum_{\mathbf{R}_1,\mathbf{R}_2}P_\mu(\mathbf{r}-\mathbf{R}_1-\boldsymbol{\tau}_{\mu})\chi_{\mu\nu}^0(\mathbf{R}_2-\mathbf{R}_1,\mathrm{i}\tau)P_\nu(\mathbf{r}^{\prime}-\mathbf{R}_2-\boldsymbol{\tau}_{\nu}),
\end{equation}
where $\boldsymbol{\tau}_{\mu}$ and $\boldsymbol{\tau}_{\nu}$ denote the centers of the ABFs $\mu$ and $\nu$, respectively, and the polarizability matrix in the ABF representation is explicitly given by
\begin{align}
  \chi^{0}_{\mu\nu}(\mathbf{R},\mathrm{i}\tau)
  &= \sum_{ij}\sum_{kl}\sum_{\mathbf{R}_1,\mathbf{R}_2}
  C^{\mu(\mathbf{0})}_{i(\mathbf{0}),k(\mathbf{R}_1)}\,
  C^{\nu(\mathbf{R})}_{j(\mathbf{R}),l(\mathbf{R}_2)}
  \nonumber\\
  &\quad\times\Big[
  G^{0}_{ij}(\mathbf{R},\mathrm{i}\tau)\,
  G^{0}_{lk}(\mathbf{R}_1-\mathbf{R}_2,-\mathrm{i}\tau)
  \nonumber\\
  &\qquad
  +G^{0}_{il}(\mathbf{R}_2,\mathrm{i}\tau)\,
  G^{0}_{jk}(\mathbf{R}_1-\mathbf{R},-\mathrm{i}\tau)
  \nonumber\\
  &\qquad
  +G^{0}_{ji}(-\mathbf{R},-\mathrm{i}\tau)\,
  G^{0}_{kl}(\mathbf{R}_2-\mathbf{R}_1,\mathrm{i}\tau)
  \nonumber\\
  &\qquad
  +G^{0}_{li}(-\mathbf{R}_2,-\mathrm{i}\tau)\,
  G^{0}_{kj}(\mathbf{R}-\mathbf{R}_1,\mathrm{i}\tau)
  \Big].
  \label{eq:chi0-mn}
\end{align}

To facilitate the computation of the screened Coulomb interaction, the response function in reciprocal space and frequency domain is obtained by applying Fourier and cosine transformations, given by 
\begin{equation}
    \chi_{\mu\nu}^0(\mathbf{q},\mathrm{i}\omega_k)=\frac{1}{N_\mathbf{q}}\sum_\mathbf{R}e^{\mathrm{i}\mathbf{q}\cdot\mathbf{R}}\sum_j\gamma_{kj}\chi_{\mu\nu}^0(\mathbf{R},\mathrm{i}\tau_j)\cos(\tau_j\omega_k),
    \label{eq:chi0_qw_from_Rtau}
\end{equation}
where $\{\gamma_{kj}\}$ are the quadrature weights associated with the forward minimax cosine transform from imaginary time to imaginary frequency.
The use of the cosine transform owes to the symmetry property $\chi^0(\mathbf{r},\mathbf{r}^{\prime},\mathrm{i}\tau)=\chi^0(\mathbf{r}^{\prime},\mathbf{r},-\mathrm{i}\tau)$, which ensures that the imaginary-time dependence of $\chi^0$ is an even function.
For computational convenience, we work with the symmetrized dielectric function $\tilde{\varepsilon}=v^{-1/2}\varepsilon v^{1/2}$, whose matrix elements are given by
\begin{equation}
    \tilde{\varepsilon}_{\mu\nu}(\mathbf{q},i\omega)=\delta_{\mu,\nu}-\sum_{\alpha\beta}V_{\mu\alpha}^{1/2}(\mathbf{q})\chi^0_{\alpha\beta}(\mathbf{q},i\omega)V_{\beta\nu}^{1/2}(\mathbf{q}),
\end{equation}
where
\begin{equation}
 V_{\mu\nu}(\mathbf{q})=\sum_{\bfR} e^{\mathrm{i}\mathbf{q}\cdot \mathbf{R}} V_{\mu\nu}(\mathbf{R}) = \sum_{\bfR} e^{\mathrm{i}\mathbf{q}\cdot \mathbf{R}} \int d\mathbf{r}d\mathbf{r}^{\prime}\frac{P_\mu^*(\mathbf{r})P_\nu(\mathbf{r}^{\prime}-\mathbf{R})}{|\mathbf{r}-\mathbf{r}^{\prime}|} 
\end{equation}
is the Coulomb matrix in the ABF representation. After inverting the symmetrized dielectric matrix $\tilde{\varepsilon}$, the correlation part of the screened Coulomb interaction is obtained as
\begin{equation}
    W^{0(c)}_{\mu\nu}(\mathbf{q},i\omega)=\sum_{\alpha,\beta}V_{\mu\alpha}^{1/2}(\mathbf{q})(\tilde{\varepsilon}_{\alpha\beta}^{-1}(\mathbf{q},i\omega)-\delta_{\alpha\beta})V_{\beta\nu}^{1/2}(\mathbf{q}).
    \label{eq:Wc in abfs}
\end{equation}
To apply the space-time algorithm for evaluating the self-energy, the screened Coulomb interaction must be transformed from the reciprocal-space and frequency domain into the real-space and imaginary-time domain. Using the symmetry relation $W^{0(c)}_{\mu\nu}(\mathbf{q},\mathrm{i}\omega)=W^{0(c)}_{\mu\nu}(\mathbf{q},-\mathrm{i}\omega)$, this transformation can also be carried out via a cosine transform 
\begin{equation}
    W^{0(c)}_{\mu\nu}(\mathbf{R},\mathrm{i}\tau_j)=\sum_\mathbf{q}e^{-i\mathbf{q}\cdot\mathbf{R}}\sum_k\xi_{jk}W^{0(c)}_{\mu\nu}(\mathbf{q},\mathrm{i}\omega_k)\cos(\tau_j\omega_k),
    \label{eq:Wc_Rtau_from_qw}
\end{equation}
where $\{\xi_{jk}\}$ are the quadrature weights associated with the inverse minimax cosine transform from imaginary frequency back to imaginary time, and are therefore distinct from the forward-transform weights $\{\gamma_{kj}\}$ introduced above.
Finally, the correlation part of the self-energy matrix in the NAO basis representation and the imaginary-time domain can be calculated as
\begin{equation}
	\begin{aligned}
		 & \Sigma^{\rm c}_{ij}(\mathbf{R},\mathrm{i}\tau) = -\sum_{k\mathbf{R}_1}\sum_{l\mathbf{R}_2} G_{kl}^{0}(\mathbf{R}_2-\mathbf{R}_1,\mathrm{i}\tau)                                                                         \\
		 & \times \left[\sum_{\mu\in I}\sum_{\nu\in J} C^{\mu(\mathbf{0})}_{i(\mathbf{0}),k(\mathbf{R}_1)} W_{\mu\nu}^{c}(\mathbf{R},\mathrm{i}\tau) C^{\nu(\mathbf{R})}_{j(\mathbf{R}),l(\mathbf{R}_2)} \right.                   \\
		 & + \,\sum_{\mu\in K}\sum_{\nu\in J} C^{\mu(\mathbf{R}_1)}_{i(\mathbf{0}),k(\mathbf{R}_1)} W_{\mu\nu}^{c}(\mathbf{R}-\mathbf{R}_1,\mathrm{i}\tau) C^{\nu(\mathbf{R})}_{j(\mathbf{R}),l(\mathbf{R}_2)}                     \\
		 & + \,\sum_{\mu\in I}\sum_{\nu\in L} C^{\mu(\mathbf{0})}_{i(\mathbf{0}),k(\mathbf{R}_1)} W_{\mu\nu}^{c}(\mathbf{R}_2,\mathrm{i}\tau) C^{\nu(\mathbf{R}_2)}_{j(\mathbf{R}),l(\mathbf{R}_2)}                                \\
		 & + \left. \,\sum_{\mu\in K}\sum_{\nu\in L} C^{\mu(\mathbf{R}_1)}_{i(\mathbf{0}),k(\mathbf{R}_1)} W_{\mu\nu}^{c}(\mathbf{R}_2-\mathbf{R}_1,\mathrm{i}\tau) C^{\nu(\mathbf{R}_2)}_{j(\mathbf{R}),l(\mathbf{R}_2)} \right].
	\end{aligned}
	\label{eq:sigmac-R}
\end{equation}
Obviously, this LRI-based $G^0W^0$ formulation relies on the accuracy of the LRI expansion [Eq.~\eqref{eq:LRI_expansion}], which in turn 
depends on the size and quality of the ABFs.
Our standard construction~\cite{lin_accuracy_2020} of the auxiliary functions $\{P_{\mu}(\mathbf{r})\}$ produces atom-centered ABFs that naturally adapt to the NAO basis and provide high accuracy for on-site orbital products. Off-site (i.e., $\varphi_i$ and $\varphi_k$ located on different atoms) products, however, are not represented exactly and require a sufficiently large auxiliary basis to achieve good accuracy~\cite{ihrig_accurate_2015}. In our scheme, this can be achieved by augmenting the orbital basis with additional functions used solely to construct the ABFs, resulting in a larger $\{P_{\mu}\}$ set. However, this increases memory and computational demands in $G^0W^0$ calculations and can even lead to numerical instabilities. To deal with this issue, we introduce a novel LRI compression scheme that reduces memory usage while improving both efficiency and accuracy, as will be discussed in Sec.~\ref{imp:compression}.

\subsection{Spin-orbit coupling in \texorpdfstring{$G^0W^0$}{GW}}
\label{ssec: soc}
Spin-orbit coupling (SOC) plays a key role in accurately predicting quasiparticle (QP) energies, particularly in systems containing heavy elements or exhibiting strong spin-momentum-locked states.\cite{aims_soc_2017,SG15}
Standard $G^0W^0$ implementations typically adopt a scalar-relativistic approximation, which cannot capture SOC-induced band splittings or topological features that are critical for many materials. Therefore, incorporating SOC into the $G^0W^0$ formalism is essential for obtaining quantitatively reliable QP corrections.

Within fully relativistic KS-DFT, the KS wavefunctions become two-component spinors, explicitly coupling spatial and spin degrees of freedom. When the SOC is included, the noninteracting Green's function $G^0$ in Eq.~\eqref{eq:G0} is correspondingly modified to reflect the spinor nature of the wavefunctions as
\begin{equation}
    G^{0}_{\sigma \sigma'}(\mathbf{r},\mathbf{r}';i\tau)=\frac{1}{N_\mathbf{k}}\sum_{n\mathbf{k}}\psi_{n\mathbf{k}}(\mathbf{r},\sigma)\psi_{n\mathbf{k}}^*(\mathbf{r}',\sigma')e^{-\epsilon_{n\mathbf{k}}\tau}\left[\Theta(\tau)(1-f_{n\mathbf{k}})-\Theta(-\tau)f_{n\mathbf{k}}\right]\, ,
    \label{eq:G0_soc}
\end{equation}
Here, $\sigma$ and $\sigma'$ are the spin coordinates, and $f_{n\mathbf{k}}$ now represents the occupation of the spinor state. Since spin is no longer a good quantum number, each formerly twofold-degenerate KS state splits into distinct spinor states, and the occupation number $f_{n\mathbf{k}}$ must be assigned to each spinor individually.

Accordingly, the independent-particle polarizability $\chi^0$ in Eq.~\eqref{eq:chi0} is generalized to
\begin{equation}
    \chi^0(\mathbf{r},\mathbf{r}';i\tau)=\sum_{\sigma\sigma'}G^{0}_{\sigma \sigma'}(\mathbf{r},\mathbf{r}';i\tau)G^{0}_{\sigma' \sigma}(\mathbf{r}',\mathbf{r};-i\tau)\, ,
    \label{eq:chi0_soc}
\end{equation}
The screened Coulomb interaction $W^{0(c)}$ in Eq.~\eqref{eq:W} remains unchanged, as it is spin-independent.
Finally, the correlation part of the self-energy in Eq.~\eqref{eq:self-energy} is modified accordingly to account for the spinor nature of the wavefunctions as
\begin{equation}
    \Sigma^c_{n\mathbf{k}}(\mathrm{i}\tau)=-\sum_{\sigma\sigma'}\int\mathrm{d}\mathbf{r}\mathrm{d}\mathbf{r^{\prime}}\psi_{n\mathbf{k}}^*(\mathbf{r},\sigma)G^{0}_{\sigma\sigma'}(\mathbf{r},\mathbf{r^{\prime}},\mathrm{i}\tau)W^{0(c)}(\mathbf{r},\mathbf{r^{\prime}},\mathrm{i}\tau)\psi_{n\mathbf{k}}(\mathbf{r^{\prime}},\sigma').
\end{equation}
Thus, by incorporating SOC into the $G^0W^0$ formalism via spinor wavefunctions and the corresponding modified Green's functions, the effects of SOC on quasiparticle energies and the electronic structure can be naturally captured.

\section{IMPLEMENTATION DETAILS}
\label{sec:implement}

In this section, we describe certain key features of our implementation aimed at enabling efficient and accurate $G^0W^0$ calculations. First, the $\Gamma$-point singularity is properly treated by introducing head and wing corrections to the dielectric function. Second, a novel LRI compression scheme is employed to reduce memory usage while simultaneously improving computational efficiency and accuracy. Finally, we propose a practical strategy for selecting a suitable KS-DFT pseudopotential prior to $G^0W^0$ calculations, further enhancing overall robustness and reliability.

\subsection{\texorpdfstring{The $\Gamma$-point singularity treatment}{The Gamma-point singularity treatment}}
\label{imp:headwing}
 
To address the 1/$q^2$ divergence of the bare Coulomb potential as $\mathbf{q}\rightarrow 0$ in reciprocal space for 3D systems, the dielectric function can be expressed analytically, allowing the $\Gamma$-point contribution to be naturally included in $G^0W^0$ calculations. Similar to the plane-wave basis, the Coulomb matrix in the auxiliary-basis representation exhibits a 1/$q^2$ divergence between two nodeless $s$-type ABFs, and a 1/$q$ divergence between one nodeless $s$-type ABF and one nodeless $p$-type ABF \cite{Ren_gw_2021}.

For semiconductors and insulators, since the dielectric function remains finite in the vicinity of $\mathbf{q}=0$, the screened Coulomb $W_{\mu\nu}(\mathbf{q},i\omega)$ inherits the same divergence as the bare Coulomb $V_{\mu\nu}(\mathbf{q})$ in the limit $\mathbf{q}\rightarrow 0$. To ensure rapid convergence with respect to the $\mathbf{q}$-point sampling, we adopt the truncated Coulomb operator introduced by Spencer and Alavi~\cite{Spencer_coulomb_2008}, which suppresses the long-range tail of the Coulomb potential and strictly localizes it within the BvK supercell. The corresponding truncated screened Coulomb matrix $W_{\mu\nu}^{cut}(\mathbf{q},i\omega)$ is therefore given by
\begin{equation}
    W_{\mu\nu}^\text{cut}(\mathbf{q},i\omega)=\varepsilon^{-1}_{\mu\nu}(\mathbf{q},i\omega)V^\text{cut}_{\mu\nu}(\mathbf{q}).
    \label{eq:truncated W}
\end{equation}

Additionally, we note that the Coulomb singularity also appears in the dielectric function [cf.~Eq.~\eqref{eq:truncated W}], which is however exactly canceled by the asymptotic behavior of the independent-particle polarizability $\chi^0$. Consequently, to leverage this cancellation, the full Coulomb operator should be used when computing the dielectric function. In this case, the dielectric matrix at $\mathbf{q}=0$ should be treated analytically, rather than evaluating the response function $\chi^0(\mathbf{q},i\omega)$ and the full Coulomb matrix $V(\mathbf{q})$ numerically, as is done for $\mathbf{q}\neq 0$ points.

To this end, we follow the commonly used scheme developed in the linearized augmented plane wave (LAPW) framework, where the dielectric function is expressed in terms of the eigenvectors of the Coulomb matrix \cite{Ren_gw_2021,Friedrich_hw_2009,Friedrich_hw_2010,Jiang_fhi-gap_2013}. At $\mathbf{q}=0$, we diagonalize the truncated Coulomb matrix $V^\text{cut}(\mathbf{q}=0)$ in the ABF representation 
\begin{equation}
    \sum_{\nu}V^\text{cut}_{\mu\nu}(\mathbf{q}=0)X_{\nu \lambda}=X_{\mu\lambda}v_\lambda ,
    \label{eq:diagonalize V}
\end{equation}
to obtain the eigenvectors $X_{\mu\lambda}$ and eigenvalues $v_\lambda$. We then express the symmetrized dielectric matrix at $\mathbf{q}$ point in proximity to $\Gamma$ in the basis of the Coulomb matrix eigenvectors ${\mathbf X}_\lambda$, ordered by descending eigenvalues, as
\begin{equation}  \varepsilon_v(\mathbf{q}\to0)=\begin{bmatrix}H(\hat{\mathbf{q}})&\mathbf{w}^\dagger(\hat{\mathbf{q}})\\\mathbf{w}(\hat{\mathbf{q}})&\mathbf{B}\end{bmatrix},
    \label{eq:eps in hw}
\end{equation}
where the head term $H(\hat{\mathbf{q}})$ is a scalar, the wing term $\mathbf{w}(\hat{\mathbf{q}})$ is a column vector of length $n_v-1$, and the body term $\mathbf{B}$ is a $(n_v-1) \times (n_v-1)$ matrix. These components correspond respectively to the $1/q^2$, $1/q$ and regular contributions arising from the full Coulomb matrix. The body term is evaluated in the usual manner and exhibits no dependence on $\mathbf{q}$. In contrast, the head and wing terms are direction-dependent and are computed using $\mathbf{ k\cdot p}$ perturbation theory, reflecting the fact that their values depend on the direction $\mathbf{\hat q}$ along which $\mathbf{q}$ approaches $\Gamma$. Accordingly, the head term is written as a function of the direction $\hat{\mathbf q}$,
\begin{equation}
    \begin{aligned}
    H(\hat{\mathbf{q}})
    &= \sum_{\alpha\beta}\hat q_\alpha H^{\alpha\beta}\hat q_\beta
    = \hat{\mathbf{q}}^\mathsf{T}\mathbf{H}\hat{\mathbf{q}},
    \end{aligned}
\end{equation}
\begin{equation}
    H^{\alpha\beta}=\delta_{\alpha\beta}-\frac{4\pi}{N_{\mathbf{k}}\Omega}\sum_{\mathbf{k}\sigma}\left\{\sum_{n}\frac{f^{\prime}(\epsilon_{n\mathbf{k}\sigma})v_{\alpha,nn,\sigma}^{\mathbf{k}*}v_{\beta,nn,\sigma}^{\mathbf{k}}}{\omega^{2}}+\sum_{m<n}\frac{2(f_{m\mathbf{k}\sigma}-f_{n\mathbf{k}\sigma})v_{\alpha,mn,\sigma}^{\mathbf{k}*}v_{\beta,mn,\sigma}^{\mathbf{k}}}{\left[\left(\epsilon_{m\mathbf{k}\sigma}-\epsilon_{n\mathbf{k}\sigma}\right)^{2}+\omega^{2}\right]\left(\epsilon_{m\mathbf{k}\sigma}-\epsilon_{n\mathbf{k}\sigma}\right)}\right\},
    \label{eq:head}
\end{equation}
and
\begin{equation}
    w_\lambda(\mathbf{q})=\sum_\alpha w_\lambda^\alpha q_\alpha=[\mathbf{wq}]_\lambda,
\end{equation}
\begin{equation}
    \begin{aligned}
    w_{\lambda}^{\alpha}&=-\sqrt{\frac{4\pi}{\Omega}}\frac{2\sqrt{v_{\lambda}}}{N_{\mathbf{k}}}\sum_{\mu}X_{\mu\lambda}^{*}\sum_{\mathbf{k}\sigma}\sum_{m<n}\frac{1}{\left[\epsilon_{n\mathbf{k}\sigma}-\epsilon_{m\mathbf{k}\sigma}\right]^{2}+\omega^{2}}\times\\
    &\left[C_{m,n,\sigma}^\mu(\mathbf{k},\mathbf{k})f_{n\mathbf{k}\sigma}\left(1-f_{m\mathbf{k}\sigma}\right)v_{\alpha,nm,\sigma}^{\mathbf{k}}+C_{m,n,\sigma}^{\mu *}(\mathbf{k},\mathbf{k})f_{m\mathbf{k}\sigma}\left(1-f_{n\mathbf{k}\sigma}\right)v_{\alpha,nm,\sigma}^{\mathbf{k}*}\right]
    \end{aligned},
    \label{eq:wing}
\end{equation}
where $\epsilon_{n\mathbf{k}\sigma}$ and $f_{n\mathbf{k}\sigma}$ denote the KS eigenvalues and occupations for band index $n$ and spin index $\sigma$, $N_{\mathbf{k}}$ and $\Omega$ are the number of $\mathbf{k}$ and cell volume respectively. The quantity $C_{m,n,\sigma}^\mu(\mathbf{k},\mathbf{k})$ is the triple-coefficient tensor in the KS representation, defined as 
\begin{equation}
C_{m,n,\sigma}^\mu(\mathbf{k},\mathbf{k})=\sum_{ij}c_{im,\sigma}^*(\mathbf{k}){C}_{ij}^\mu(\mathbf{k})c_{jn,\sigma}(\mathbf{k}), 
\end{equation}
and $v_{\alpha,mn,\sigma}^{\mathbf{k}}$ is the velocity matrix element with Cartesian index $\alpha$, defined as\cite{Jin_pyatb_2023, Lee_velocity_2018}
    \begin{align}
           v_{\alpha,mn,\sigma}^{\mathbf{k}} 
    =& i\, \bracket{\psi_{m\sigma}^{\mathbf{k}}}{[H^{\mathrm{KS}}, r_\alpha]}{\psi_{n\sigma}^{\mathbf{k}}} \label{eq:commutator_matrix} \\
    =& c_{n}^{\dagger}(\mathbf{k}) \, \partial_{k_\alpha} H^{\mathrm{KS}}(\mathbf{k}) \, c_{m}(\mathbf{k})
     - \epsilon_{n\mathbf{k}} \, c_{n}^{\dagger}(\mathbf{k}) \, \partial_{k_\alpha} S(\mathbf{k}) \, c_{m}(\mathbf{k})  \label{eq:velocity_matrix} \\
    &\quad + i\bigl(\epsilon_{n\mathbf{k}} - \epsilon_{m\mathbf{k}}\bigr)\, 
       c_{n}^{\dagger}(\mathbf{k}) \, \partial_{k_\alpha} A^{R}(\mathbf{k}) \, c_{m}(\mathbf{k}) , 
        \nonumber     
    \end{align}
where $H^{\mathrm{KS}}(\mathbf{k})$ and $S(\mathbf{k})$ are the Kohn-Sham Hamiltonian and overlap matrices in $\mathbf{k}$ space, $A^{R}(\mathbf{k})$ is the matrix of the Fourier-transformed position operator in real space, and $c_{n}(\mathbf{k})$ is the KS eigenvector of band $n$ in $\mathbf{k}$. We denote the Kohn-Sham Hamiltonian by $H^{\mathrm{KS}}(\mathbf{k})$ here to distinguish it from the head term $H(\hat{\mathbf q})$ of the dielectric matrix. 
Equation~\eqref{eq:velocity_matrix} is more convenient in practice in the NAO-based pseudopotential framework, because it can be evaluated through matrix multiplications in a form consistent with the nonorthogonal basis and the nonlocal pseudopotential contribution. In the complete-basis limit, it is formally equivalent to a direct evaluation of the commutator form of the velocity operator [Eq.~\eqref{eq:commutator_matrix}]. Equation~\eqref{eq:velocity_matrix} reduces to the momentum matrix\cite{Ren_gw_2021} $p_{\alpha,mn,\sigma}^{\mathbf{k}}= -i\, \bracket{\psi_{m\sigma}^{\mathbf{k}}}{\nabla}{\psi_{n\sigma}^{\mathbf{k}}}$ only in the all-electron (AE) case or when the nonlocal component of the pseudopotential is absent.

To properly treat the divergence at the $\Gamma$ point, we partition the BZ integral of the self-energy into a small region $\gamma$ enclosing $\Gamma$, and discretize the remaining region on the regular $\mathbf{q}$-point mesh. Following the scheme of Freysoldt et al.\cite{Freysoldt_dielectric_2007}, the $\mathbf{q}$-integral over the $\gamma$ region is performed only for the screened Coulomb matrix, using the dielectric matrix at $\mathbf{q}=0$ expressed in the Coulomb eigenvector representation obtained above. This yields the $\gamma$-averaged inverse dielectric function in the Coulomb eigenvector basis as
\begin{equation}
    \begin{aligned}
        &(\overline{\varepsilon_v^{-1}})_{00}=\frac{1}{V_{q}^{\gamma}}\int_{\gamma}\mathrm{d}\mathbf{q}\frac{1}{\hat{\mathbf{q}}^{\mathsf{T}}\mathbf{L}\hat{\mathbf{q}}}=\frac{1}{V_{q}^{\gamma}}\int\mathrm{d}\hat{\mathbf{q}}\frac{1}{\hat{\mathbf{q}}^{\mathsf{T}}\mathbf{L}\hat{\mathbf{q}}}\int_{0}^{q_{\gamma}(\hat{\mathbf{q}})}\mathrm{d}qq^{2},\\
        &(\overline{\varepsilon_v^{-1}})_{\lambda\neq0,\lambda'\neq0}=B_{\lambda\lambda'}^{-1}+\frac{1}{V_{q}^{\gamma}}\int\mathrm{d}\hat{\mathbf{q}}\frac{1}{\hat{\mathbf{q}}^{\mathsf{T}}\mathbf{L}\hat{\mathbf{q}}}\left[\mathbf{B}^{-1}\mathbf{w}\hat{\mathbf{q}}\right]_{\lambda}\left[\hat{\mathbf{q}}^{\mathsf{T}}\mathbf{w}^{\dagger}\mathbf{B}^{-1}\right]_{\lambda'}\int_{0}^{q_{\gamma}(\hat{\mathbf{q}})}\mathrm{d}qq^{2},\\
        &(\overline{\varepsilon_v^{-1}})_{\lambda\neq0,0}=(\overline{\varepsilon_v^{-1}})_{0,\lambda'\neq0}=0,
    \end{aligned}
\end{equation}
where $V_{q}^{\gamma}$ is the volume of the $\gamma$ region and the tensor $\mathbf{L}$ is defined as
\begin{equation}
    \mathbf{L}=\mathbf{H}-\mathbf{w}^{\dagger}\mathbf{B}^{-1}\mathbf{w}.
\end{equation}
    
After transforming the $\gamma$-averaged inverse dielectric function back into the ABF representation, the truncated screened Coulomb $W_{\mu\nu}^\text{cut}(\mathbf{q},i\omega)$ can be evaluated across the entire BZ. This enables a rapidly convergent computation of the $G^0W^0$ self-energy using standard $\mathbf{q}$-point sampling techniques.

\subsection{The compression scheme of LRI}
\label{imp:compression}
In the conventional LRI scheme, the ABF set $\{P_{\mu}(r), \mu\in N_\text{aux}\}$ is constructed by taking all pairwise products of the standard orbital basis set (OBS) used to expand the KS orbitals in KS-DFT, followed by an orthonormalization procedure \cite{ren_resolution--identity_2012,ihrig_accurate_2015}. This construction naturally adapts the ABFs to the underlying orbital basis and ensures particularly high accuracy for on-site orbital products. In contrast, off-site products are not explicitly included in the ABF construction and therefore cannot be represented exactly. Such RI errors can be quite large in the local RI scheme. To achieve high accuracy in computing integrals involving such off-site contributions, the auxiliary basis needs to be enlarged to $\{P_{\mu}(r), \mu\in N_\text{aux}+N_\text{extra}\}$ by augmenting the OBS with additional higher angular momentum functions (OBS+), which are used solely for generating the ABFs.

However, the ABF set constructed from OBS+ is typically more than twice as large as the one generated from the standard OBS. Since the dynamical response function $\chi^0_{\mu\nu}(\mathbf{q},i\omega)$ expanded in this larger ABF set depends on both $\bfk$ vectors and the frequency, it requires storing a substantial number of large matrices, resulting in significant memory consumption and even numerical instability in subsequent operations such as matrix inversion. To alleviate these issues, we introduce a compression scheme that substantially reduces the size of the ABFs while retaining the accuracy of the LRI expansion.

Formally, a non-orthogonal ABF set represents real-space quantities through the completeness relation
\begin{equation}
    \sum_{\mu,\nu}\sum_{\mathbf{R},\mathbf{R}^{\prime}}\left|P_{\mu\mathbf{R}}\right\rangle\left({\cal S}^{-1}\right)_{\mu\mathbf{R},\nu\mathbf{R}^{\prime}}\langle P_{\nu\mathbf{R}^{\prime}}|=\hat I,
    \label{eq:completeness}
\end{equation}
where ${\cal S}_{\mu\mathbf{R},\nu\mathbf{R}^{\prime}}$ is the overlap matrix between two non-orthogonal ABFs $\mu$ and $\nu$ located at atomic sites $\tau_\mu$ and $\tau_\nu$ in the supercell $\mathbf{R}$ and $\mathbf{R}^{\prime}$, respectively. The overlap matrix element is given by
\begin{equation}
    {\cal S}_{\mu\mathbf{R},\nu\mathbf{R}^{\prime}}=\left\langle P_\mu(\mathbf{r}-\mathbf{R}-\tau_\mu)\right| P_\nu(\mathbf{r}-\mathbf{R}^{\prime}-\tau_\nu)\rangle={\cal S}_{\mu\nu}(\mathbf{R'}-\mathbf{R}).
\end{equation}

To motivate the compression scheme, it is useful to distinguish two different auxiliary-basis representations of the same response function. The standard OBS-generated ABFs $\{P_\rho\}$ provide a compact RI representation of $\chi^0(\mathbf{r},\mathbf{r}',i\omega)$ in real space. In contrast, the enlarged OBS+-generated ABFs $\{P_\mu\}$ are required in the LRI construction stage to reduce the local-RI error for interatomic AO products. Because the OBS+ set is much larger, the corresponding response matrix in the enhanced ABF representation becomes the main memory bottleneck.

Once $\chi^0(\mathbf{r},\mathbf{r}',i\omega)$ has been constructed accurately, however, the RI and LRI matrices are simply two representations of the same real-space response function. For a non-periodic system, one may therefore write
\begin{equation}
    \chi^0(\mathbf{r},\mathbf{r}',i\omega)
    = \sum_{\rho\lambda}
    P_\rho(\mathbf{r})\,
    \chi^{0,\mathrm{RI}}_{\rho\lambda}(i\omega)\,
    P_\lambda(\mathbf{r}')
    = \sum_{\mu\nu}
    P_\mu(\mathbf{r})\,
    \chi^{0,\mathrm{LRI}}_{\mu\nu}(i\omega)\,
    P_\nu(\mathbf{r}').
\end{equation}
Here $\chi^{0,\mathrm{RI}}$ denotes the response matrix in the standard OBS-generated ABFs, while $\chi^{0,\mathrm{LRI}}$ denotes the response matrix in the enlarged OBS+-generated ABFs. Projecting the equality above onto the standard ABFs, i.e., multiplying it from the left by $\langle P_\rho^{\mathrm{std}}|$ and from the right by $|P_\lambda^{\mathrm{std}}\rangle$, gives
\begin{equation}
    \begin{aligned}
    \chi^{0,\mathrm{proj}}_{\rho\lambda}(i\omega)
    &\equiv
    \langle P^{\mathrm{std}}_\rho|\chi^0(i\omega)|P^{\mathrm{std}}_\lambda\rangle \\
    &=
    \sum_{\rho'\lambda'}
    ({\cal S}^{\mathrm{std,std}})_{\rho\rho'}\,
    \chi^{0,\mathrm{RI}}_{\rho'\lambda'}(i\omega)\,
    ({\cal S}^{\mathrm{std,std}})_{\lambda'\lambda} \\
    &=
    \sum_{\mu\nu}
    ({\cal S}^{\mathrm{std,enh}})_{\rho\mu}\,
    \chi^{0,\mathrm{LRI}}_{\mu\nu}(i\omega)\,
    ({\cal S}^{\mathrm{enh,std}})_{\nu\lambda}.
    \end{aligned}
\end{equation}
Here $({\cal S}^{\mathrm{std,std}})_{\rho\lambda}=\langle P^{\mathrm{std}}_\rho|P^{\mathrm{std}}_\lambda\rangle$ is the overlap matrix within the standard ABFs. The mixed overlaps are $({\cal S}^{\mathrm{std,enh}})_{\rho\mu}=\langle P^{\mathrm{std}}_\rho|P^{\mathrm{enh}}_\mu\rangle$ and $({\cal S}^{\mathrm{enh,std}})_{\nu\lambda}=\left[({\cal S}^{\mathrm{std,enh}})^\dagger\right]_{\nu\lambda}$. Multiplying the inverse of $({\cal S}^{\mathrm{std,std}})$ from both sides then yields the compressed response matrix in the compact RI representation,
\begin{equation}
    \chi^{0,\mathrm{RI}}_{\rho\lambda}(i\omega)
    = \sum_{\rho'\lambda'\mu\nu}
    \left(({\cal S}^{\mathrm{std,std}})^{-1}\right)_{\rho\rho'}
    ({\cal S}^{\mathrm{std,enh}})_{\rho'\mu}\,
    \chi^{0,\mathrm{LRI}}_{\mu\nu}(i\omega)\,
    ({\cal S}^{\mathrm{enh,std}})_{\nu\lambda'}\,
    \left(({\cal S}^{\mathrm{std,std}})^{-1}\right)_{\lambda'\lambda}.
\label{eq:chi0_compression}
\end{equation}

This equation is the basis of our compression scheme: the enhanced OBS+-generated ABFs are used only to construct the accurate LRI response matrix $\chi^{0,\mathrm{LRI}}$, while the subsequent storage, inversion, and frequency-domain manipulations are carried out in the compact RI representation $\chi^{0,\mathrm{RI}}$. The transformation in periodic systems follows the same logic and is summarized in Appendix~\ref{ap:periodic compression}. In our implementation, the OBS+ is generated by augmenting the OBS with an additional set of higher angular momentum functions, typically including $f$- and $g$-type components. The on-site ABF set is then constructed using principal component analysis (PCA) with a controllable threshold \cite{lin_accuracy_2020}, and the compressed ABF set is obtained by applying PCA to the OBS with the same threshold.

It is worth noting that the inverse of the standard-basis overlap matrix, $\left(({\cal S}^{\mathrm{std,std}})^{-1}\right)_{\rho\rho'}$ in Eq.~\eqref{eq:chi0_compression}, may become ill-conditioned due to the near-linear dependence among the ABF functions constructed from the OBS. To avoid numerical instability, we employ the pseudo-inverse of this overlap matrix by discarding eigenvalues smaller than a prescribed threshold (e.g., $10^{-3}$) in its eigen-decomposition. We have also tested Tikhonov regularization \cite{tikhonov_ill_posed_1995}, which yields similar results. 

Algorithm~\ref{alg:compression} summarizes the workflow of the compression scheme for the LRI approach in our $G^0W^0$ calculations. The key difference from the conventional LRI scheme \cite{rpa_shi_2024,Zhang2026arXiv} is that the response function $\chi^0_{\mu\nu}(\mathbf{q},i\tau)$ at each imaginary-time point is first computed in the enhanced ABF set constructed from OBS+, and then projected onto the standard (small) ABF set constructed from OBS. Conversely, the screened Coulomb matrix $W^c_{\mu\nu}(\mathbf{q},i\tau)$ at each imaginary-time point is upfolded back to the enhanced ABFs when evaluating the correction self-energy $\Sigma^c_{ij}(R, i\tau)$. 
As a result, the peak memory usage is determined by the larger of (i) $\chi^0_{\mu\nu}(\mathbf{q},i\tau)$ stored for a single imaginary-time point in the large ABFs representation and (ii) $\chi^0_{\rho\lambda}(\mathbf{q},i\tau)$ stored for all imaginary-time points in the small ABF representation. This scheme significantly reduces memory consumption while simultaneously improving computational efficiency.
   
Additionally, the Coulomb matrix $V_{\mu\nu}(\mathbf{q})$ in the ABF representation only needs to be stored in the small ABF set constructed from OBS when computing the exact-exchange and screened Coulomb matrices, which further reduces memory usage and I/O operations. Moreover, evaluating the inverse of the dielectric matrix in Eq.~\eqref{eq:Wc in abfs} within the small ABF representation is more stable and efficient than in the large ABF set, yielding accurate and smooth quasiparticle band structures, particularly for high-lying virtual states.

\begin{algorithm}[t]
    \caption{New framework: compression scheme of LRI in NAO-based space-time $G^0W^0$ calculations. 
    }
    \label{alg:compression}
    \KwIn{$\chi^0_{\mu\nu}(R, i\tau)$}
    \KwOut{$\Sigma^c_{ij}(R, i\tau)$}
    
    \ForEach{$\tau$}{
        Compute $\chi^0_{\mu\nu}(R, i\tau)$ for all $R$\;
        Fourier transform: $\chi^0_{\mu\nu}(R, i\tau) \rightarrow \chi^0_{\mu\nu}(q, i\tau)$\;
        Compress basis: $\chi^0_{\mu\nu}(q, i\tau) \rightarrow \chi^0_{\rho\lambda}(q, i\tau)$\;
        Free $\chi^0_{\mu\nu}(q, i\tau)$\;
    }
    
    Gather $\chi^0_{\rho\lambda}(q, i\tau)$ and Fourier transform: $\chi^0_{\rho\lambda}(q, i\tau) \rightarrow \chi^0_{\rho\lambda}(q, i\omega)$\;
    Compute $W^{0(c)}_{\rho\lambda}(q, i\omega)$\;
    Fourier transform: $W^{0(c)}_{\rho\lambda}(q, i\omega) \rightarrow W^{0(c)}_{\rho\lambda}(q, i\tau)$\;
    \ForEach{$\tau$}{
        Unfold basis: $W^{0(c)}_{\rho\lambda}(q, i\tau) \rightarrow W^{0(c)}_{\mu\nu}(q, i\tau)$\;
        Fourier transform: $W^{0(c)}_{\mu\nu}(q, i\tau) \rightarrow W^{0(c)}_{\mu\nu}(R, i\tau)$\;
        Free $W^{0(c)}_{\mu\nu}(q, i\tau)$\;
        Compute $\Sigma^c_{ij}(R, i\tau)$\;
        Free $W^{0(c)}_{\mu\nu}(R, i\tau)$\;
    }
    
\end{algorithm}

\subsection{Selection of a suitable KS-DFT pseudopotential}

\label{imp:select pp}
In $G^0W^0$ calculations based on pseudopotentials, the quality of the underlying KS-DFT pseudopotential (PP) directly affects the frozen-core Green's function in Eq.~\eqref{eq:G0} and, consequently, the resulting quasiparticle energies. A PP that is adequate for ground-state KS-DFT may still be unreliable for $G^0W^0$ \cite{Li_gwpp_2012, Gomez_gwpp_2008}. Since $G^0W^0$ is far more demanding than KS-DFT and there is currently no systematic low-cost protocol for assessing a KS-DFT PP in advance, we propose here a practical pre-screening strategy before performing the full $G^0W^0$ calculation.

It is worth noting that the accuracy of $G^0W^0$ calculations is fundamentally controlled by the quality of the dynamic response function $\chi^0(\mathbf{q},i\omega)$, which determines the screened Coulomb interaction $W^0$ and hence the quasiparticle self-energy. Because core electrons are frozen at the mean-field level and thus do not contribute to density fluctuations $\chi^0$ in pseudopotential-based calculations, the choice of the pseudopotential directly affects its accuracy. As a result, the macroscopic dielectric function, which is determined by $\chi^0$, provides a direct and practical measure for assessing whether a given KS-DFT pseudopotential is suitable for reliable $G^0W^0$ calculations or not. It can be obtained by
\begin{equation}
    \varepsilon_M(i\omega)=\lim_{\mathbf{q}\to0}\varepsilon_{00}(\mathbf{q},i\omega)=\hat{\mathbf{q}}^\mathsf{T}\mathbf{H}(i\omega)\hat{\mathbf{q}},
\end{equation}
which corresponds to the $\mathbf{q}$-averaged head element of the microscopic dielectric matrix and can be evaluated directly from the KS-DFT eigensystems using Eq.~\eqref{eq:head}.
Therefore, we can employ the frequency dependence of the macroscopic dielectric constant as a criterion for selecting a suitable KS-DFT PP for $G^0W^0$ calculations.

The physical reason why the low-frequency part of the macroscopic dielectric function is most sensitive to the pseudopotential choice can be understood directly from Eq.~\eqref{eq:head}. In the head term, each contribution is weighted by the factor $\left[(\epsilon_{m\mathbf{k}\sigma}-\epsilon_{n\mathbf{k}\sigma})^2+\omega^2\right]^{-1}$. At low imaginary frequency, transitions with relatively small excitation energies dominate the summation, and these are precisely the transitions that are most sensitive to the treatment of semicore states in the pseudopotential construction. As $\omega$ increases, the denominator suppresses these differences, so the influence of the pseudopotential choice becomes much weaker. This explains why the low-frequency behavior of $\varepsilon_M(\omega)$ provides a sensitive criterion for selecting pseudopotentials for $G^0W^0$ calculations.

\begin{figure}[htbp]
    \centering
    \includegraphics[width=0.9\textwidth]{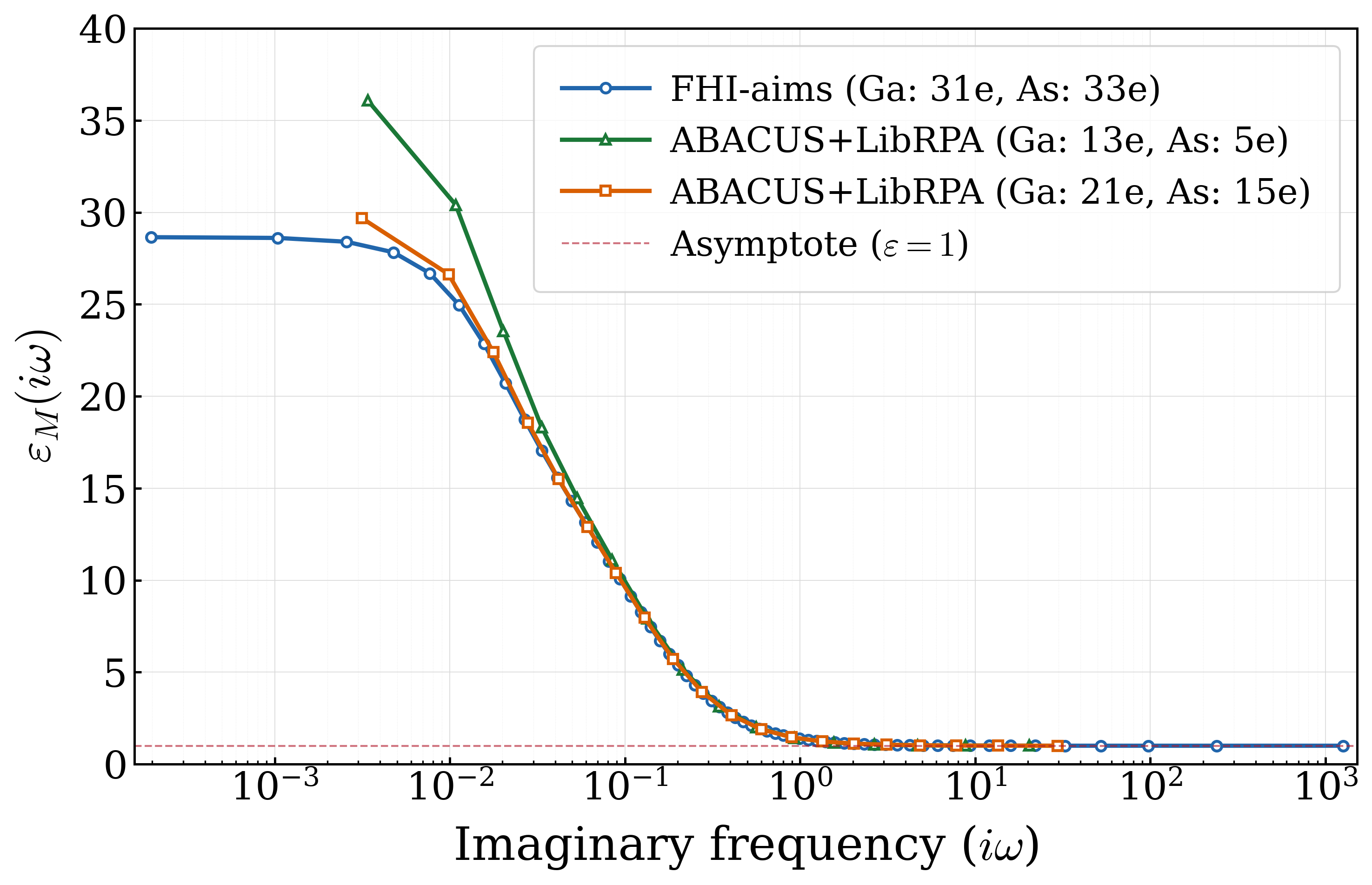}
    \caption{Macroscopic dielectric constant of GaAs computed with two Dojo pseudopotentials, the standard and stringent versions, as a function of imaginary frequency. All-electron results obtained with FHI-aims are included for comparison.}
    \label{fig:head_GaAs}
\end{figure}

Using GaAs as a representative semiconductor consisting of heavy elements, Fig.~\ref{fig:head_GaAs} presents the macroscopic dielectric constant as a function of imaginary frequency, computed with two Dojo pseudopotentials: the standard and the stringent versions \cite{van_setten_pseudodojo_2018}.
It is clear that pseudopotentials containing more valence electrons, such as Ga~($3s^2 3p^6 3d^{10} 4s^2 4p^1$, 21~electrons) and As~($3d^{10} 4s^2 4p^3$, 15~electrons) Dojo potentials, yield dielectric constants that closely follow the all-electron reference obtained from FHI-aims. In contrast, pseudopotentials with fewer valence electrons --- Ga~($3d^{10} 4s^2 4p^1$, 13~electrons) and As~($4s^2 4p^3$, 5~electrons) --- show pronounced deviations. This demonstrates that including more semicore states in the valence space improves the description of the independent-particle response function and thus provides a more reliable foundation for subsequent $G^0W^0$ calculations.

\begin{figure}[htbp]
    \centering
    \includegraphics[width=0.9\textwidth]{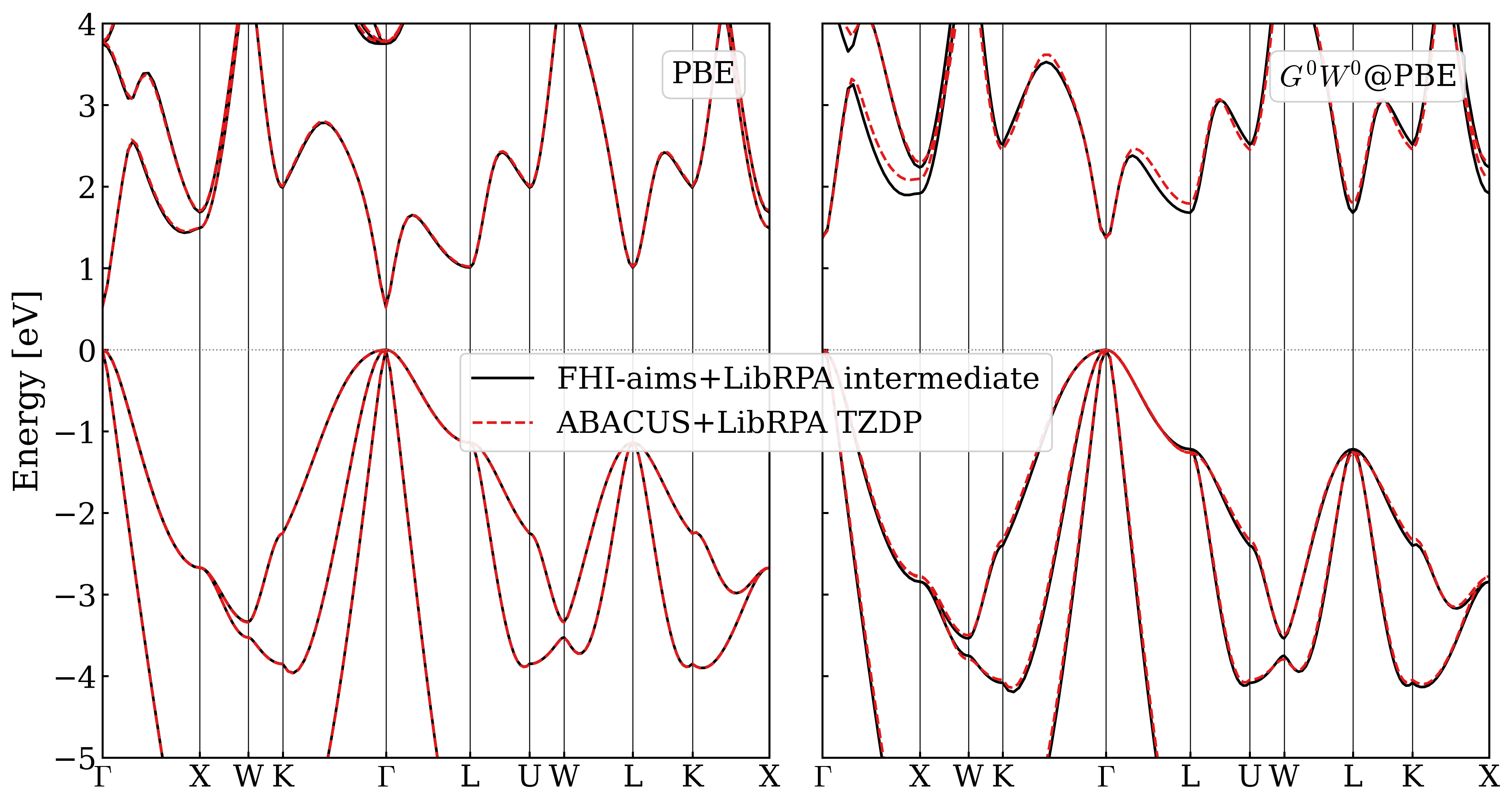}
    \caption{$G^0W^0$ quasiparticle band structure of GaAs computed using Dojo pseudopotentials with Ga~($3s^2 3p^6 3d^{10} 4s^2 4p^1$) and As~($3d^{10} 4s^2 4p^3$). All-electron results obtained with FHI-aims are included for comparison.
    }
    \label{fig:band_GaAs}
\end{figure}

Consistently, our explicit $G^0W^0$ benchmarks confirm that only the Ga~($3s^2 3p^6 3d^{10} 4s^2 4p^1$) and As~($3d^{10} 4s^2 4p^3$) pseudopotentials reproduce the all-electron quasiparticle band structure with high fidelity, as shown in Fig.~\ref{fig:band_GaAs}. For completeness, a broader comparison of the $G^0W^0$ band structures obtained with different Dojo pseudopotentials is provided in Appendix~\ref{ap:pp_comparison}.
These observations validate that the frequency dependence of the macroscopic dielectric function serves as an effective and inexpensive criterion for identifying suitable KS-DFT pseudopotentials prior to performing $G^0W^0$ calculations.

\section{RESULTS AND DISCUSSIONS}
\label{sec:result}

We have implemented the algorithms described above in the LibRPA code \cite{rpa_shi_2024,librpa_shi_2025,Zhang2026arXiv} integrated with ABACUS \cite{chen_systematically_2010,li_large-scale_2016,lin_ab_2024}, enabling efficient $G^0W^0$ calculations. In this section, we first examine the computational efficiency of the compression scheme and real-space tensor filtering strategies, then present systematic convergence tests, and finally benchmark our implementation for a set of prototypical semiconductors against the results obtained from VASP and FHI-aims.

\subsection{Efficiency benchmark}
We start by assessing the efficiency of our implementation, highlighting the impact of the LRI compression scheme and real-space tensor filtering.

\subsubsection{Compression scheme of LRI}
\label{res:compression}
To demonstrate the effectiveness of the compression scheme proposed in Section~\ref{imp:compression}, we perform $G^0W^0$ calculations on MgO using both the conventional LRI scheme and the new compression scheme. A triple-zeta double-polarized (TZDP) basis set of the second generation developed for the ABACUS code\cite{lin_strategy_2021} is employed for all elements, with the $\bfk$-point mesh set to $8\times 8\times 8$. The lattice constant is fixed at 4.211 \AA.

\begin{figure}[htbp]
    \centering
    \includegraphics[width=0.9\textwidth]{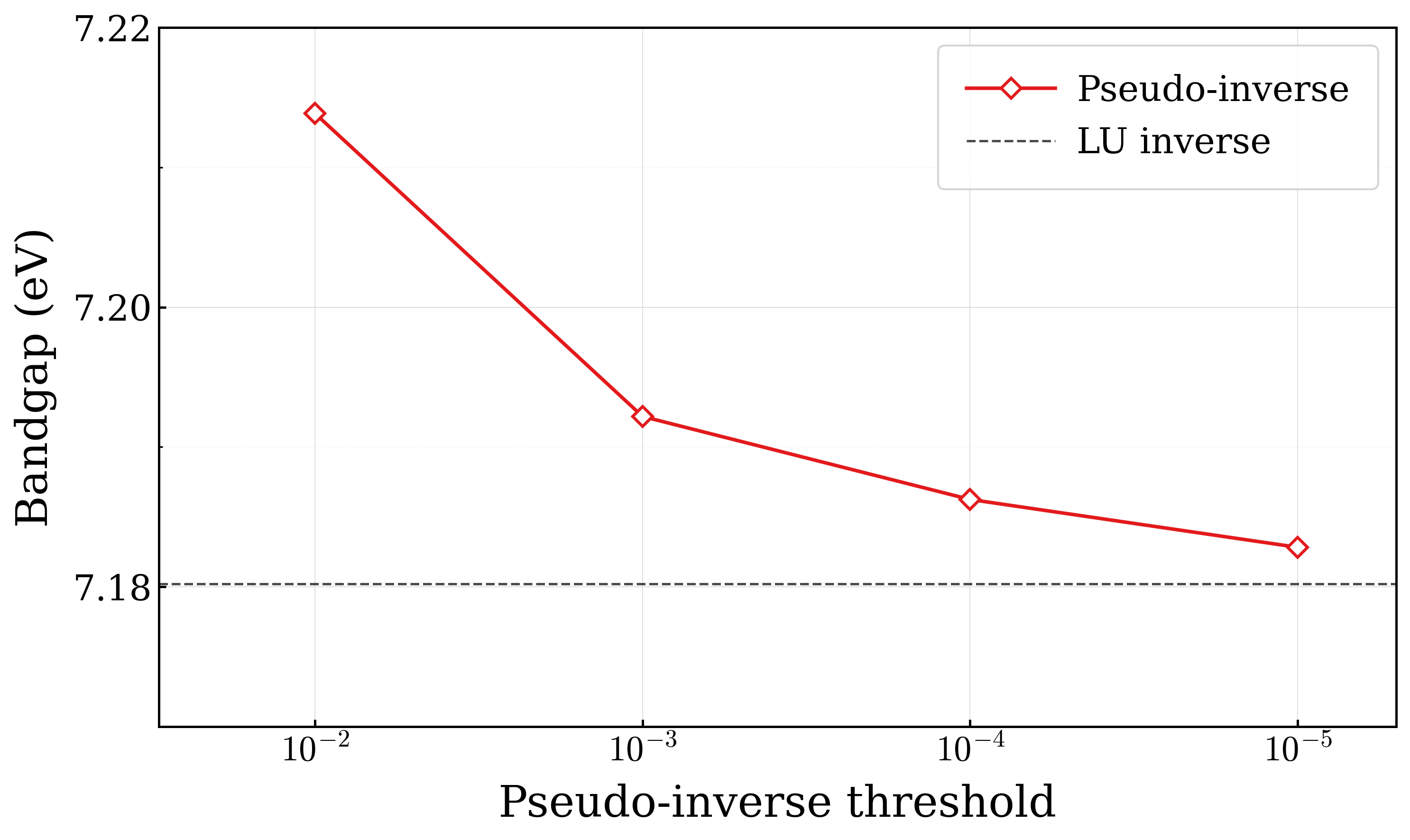}
    \caption{$G^0W^0$@PBE band gap of MgO as a function of the pseudo-inverse threshold used for the inversion of the overlap matrix $({\cal S}^{\mathrm{std,std}})_{\rho\rho'}$ in the compression transformation. The horizontal line denotes the result obtained from direct LU-based inversion of the same overlap matrix.}
    \label{fig:MgO_gap_inv_thr}
\end{figure}

Figure~\ref{fig:MgO_gap_inv_thr} shows the convergence of the $G^0W^0$ band gap of MgO with respect to the pseudo-inverse threshold used for the inversion of the overlap matrix $({\cal S}^{\mathrm{std,std}})_{\rho\rho'}$ entering the compression transformation. As the threshold decreases, the band gap approaches the value obtained from a direct LU-based inversion of the same overlap matrix, i.e., without discarding small eigenvalues. While the LU decomposition can become numerically unstable for heavily linearly dependent basis sets, the pseudo-inverse method ensures robust and reliable results in our calculations. A threshold of $10^{-5}$ yields a band gap of 7.183 eV, in close agreement with the LU inversion result of 7.180 eV, demonstrating the convergence and reliability of the inversion procedure within the compression scheme.

\begin{figure}[htbp]
    \centering
    \includegraphics[width=0.9\textwidth]{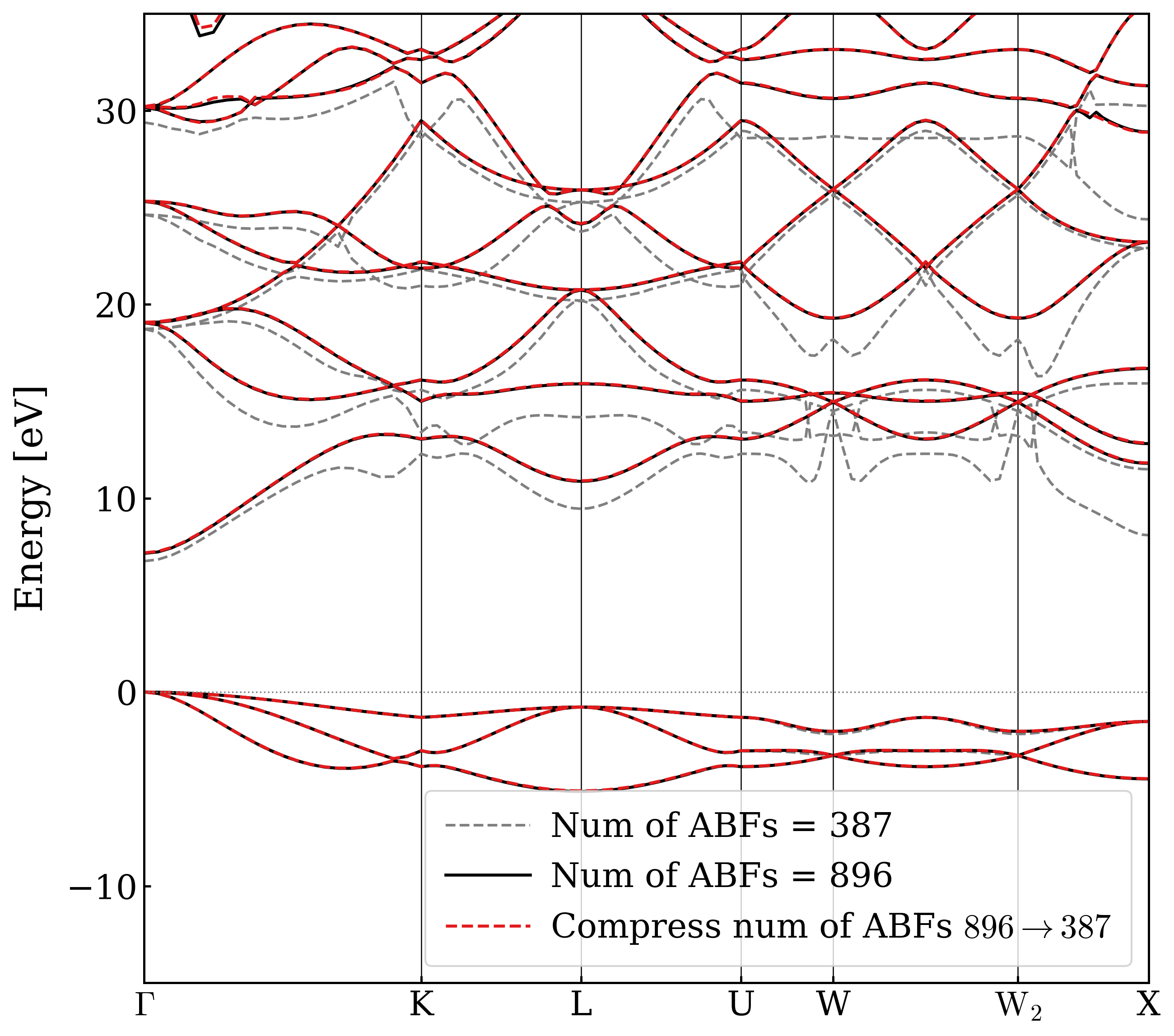}
    \caption{Comparison of the $G^0W^0$@PBE band structures of MgO obtained with the compression scheme and with the conventional LRI scheme using the small and enhanced ABF sets. The gray dashed curve labeled ``Num of ABFs = 387'' denotes the conventional LRI calculation with the small ABF set, the black solid curve labeled ``Num of ABFs = 896'' denotes the conventional LRI calculation with the enhanced ABF set, and the red dashed curve labeled ``Shrink num of ABFs 896$\rightarrow$387'' denotes the present compression scheme in which the enlarged 896-dimensional ABF space is compressed to the standard 387-dimensional one. }
    \label{fig:compression_MgO}
\end{figure}

Building on this convergence, Fig.~\ref{fig:compression_MgO} presents a comparison between LRI schemes with and without compression for the $G^0W^0$ band structure of MgO. While the original LRI scheme with a small ABF set (constructed from the OBS, see Sec.~\ref{imp:compression}) suffers from inaccuracies in the conduction band, the compression scheme with a pseudo-inverse threshold of $10^{-5}$ produces results in excellent agreement with the original LRI scheme using an enhanced large ABF set. As such, the compression scheme can achieve comparable accuracy while significantly reducing memory and computational requirements. As summarized in Table~\ref{tab:abf-comparison}, the compression scheme lowers memory consumption from 80.7 GB to approximately 55.8 GB (a 31\

Interestingly, we find that for certain systems such as MgO, the LRI error is much less pronounced for $\chi^0$ than for the self-energy, allowing one to compute $\chi^0$ exclusively with the small ABF set without appreciable loss of accuracy (Fig.~\ref{fig:MgO_chi0_comparison}). 
This observation is particularly valuable for SOC-$G^0W^0$ calculations, where the non-SOC $G^0W^0$ result can first be used to assess whether $\chi^0$ compression is necessary before carrying out the more expensive SOC calculation. A more detailed analysis of this behavior is presented in Appendix~\ref{ap:chi0-compression}.

These results clearly demonstrate that the compression scheme is an effective approach for enhancing both the efficiency and accuracy of LRI-based $G^0W^0$ calculations, particularly for large systems where memory usage is a critical bottleneck.

\begin{table}[!ht]
    \centering
    \caption{Performance and accuracy comparison between the compression scheme and the conventional LRI scheme for MgO. In the wall time column, values in parentheses denote the time required for $\chi^0$ calculation.}
    \label{tab:abf-comparison}
    \renewcommand{\arraystretch}{1.4}

    \begin{threeparttable}
        \begin{tabular}{cccc}
            \toprule
            Number of ABFs & Band gap (eV) & Wall time (h) & Max memory (GB) \\
            \midrule
            387 & 6.754 & 3.8 & 21.7 \\
            896 & 7.210 & 19.9 & 80.7 \\
            $896 \rightarrow 387^{\mathrm{a}}$ & 7.183 & 11.3 (0.9) & 55.8 \\
            $896 \rightarrow 387^{\mathrm{b}}$ & 7.154 & 10.6 (0.2) & 32.7 \\
            \bottomrule
        \end{tabular}

        \begin{tablenotes}
            \footnotesize
            \item[a] $\chi^0(R,i\tau)$ is calculated using the large (896) auxiliary basis set.
            \item[b] $\chi^0(R,i\tau)$ is calculated using the small (387) auxiliary basis set.
        \end{tablenotes}
    \end{threeparttable}
\end{table}

\subsubsection{\texorpdfstring{Filtering of real-space tensors}{Filtering of real-space tensors}}
\label{res:filtering}
Due to the spatial locality of NAOs and the corresponding real-space representation of the quantities entering the $G^0W^0$ workflow, the sparsity of the relevant atom-pair tensor blocks can be exploited to accelerate the computation. Following Ref.~\citenum{Zhang2026arXiv}, we apply a block-wise filtering algorithm to the dominant real-space tensors and discard the entire block if its maximum absolute element is smaller than a prescribed threshold. In the present subsection, we focus on the practical choice of these thresholds and their impact on the accuracy and efficiency of the calculation.

The relevant filtered quantities are the response function $\chi^0$, the exact-exchange operator $\Sigma^{\mathrm{x}}$, and the correlation self-energy $\Sigma^{\mathrm{c}}$. Each target quantity involves its own set of input quantities, as discussed in Sec.~\ref{ssec:GW in abfs}. In particular, the RI coefficients appear in all three quantities; the Green's function is involved in the evaluation of $\chi^0$ and $\Sigma^{\mathrm{c}}$; the density matrix and bare Coulomb matrix are relevant to $\Sigma^{\mathrm{x}}$; and the screened Coulomb matrix enters $\Sigma^{\mathrm{c}}$.

\begin{figure}[htbp]
    \centering

    \hspace*{\fill}%
    \begin{subfigure}[b]{0.32\textwidth}
        \centering
        \includegraphics[width=\textwidth]{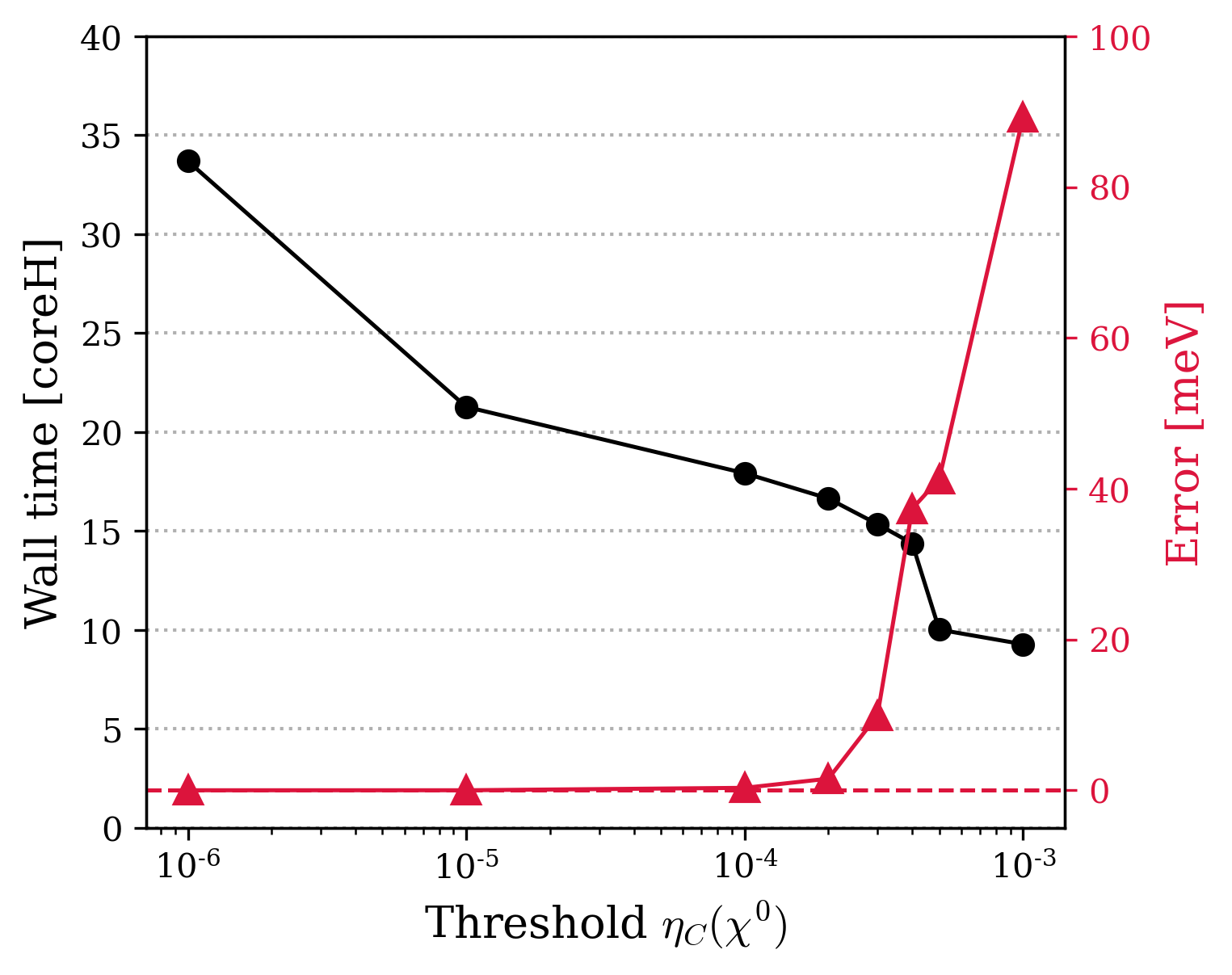}
        \subcaption{RI coefficients for $\chi^0$}
        \label{fig:threshold_chi_C}
    \end{subfigure}%
    \hfill%
    \begin{subfigure}[b]{0.32\textwidth}
        \centering
        \includegraphics[width=\textwidth]{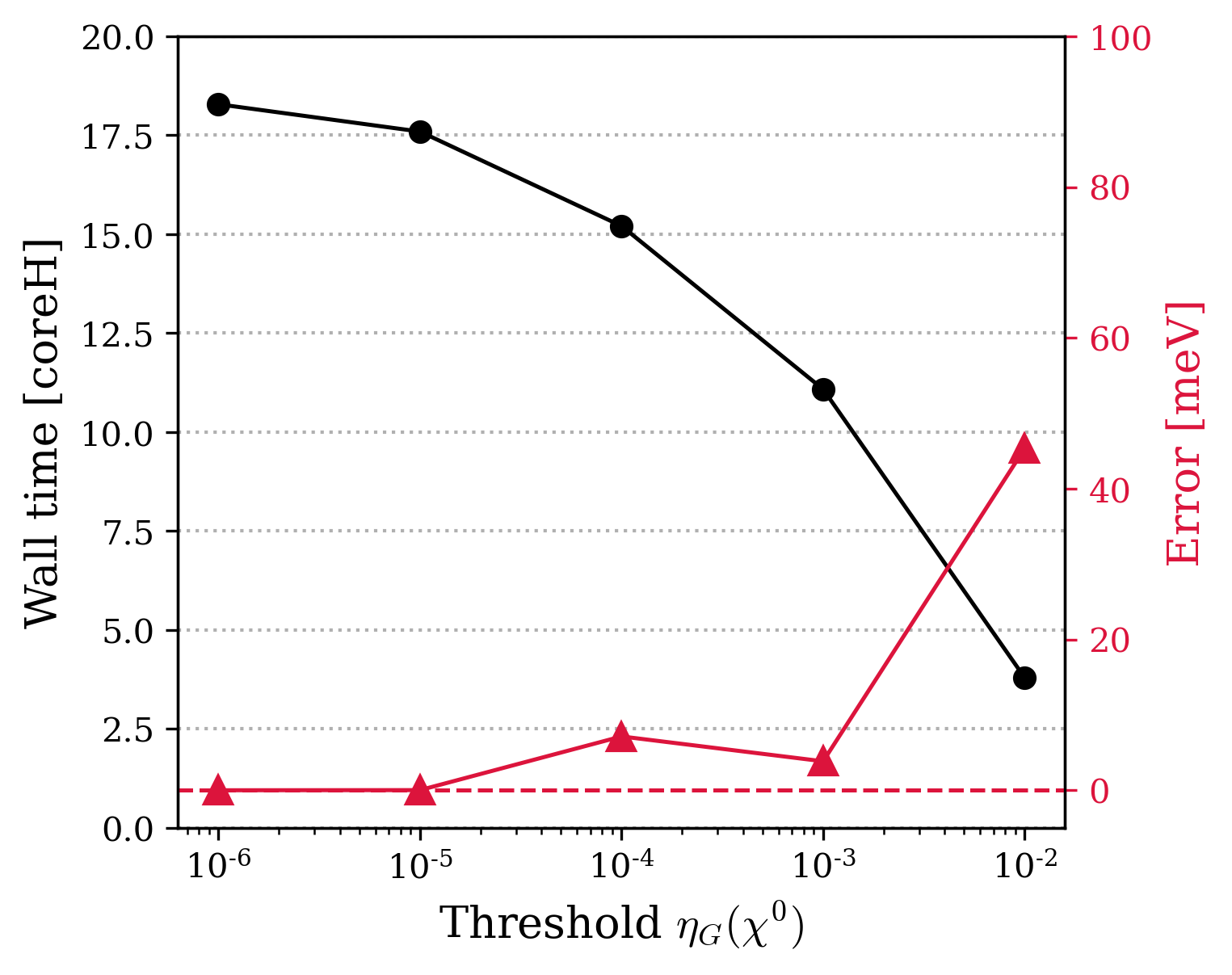}
        \subcaption{Green's function for $\chi^0$}
        \label{fig:threshold_chi_G}
    \end{subfigure}%
    \hspace*{\fill}

    \bigskip

    \begin{subfigure}[b]{0.32\textwidth}
        \centering
        \includegraphics[width=\textwidth]{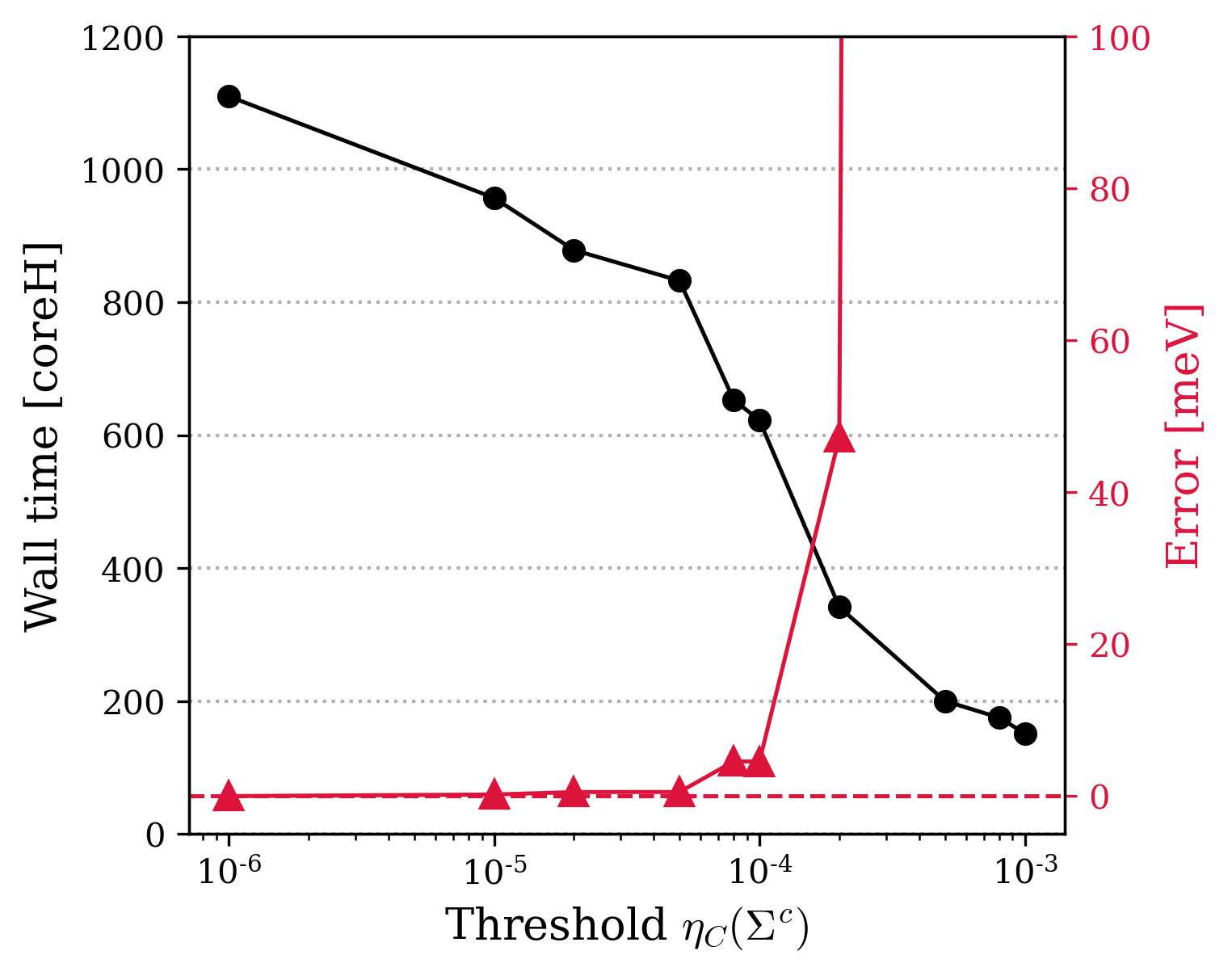}
        \subcaption{RI coefficients for $\Sigma^c$}
        \label{fig:threshold_sig_C}
    \end{subfigure}
    \hfill
    \begin{subfigure}[b]{0.32\textwidth}
        \centering
        \includegraphics[width=\textwidth]{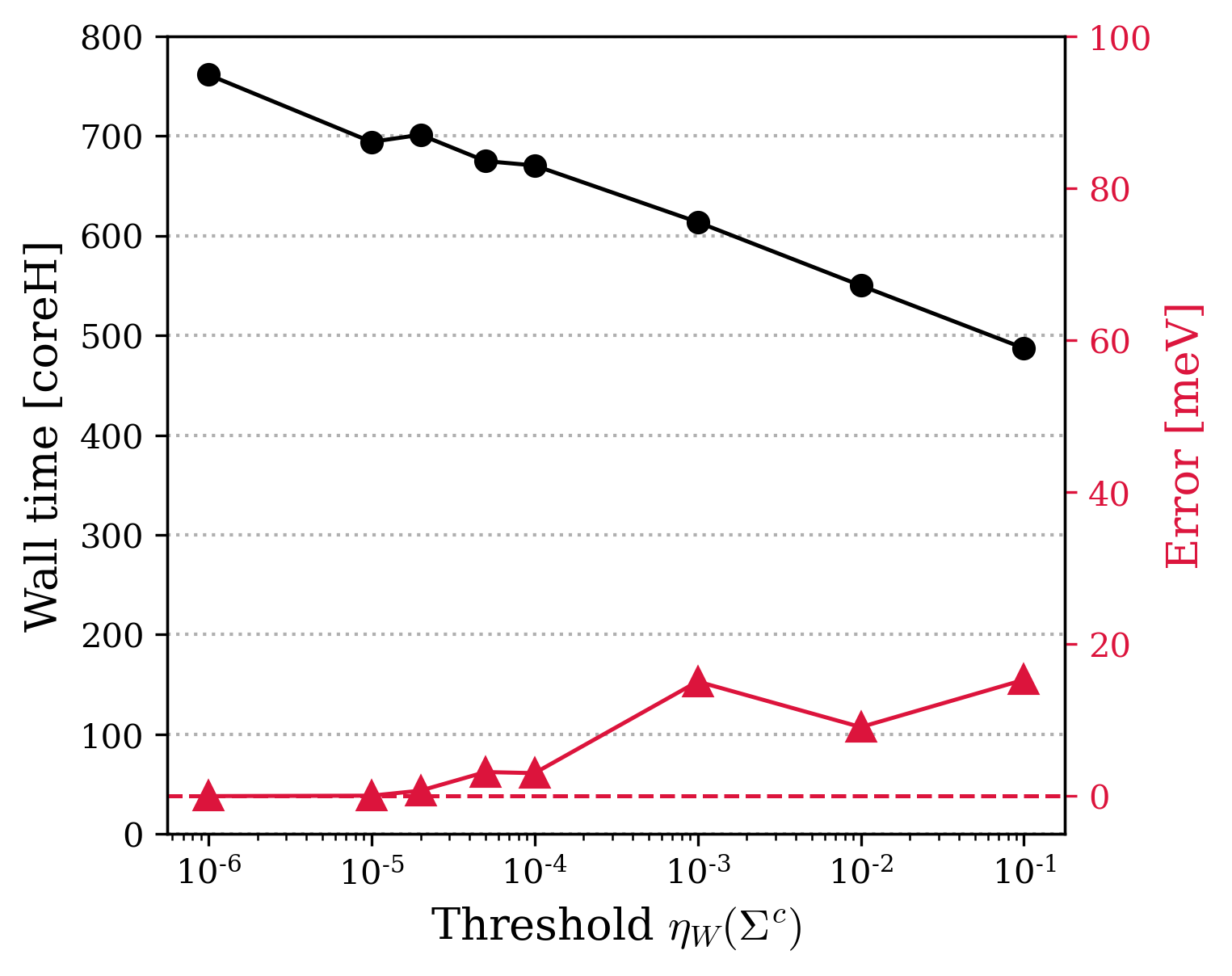}
        \subcaption{Screened Coulomb $W$ for $\Sigma^c$}
        \label{fig:threshold_sig_W}
    \end{subfigure}
    \hfill
    \begin{subfigure}[b]{0.32\textwidth}
        \centering
        \includegraphics[width=\textwidth]{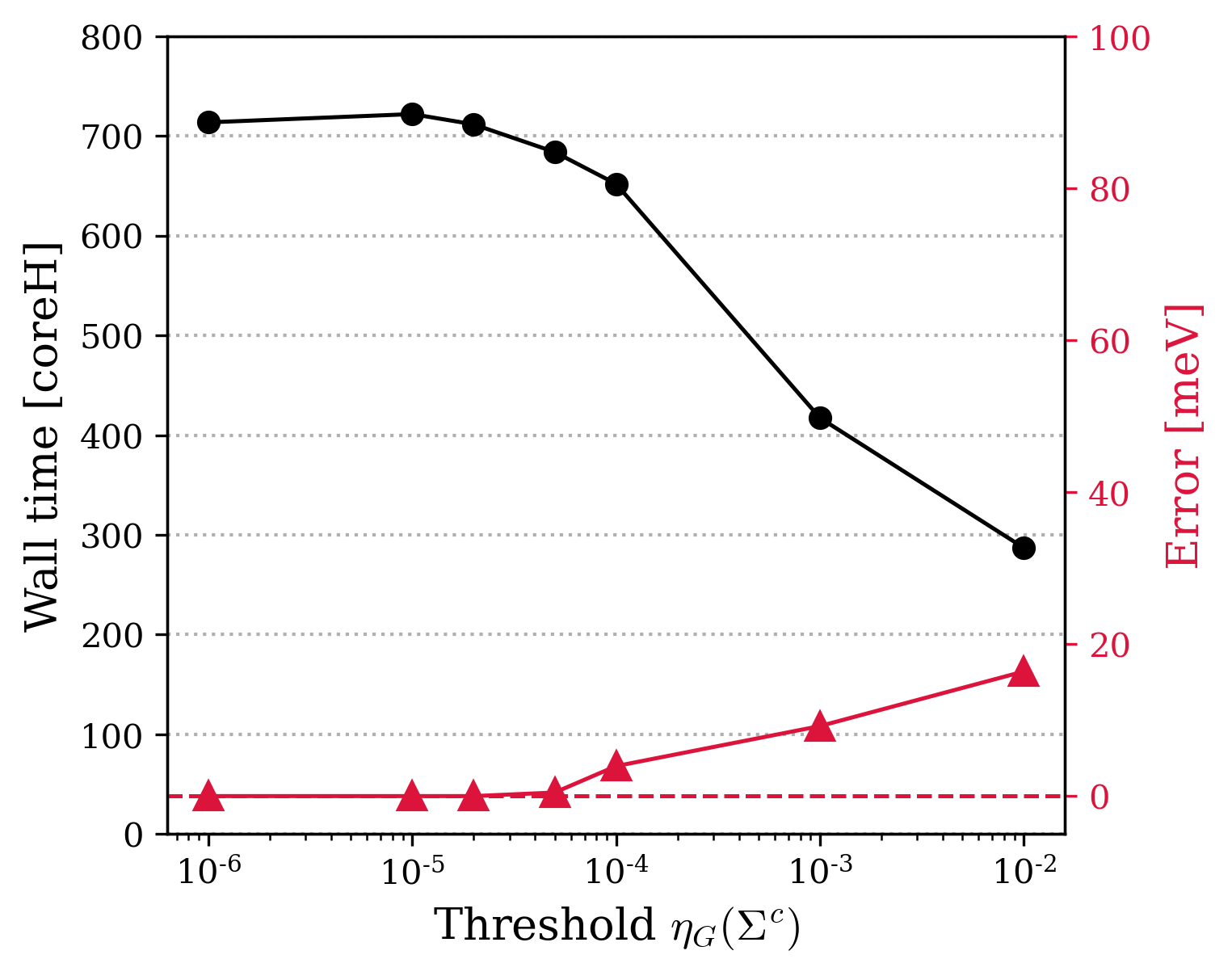}
        \subcaption{Green's function for $\Sigma^c$}
        \label{fig:threshold_sig_G}
    \end{subfigure}

    \caption{Wall time (black, in core hours) and error (red, in meV) in the fundamental band gap of bulk cubic BN computed by $G^0W^0$@PBE with respect to the filtering thresholds of (a) RI coefficients for response function $\eta_C(\chi^0)$, (b) Green's function for response function $\eta_G(\chi^0)$, (c) RI coefficients for the self-energy $\eta_C(\Sigma^c)$, (d) screened Coulomb matrix for self-energy $\eta_W(\Sigma^c)$, and (e) Green's function for self-energy $\eta_G(\Sigma^c)$. All calculations are performed with MPI+OpenMP parallelization (4 MPI tasks, each with 32 threads) on 4 Intel(R) Xeon(R) Platinum 6162 CPU nodes (32 cores on each node).}
    \label{fig:threshold_all}
\end{figure}

Since the computational cost of the exact-exchange operator is significantly smaller than that of the response function and the correlation self-energy, and the corresponding thresholds have already been studied in previous works on hybrid functionals~\cite{lin_efficient_2021}, we focus here on the filtering of the response function and the correlation part of the self-energy. For a detailed discussion of the filtering algorithm, we refer the reader to Ref.~\citenum{Zhang2026arXiv}. As demonstrated there, the block-wise filtering algorithm drives the two dominant steps in large-scale $G^0W^0$ calculations, namely the evaluations of $\chi^0$ and $\Sigma^{\mathrm{c}}$, toward quadratic scaling with system size. This is because the number of tensor blocks that survive the filtering tends to saturate as the system becomes larger instead of increasing proportionally with the full system size. Here we use BN as an example to demonstrate the practical impact of the filtering thresholds. The results are shown in Fig.~\ref{fig:threshold_all}.

Each subfigure in Fig.~\ref{fig:threshold_all} presents the wall time (in core hours) and the error (in meV) in the fundamental band gap of bulk cubic BN computed by $G^0W^0$@PBE with respect to the filtering threshold of a specific component. Figures~\ref{fig:threshold_chi_C} and \ref{fig:threshold_chi_G} show the results for the response function, while Figs.~\ref{fig:threshold_sig_C}, \ref{fig:threshold_sig_W}, and \ref{fig:threshold_sig_G} correspond to the correlation self-energy. All calculations are performed with MPI+OpenMP parallelization (4 MPI tasks, each with 32 threads) on 4 Intel(R) Xeon(R) Platinum 6162 CPU nodes (32 cores on each node). A comprehensive investigation of filtering thresholds reveals an optimal configuration that achieves an excellent balance between accuracy and computational efficiency, cutting the wall time by 50\

\subsection{Accuracy benchmark results}
\label{res:benchmark}
In this section, we compare our $G^0W^0$ results with those obtained from the plane-wave projector-augmented-wave (PAW) method implemented in the VASP code \cite{vasp_gw_2006,vasp_gw_2014}.
and the all-electron $\bfk$-space method implemented in the FHI-aims code \cite{Ren_gw_2021}.

\subsubsection{Convergence study}
\label{res:3d convergence}
Using Si as an example, we systematically examine the convergence behavior of the quasiparticle (QP) band gap with respect to five key computational parameters: (i)  cutoff radius of basis functions $r_{\text{cut}}$, (ii) $\bfk$-point sampling density, (iii) size of the auxiliary basis set, (iv) size of the NAO basis set, and (v) number of minimax points for frequency/time transformation. All KS-DFT calculations employ the PBE exchange-correlation functional with norm-conserving pseudopotentials (four valence electrons for Si). The lattice constant of
Si is fixed at its experimental value: 5.431~{\AA}.
The convergence tests are carried out separately for each of these five parameters.
For each test, only one parameter is varied while all others are held fixed at the following default values: 
\begin{itemize}
    \item \textbf{$\bfk$-point sampling:} $8\times8\times8$ Monkhorst-Pack grid.
    \item \textbf{Basis set cutoff:} $r_{\text{cut}} = 8.0$ Bohr.
    \item \textbf{Auxiliary basis:} Generated via the OBS+1f1g scheme (labeled "for\_aux 1f1g").
    \item \textbf{Orbital basis set:} Triple-zeta double-polarized (TZDP).
    \item \textbf{Time/frequency grid:} 16 minimax points.
\end{itemize}
The quasiparticle band gap considered here is the indirect $\Gamma$-$X$ gap, i.e., $E_{\mathrm{CBM}}(X)-E_{\mathrm{VBM}}(\Gamma)$.

\begin{figure}[htbp]
    \centering
    \includegraphics[width=0.9\textwidth]{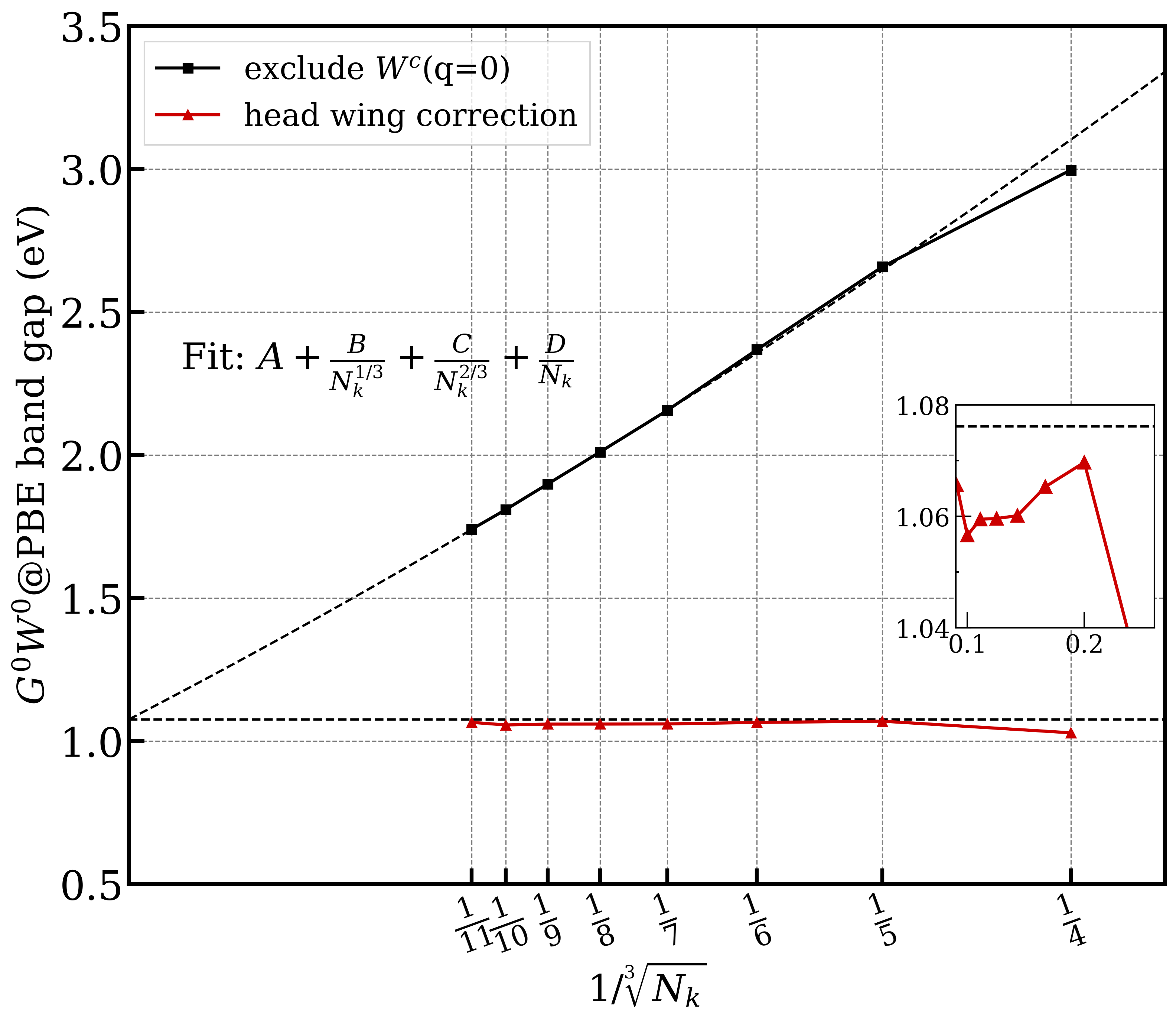}
    \caption{Convergence test of the $G^0W^0$@PBE band gap of Si with respect to $\bfk$-point sampling. The black squares denote calculations in which the $W^c(\mathbf{q}=0)$ contribution is excluded from the Brillouin-zone summation of $\Sigma$, while the red triangles include the head/wing correction.}
    \label{fig:k_convg}
\end{figure}

\textbf{$\bfk$-point sampling.}
The convergence behavior with respect to $\bfk$-point sampling is shown in Fig.~\ref{fig:k_convg} by comparing the calculations with the head/wing corrections and those in which the $W^c(\mathbf{q}=0)$ contribution is excluded from the Brillouin-zone summation of $\Sigma$.
For the latter case, the band gap varies approximately linearly with respect to $N_k^{-1/3}$ and remains rather far from convergence on the meshes considered here.
A cubic fit to the four densest meshes ($8\times8\times8$ to $11\times11\times11$) yields an extrapolated gap of 1.076 eV in the dense $\bfk$-point limit ($N_k^{-1/3}\rightarrow 0$).
In contrast, with the head/wing corrections included, the convergence is dramatically accelerated and the band gap is already essentially converged on the coarsest $4\times4\times4$ mesh, with variations below 0.01 eV from $k=6$ onward.
The comparison shows that explicitly treating the $\mathbf{q}\rightarrow 0$ head and wing terms is crucial for obtaining reliable $\bfk$-point convergence at modest computational cost.

\textbf{Cutoff radius of atomic basis sets.}
 The NAO basis sets are strictly localized in real space, which facilitates the design of low-scaling algorithms. The computational cost can be substantially lowered by reducing the cutoff radii of the atomic basis functions, but this may lead to a sacrifice in accuracy. Hence, it is crucial to determine the suitable cutoff radius for the NAOs in both DFT and $G^0W^0$ calculations.

\begin{figure}[htbp]
    \centering
    \begin{subfigure}[b]{0.48\textwidth}
        \centering
        \includegraphics[width=\textwidth]{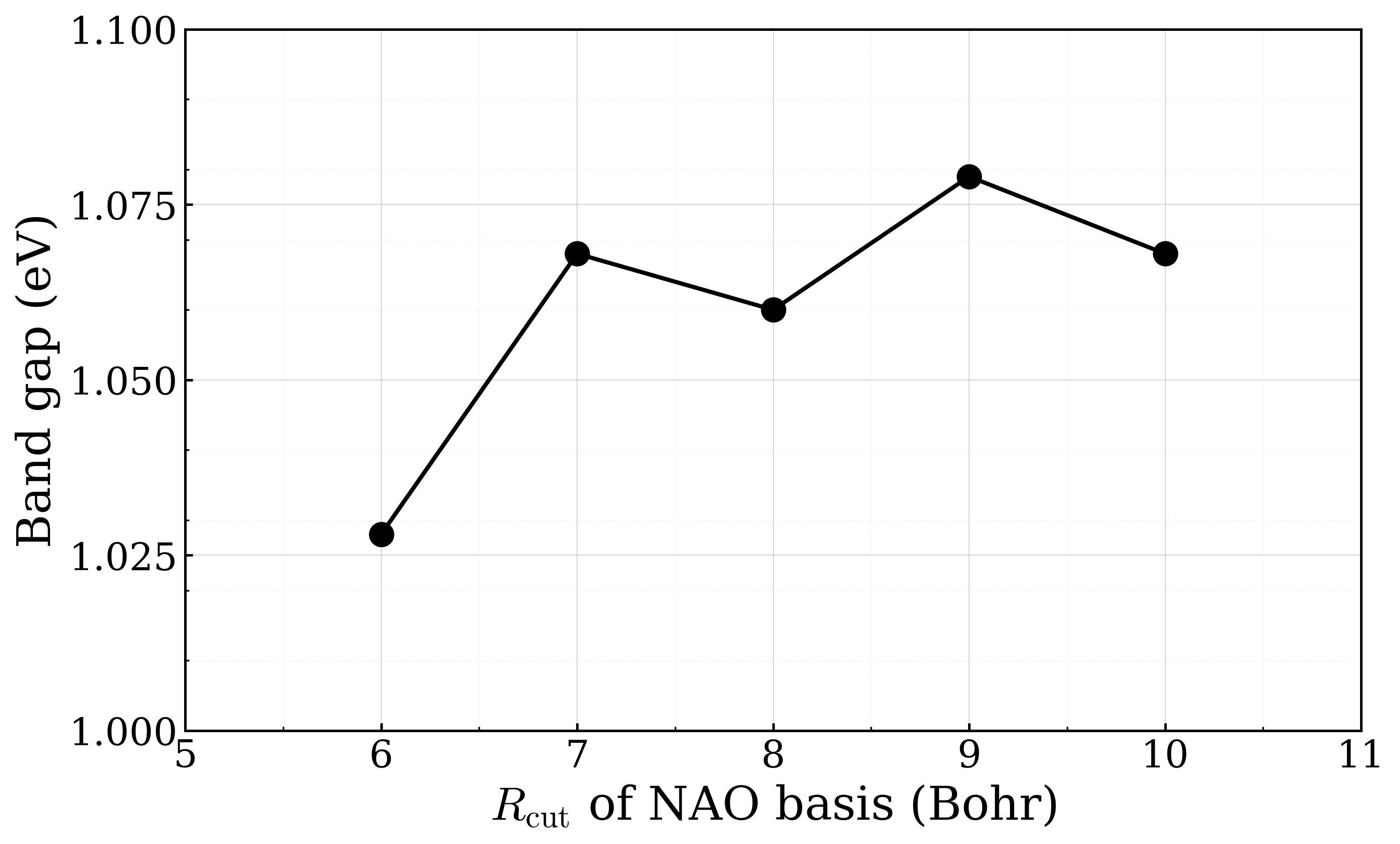}
        \subcaption{Basis cutoff radius}
        \label{fig:rcut_convg}
    \end{subfigure}
    \hfill
    \begin{subfigure}[b]{0.48\textwidth}
        \centering
        \includegraphics[width=\textwidth]{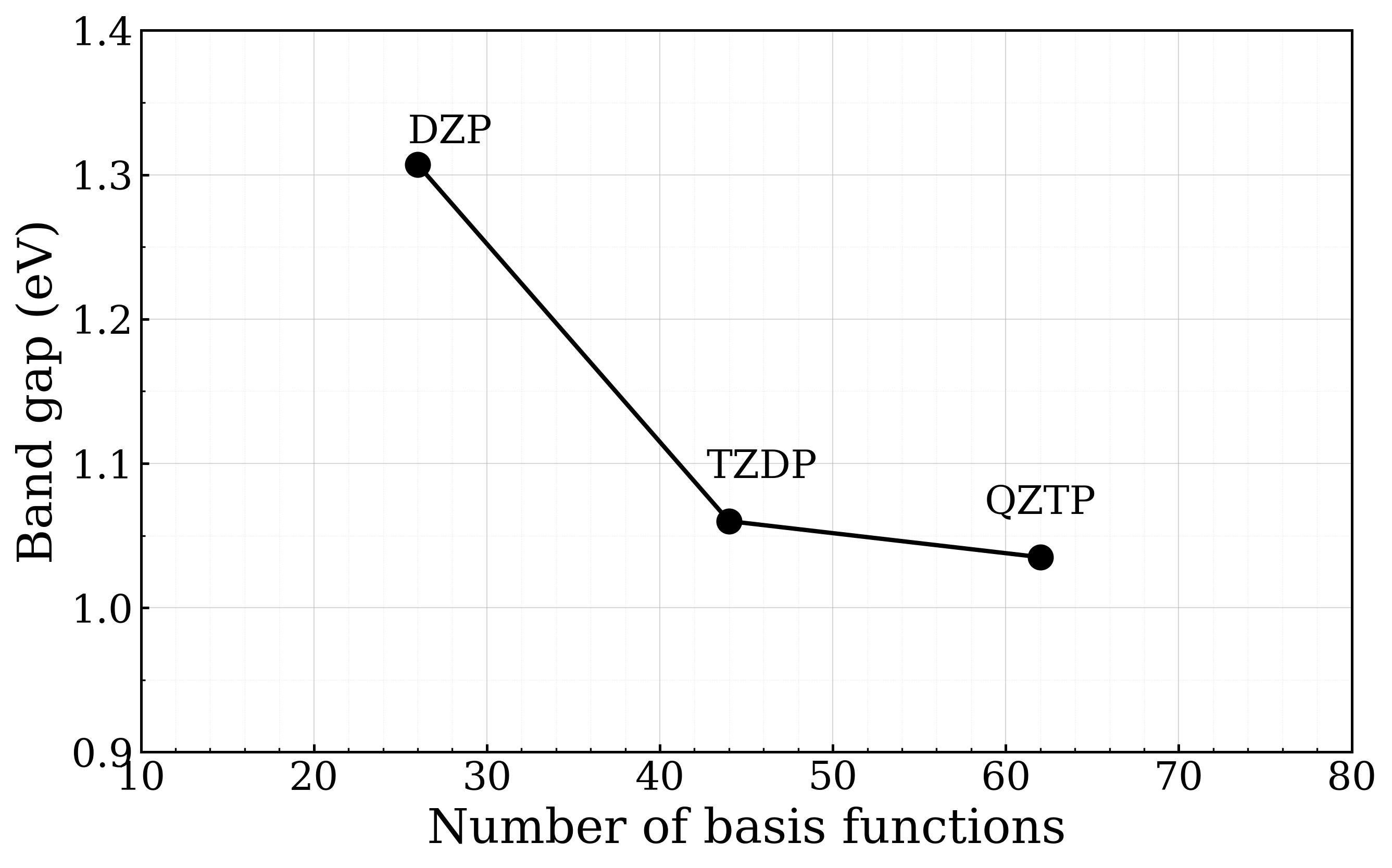}
        \subcaption{NAO basis}
        \label{fig:basis_convg}
    \end{subfigure}

    \bigskip

    \begin{subfigure}[b]{0.48\textwidth}
        \centering
        \includegraphics[width=\textwidth]{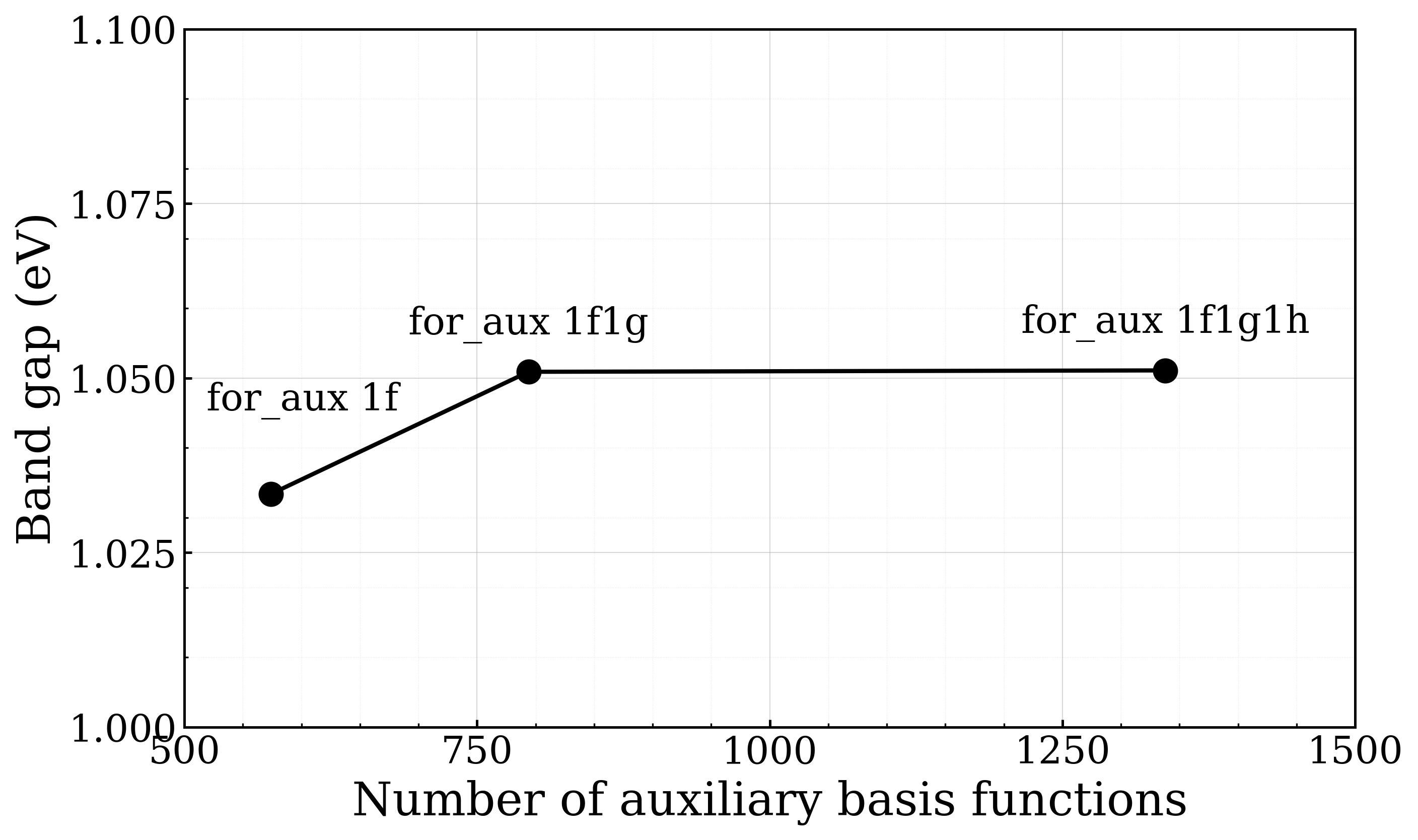}
        \subcaption{ABFs}
        \label{fig:abfs_convg}
    \end{subfigure}
    \hfill
    \begin{subfigure}[b]{0.48\textwidth}
        \centering
        \includegraphics[width=\textwidth]{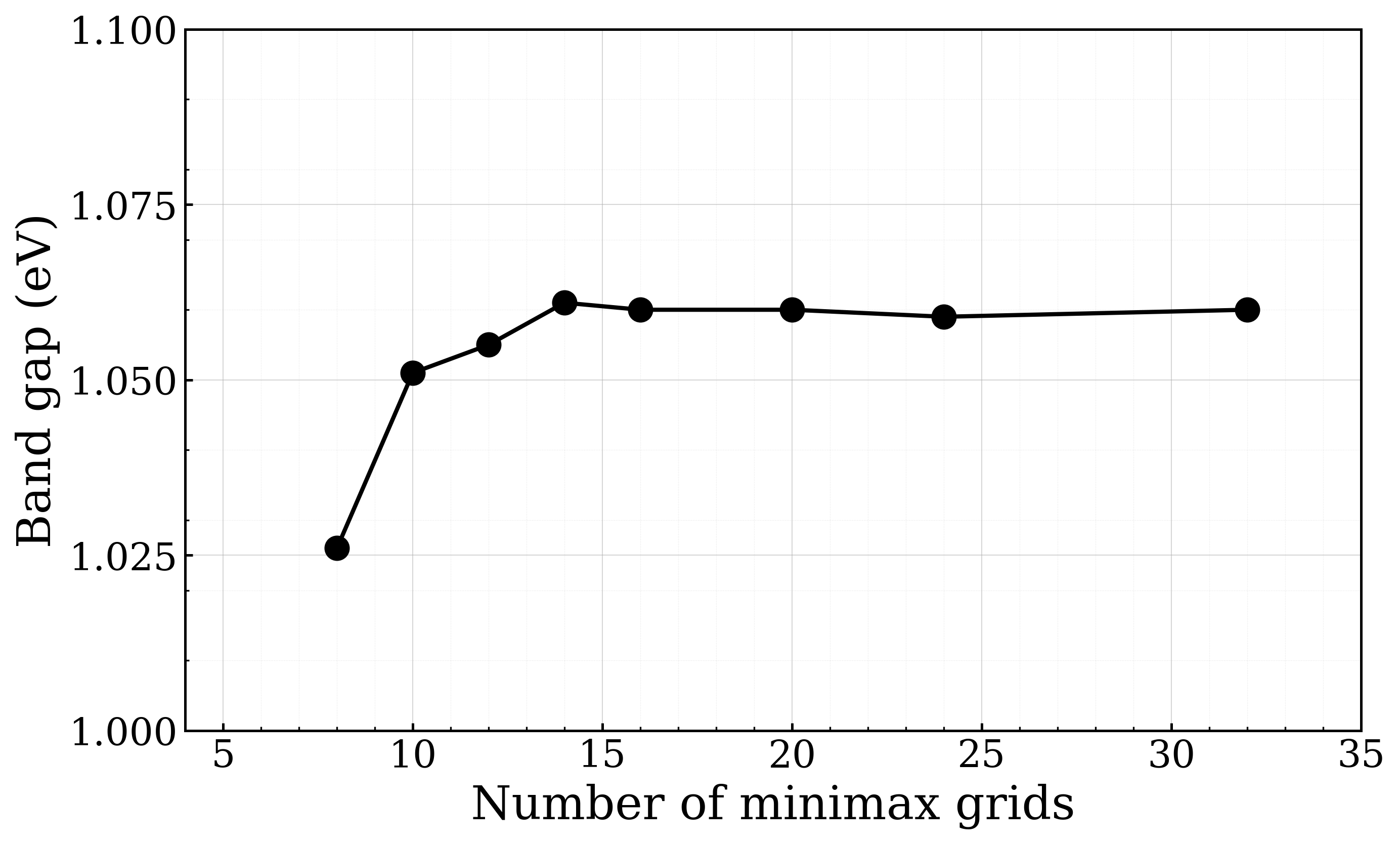}
        \subcaption{Minimax points}
        \label{fig:grid_convg}
    \end{subfigure}

    \caption{Convergence tests of the $G^0W^0$@PBE band gap of Si with respect to (a) the basis set cutoff radius $r_{\text{cut}}$, (b) the number of NAOs in the OBS, (c) the number of ABFs where the tag "for\_aux" signifies that additional NAOs are added to the OBS to generate ABFs only, and (d) the number of minimax frequency/time points. }
    \label{fig:convergence_all}
\end{figure}

In Fig.~\ref{fig:rcut_convg}, the convergence of the $G^0W^0$@PBE band gap with respect to the cutoff radius  $r_{\text{cut}}$ is presented.
Unlike the other convergence parameters that exhibit monotonic behavior, the band gap shows small oscillations as a function of $r_{\text{cut}}$. This oscillatory behavior arises from the discrete nature of the NAO basis set: as $r_{\text{cut}}$ increases, the basis functions are truncated at different radial distances, which can cause discontinuous changes in the completeness of the basis set for certain $r_{\text{cut}}$ values. Nevertheless, the amplitude of these oscillations is rather small, typically within 0.02 eV for $r_{\text{cut}} \geq 7.0$ Bohr. This indicates that even with moderate cutoff radii, the NAO basis set provides a sufficiently complete variational space for accurate $G^0W^0$ calculations. The converged value stabilizes around 1.06 eV for $r_{\text{cut}} \geq 8.0$ Bohr, which is adopted as the default parameter in our calculations.

\textbf{Size of the NAO basis set.} Fig.~\ref{fig:basis_convg} shows the convergence of the band gap with respect to the NAO basis set. While there is a substantial change in the band gap from DZP to TZDP, the additional change from TZDP to QZTP is already below $\sim 0.1$ eV. Here, DZP, TZDP, and QZTP correspond to $2s2p1d$, $3s3p2d$, and $4s4p3d$ basis functions, respectively, i.e., to double-, triple-, and quadruple-zeta descriptions of the valence $s/p$ channels supplemented by one, two, and three $d$ polarization shells. This indicates that a TZDP basis captures most of the basis-set space while retaining a favorable computational cost for silicon.

\textbf{Auxiliary basis functions.} Within the LRI scheme, the quality of the ABF set is a crucial factor. By augmenting the standard OBS with additional ``for\_aux" functions, one can systematically increase the size and quality of the ABF set. In Fig.~\ref{fig:abfs_convg}, it can be seen that the convergence with respect to the number of ABFs follows a monotonic trend. The band gap approaches its converged value as the ABF set goes beyond the ``OBS+1f1g" level, with variations below 1 meV observed when including additional high angular momentum functions. This systematic improvement ensures that highly accurate results can be obtained with the LRI-based $G^0W^0$ approach by using the enhanced ABF set. 

\textbf{Number of minimax frequency/time points.} Fig.~\ref{fig:grid_convg} presents the convergence behavior of the band gap with respect to the minimax grid size.  It can be seen that using 12 or more minimax grid points, the band gap is converged within 0.01 eV. This shows that the minimax grid provides an efficient sampling strategy for the Fourier transform between the imaginary time and frequency axes, and delivers high accuracy for the quasiparticle energies when combined with  the analytic continuation of the self-energy from imaginary frequency to real frequencies.

Based on the convergence tests above, unless otherwise specified, all subsequent ABACUS+LibRPA benchmark calculations use an $8\times 8\times 8$ $\bfk$-point mesh, a pseudo-inverse threshold of $10^{-3}$ for the overlap-matrix inversion in the compression scheme, a PCA threshold of $10^{-6}$ for constructing the auxiliary basis set, a TZDP orbital basis, norm-conserving PseudoDojo pseudopotentials, and experimental lattice constants.

\subsubsection{Benchmarking against plane-wave PAW \texorpdfstring{$G^0W^0$}{GW} results (VASP)}
\label{res:benchmark vasp}
To validate our implementation, we benchmark our $G^0W^0$ results against those obtained from the plane-wave projector-augmented-wave (PAW) method implemented in the VASP code.
In the work of Klimeš et al.\cite{vasp_gw_2014}, $G^0W^0$@PBE calculations were carried out using a plane-wave basis and norm-conserving PAW potentials. 
For all materials calculated in VASP, a $6\times 6\times 6$ k-point mesh, 200 frequency points, and energy cutoffs of 600--1200 eV for different atoms are used to ensure the convergence of quasiparticle energies.
Following the convergence tests in the last section, we compare our converged $G^0W^0$@PBE results with theirs, as summarized in Table~\ref{tab:vasp_benchmark}.

The results in Table~\ref{tab:vasp_benchmark} demonstrate that the quasiparticle energies computed with ABACUS+LibRPA are in good agreement with the VASP results for a wide range of crystalline materials. For most systems and most high-symmetry points, the deviations are within 0.1--0.2 eV, underscoring the reliability of our $G^0W^0$ implementation.

Notably, several materials exhibit larger deviations at the $X_c$ point. AlAs and AlSb show the largest deviations of 0.30 and 0.29~eV at $X_c$, respectively. GaN also displays a deviation of 0.32~eV at $X_c$, although its direct gap at $\Gamma$ ($\Gamma_c$) agrees excellently with the VASP result (difference of only 0.01~eV). CdS exhibits a deviation of 0.25~eV at $X_c$ and 0.23~eV at $X_v$. These discrepancies presumably originate from four sources: (1) The incompleteness error of one-electron orbital basis set. In the present work, the optimized NAO basis set at the TZDP level is used, while VASP employs the plane-wave basis set instead. (2) The LRI approximation. This directly affects the description of high-energy unoccupied states in the polarizability and screened Coulomb interactions, which may affect the low-lying conduction bands in an indirect way. (3) Different treatments of core electrons -- VASP employs the PAW method while ABACUS directly uses norm-conserving pseudopotentials, whereby the impact of the different treatments becomes more pronounced for heavier elements such as Ga and Cd. (4) Different $\mathbf{k}$-point samplings -- the VASP calculations use a $6\times6\times6$ mesh, while the present ABACUS+LibRPA calculations employ an $8\times8\times8$ mesh, which can lead to residual differences in the quasiparticle energies. Identifying the decisive factor that causes the difference requires dedicated efforts and will be deferred to future work.

Overall, however, except for these few examples, the agreement is rather close, with most deviations within 0.2 eV, especially for materials such as C, Si, and AlP, highlighting the robustness of both implementations for systems with moderate band gaps and lighter elements. These results validate our implementation and support its use for studies of quasiparticle properties in semiconductors and insulators.

\begin{table}[h!]
    \centering
    \caption{Comparison of $G^0W^0$@PBE quasiparticle energies (in eV) at the valence- and conduction-band edges at the $\Gamma$ and X points obtained from VASP~\cite{vasp_gw_2014} and ABACUS+LibRPA. The valence-band maximum at $\Gamma$ is aligned to 0 for all systems. All materials are considered in either the diamond or zinc-blende crystal structure. Values in parentheses indicate the absolute differences with respect to the VASP results. }
    \label{tab:vasp_benchmark}
    \begin{tabular}{lrrrrrr}
        \toprule
        \multirow{2}{*}{Material}
        & \multicolumn{3}{c}{VASP~\cite{vasp_gw_2014}}
        & \multicolumn{3}{c}{ABACUS+LibRPA} \\
        \cmidrule(lr){2-4} \cmidrule(lr){5-7}
        & $\Gamma_c$ & $X_v$ & $X_c$
        & $\Gamma_c$ & $X_v$ & $X_c$ \\
        \midrule
        C    & 7.43 & -6.58 & 6.23 & 7.43(0.00) & -6.69(0.11) & 6.17(0.06) \\
        SiC  & 7.35 & -3.30 & 2.42 & 7.30(0.05) & -3.43(0.13) & 2.26(0.16) \\
        Si   & 3.25 & -2.86 & 1.28 & 3.22(0.03) & -2.97(0.11) & 1.23(0.05) \\
        BN   & 11.33& -5.18 & 6.40 & 11.11(0.22)& -5.27(0.09) & 6.32(0.08) \\
        AlP  & 4.23 & -2.13 & 2.48 & 4.18(0.05) & -2.17(0.04) & 2.37(0.11) \\
        AlAs & 2.99 & -2.17 & 2.31 & 2.79(0.20) & -2.30(0.13) & 2.01(0.30) \\
        AlSb & 2.40 & -2.20 & 1.87 & 2.27(0.13) & -2.32(0.12) & 1.58(0.29) \\
        GaN  & 2.85 & -2.70 & 4.50 & 2.86(0.01) & -2.69(0.01) & 4.18(0.32) \\
        GaP  & 2.62 & -2.71 & 2.30 & 2.76(0.14) & -2.60(0.11) & 2.28(0.02) \\
        GaAs & 1.23 & -2.66 & 2.04 & 1.38(0.15) & -2.78(0.12) & 2.10(0.06) \\
        MgO  & 7.55 & -1.45 & 11.82& 7.39(0.16) & -1.47(0.02) & 11.92(0.10) \\
        CdS  & 2.15 & -1.93 & 4.55 & 2.08(0.07) & -2.16(0.23) & 4.30(0.25) \\
        \bottomrule
    \end{tabular}
\end{table}

\subsubsection{Benchmarking against NAO all-electron $\bfk$-space \texorpdfstring{$G^0W^0$}{GW} results (FHI-aims)}
\label{res:benchmark fhiaims}

Previous studies suggest that different $G^0W^0$ implementations still yield noticeable differences in the calculated band gaps for certain materials \cite{vasp_gw_2014,PhysRevB.93.115203,exciting_Gulans_2014,Rangel/etal:2020,ZhuT21,Ren_gw_2021}. As previously alluded to, the differences may stem from different choices of basis sets, different treatments of valence-core interactions, different handling of the frequency dependence, and/or other numerical factors in each implementation. Pinpointing the precise factors that lead to the observed discrepancies in the band gaps and band structures is a highly nontrivial task. In this section, as a further benchmark of our NAO-PP $G^0W^0$ implementation, we compare our band gap results with those produced by the FHI-aims code \cite{blum_ab_2009, ren_resolution--identity_2012,Ren_gw_2021}.
This comparison is particularly meaningful because both codes employ NAO basis sets and the LRI technique for evaluating the four-orbital integrals, so the differences arising from the basis set representation and the RI approximation are largely eliminated. The remaining discrepancy, therefore, primarily reflects the pseudopotential versus all-electron treatment of core-valence interactions.

\begin{figure}[htbp]
    \centering
    \includegraphics[width=\textwidth]{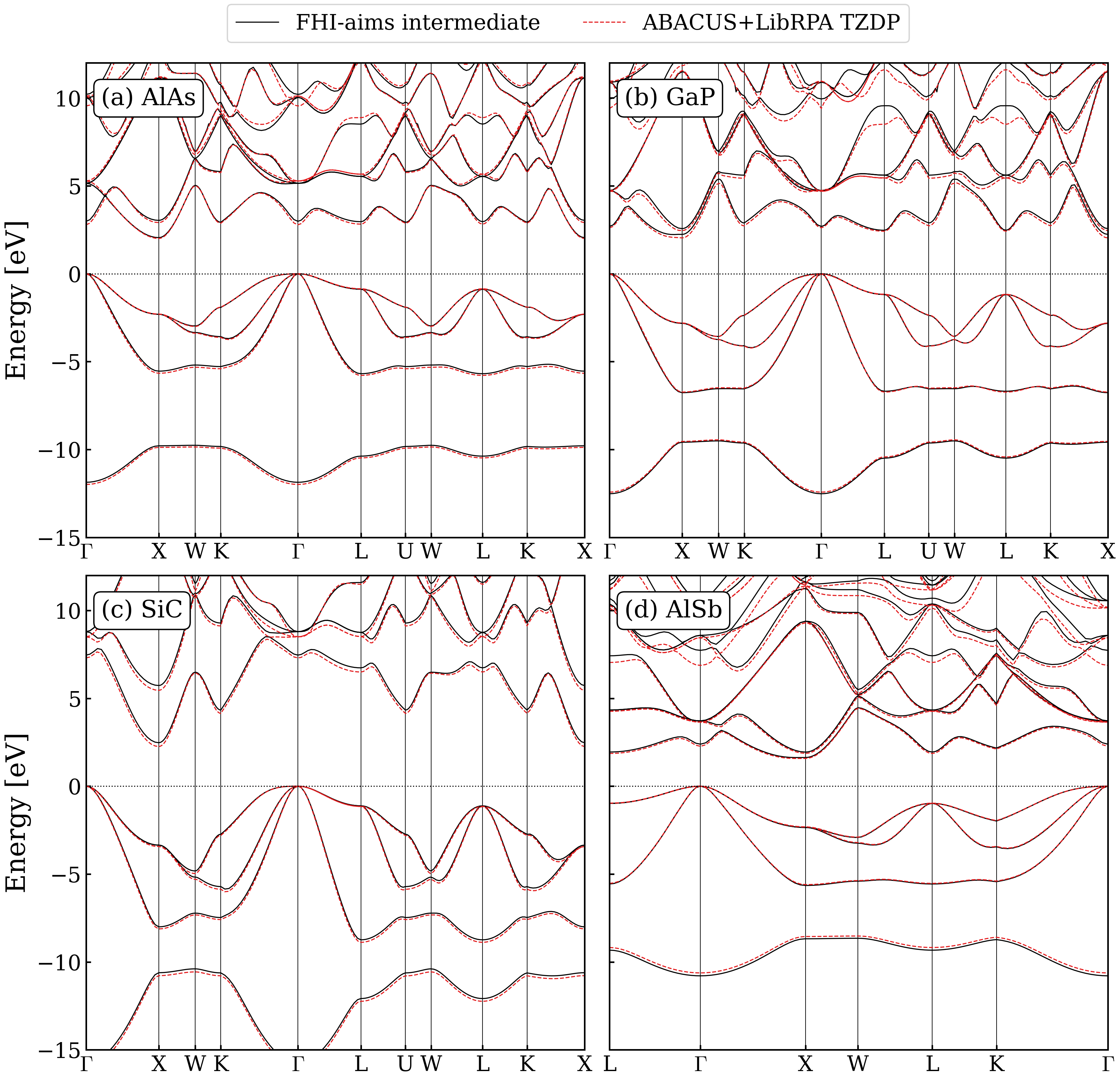}
    \caption{$G^0W^0$@PBE quasiparticle band structures of AlAs, GaP, SiC, and AlSb computed using Dojo pseudopotentials within ABACUS+LibRPA. All-electron reference results from FHI-aims are shown for comparison.}
    \label{fig:compare_librpa_aims}
\end{figure}

For all materials calculated in FHI-aims, we employ an $8\times 8\times 8$ $\bfk$-point mesh, the "intermediate\_gw" basis settings, head and wing corrections, 32 Gauss-Legendre frequency points, and a Padé approximant for analytic continuation, ensuring well-converged quasiparticle energies. Fig.~\ref{fig:compare_librpa_aims} compares the $G^0W^0$@PBE band structures of AlAs, GaP, SiC, and AlSb obtained from ABACUS+LibRPA and FHI-aims, demonstrating excellent agreement between the two implementations. Similar consistency is observed for the other semiconductors, whose band-gap values are summarized in Table~\ref{tab:GW_comparison}.

The combined PBE and $G^0W^0$@PBE comparison in Table~\ref{tab:GW_comparison} reveals two key features. First, the overall agreement with FHI-aims is notably closer than with VASP. For the twelve materials common to both benchmark sets (Si, SiC, C, BN, GaN, GaP, GaAs, AlP, AlAs, AlSb, MgO, CdS), the typical $G^0W^0$ deviation from FHI-aims is in the range of 0.04--0.22~eV, whereas the corresponding deviations from VASP range from 0.05 to 0.32~eV. This improvement is expected: both codes share the same NAO basis framework and the LRI approximation, so the remaining differences arise from a combination of the pseudopotential versus all-electron treatment and the residual basis-set incompleteness in both calculations (TZDP in ABACUS and ``intermediate\_gw'' in FHI-aims).  In particular, materials containing heavier elements with significant semicore contributions, such as GaN (0.039~eV), GaP (0.015~eV), and CdS (0.022~eV), exhibit excellent agreement, indicating that the Dojo pseudopotentials, which explicitly include the relevant semicore electrons in the valence partition, adequately capture the valence and semicore physics.

Second, the largest residual discrepancies occur for wide-gap materials where the $GW$ correction is large and the description of high-energy unoccupied states becomes critical. Diamond exhibits the largest deviation (0.285~eV), while its PBE gap differs by only 0.004~eV. Notably, a previous all-electron $G^0W^0$ study using the FHI-gap code \cite{PhysRevB.93.115203} reports the diamond gap as 5.49~eV without additional local orbitals, which is already close to our value of 5.549~eV, whereas including five high-energy local orbitals increases the FHI-gap result to 5.69~eV. This suggests that the deviation for diamond is mainly due to basis-set incompleteness in the description of high-energy states rather than the pseudopotential approximation. A similar trend is observed for SiC (0.217~eV) and LiF (0.172~eV). Despite these few cases, the overall agreement across all 17 materials confirms that ABACUS+LibRPA can reproduce all-electron quasiparticle gaps with good accuracy while retaining the computational advantages of the pseudopotential framework.

\begin{figure}[htbp]
    \centering
    \includegraphics[width=\textwidth]{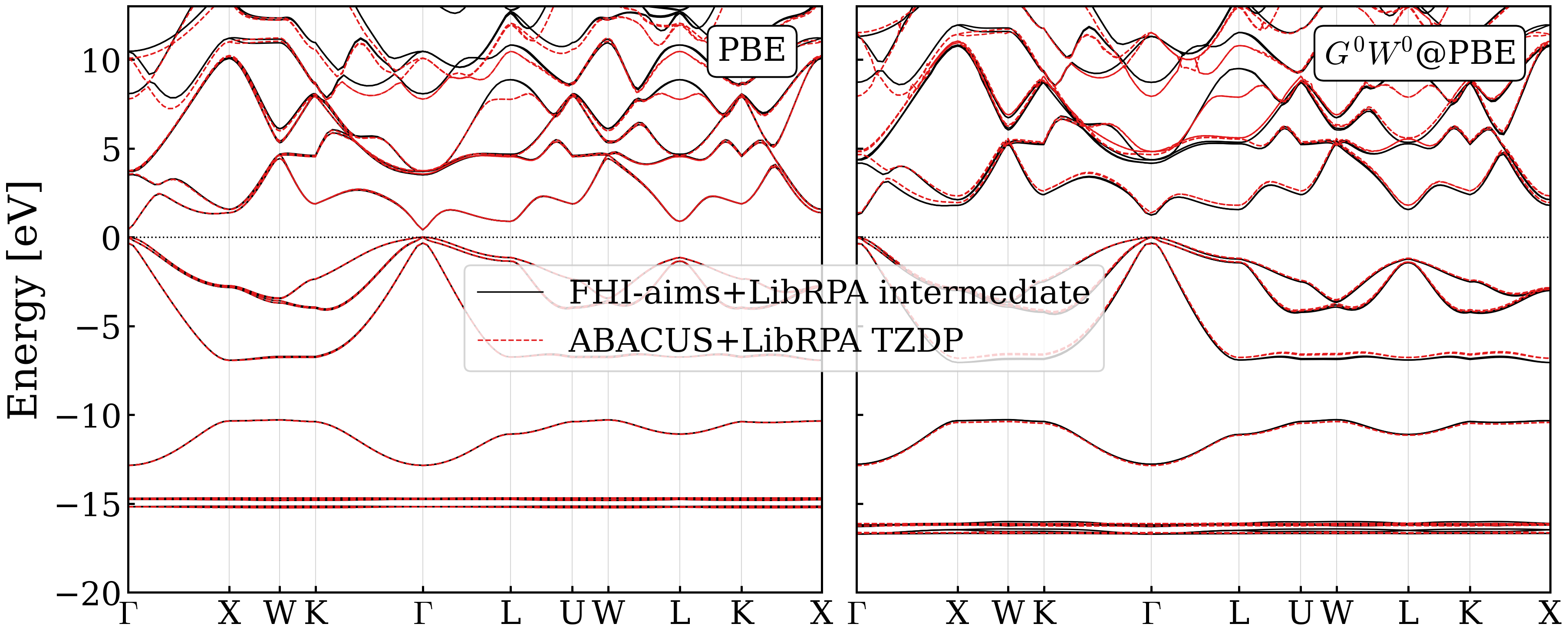}
    \caption{Comparison of $G^0W^0$@PBE band structures for GaAs with spin-orbit coupling (SOC) computed by ABACUS+LibRPA and FHI-aims+LibRPA.}
    \label{fig:soc_GaAs_comparison}
\end{figure}

To further validate our implementation for systems with SOC, we compare the $G^0W^0$@PBE band structures of GaAs with SOC computed by ABACUS+LibRPA and FHI-aims+LibRPA. As shown in Fig.~\ref{fig:soc_GaAs_comparison}, the two calculations exhibit excellent agreement, particularly for the spin-orbit splitting at the valence band maximum ($0.36$~eV from ABACUS+LibRPA vs.\ $0.34$~eV from FHI-aims+LibRPA, both close to the experimental value of $0.34$~eV~\cite{Vurgaftman/etal:2001}). This confirms that our implementation accurately captures relativistic effects within the $G^0W^0$ approximation, which is crucial for materials with significant spin-orbit interactions.

In summary, the comparison with FHI-aims shows systematically closer agreement than the VASP benchmark, consistent with the shared use of NAO basis sets and the LRI technique. The remaining differences are attributable to the pseudopotential versus all-electron treatment as well as residual basis-set incompleteness, with the largest deviations occurring for wide-gap materials where basis-set completeness for high-energy states is critical. The additional validation for spin-orbit coupling in GaAs further demonstrates the capability to handle relativistic effects essential for heavy-element systems.

\begin{table}[h!]
    \centering
    \small
    \setlength{\tabcolsep}{4.5pt}
    \caption{Comparison of PBE and $G^0W^0$@PBE band gaps (in eV) calculated by FHI-aims and ABACUS+LibRPA for various materials.}
    \label{tab:GW_comparison}
    \begin{tabular}{lcccccc}
        \toprule
        System & \multicolumn{3}{c}{PBE} & \multicolumn{3}{c}{$G^0W^0$@PBE} \\
        \cmidrule(lr){2-4} \cmidrule(lr){5-7}
         & FHI-aims & ABACUS & $\Delta$ & FHI-aims & ABACUS+LibRPA & $\Delta$ \\
        \midrule
        AlAs & 1.440 & 1.427 & 0.013 & 2.054 & 2.008 & 0.046 \\
        AlP  & 1.584 & 1.607 & 0.023 & 2.310 & 2.374 & 0.064 \\
        AlSb & 1.205 & 1.191 & 0.014 & 1.628 & 1.576 & 0.052 \\
        BAs  & 1.190 & 1.186 & 0.004 & 1.903 & 1.762 & 0.141 \\
        BN   & 4.501 & 4.584 & 0.083 & 6.447 & 6.324 & 0.123 \\
        BP   & 1.246 & 1.290 & 0.044 & 2.018 & 2.030 & 0.012 \\
        C    & 4.183 & 4.187 & 0.004 & 5.834 & 5.549 & 0.285 \\
        Si   & 0.590 & 0.561 & 0.029 & 1.099 & 1.060 & 0.039 \\
        SiC  & 1.380 & 1.373 & 0.007 & 2.472 & 2.255 & 0.217 \\
        GaAs & 0.527 & 0.518 & 0.009 & 1.269 & 1.378 & 0.109 \\
        GaN  & 1.621 & 1.662 & 0.041 & 2.896 & 2.857 & 0.039 \\
        GaP  & 1.603 & 1.622 & 0.019 & 2.230 & 2.245 & 0.015 \\
        MgO  & 4.737 & 4.721 & 0.016 & 7.528 & 7.392 & 0.136 \\
        NaCl & 5.087 & 5.088 & 0.001 & 7.962 & 7.832 & 0.130 \\
        CdS  & 1.170 & 1.177 & 0.007 & 2.054 & 2.076 & 0.022 \\
        CaO  & 4.703 & 4.634 & 0.069 & 6.307 & 6.368 & 0.061 \\
        LiF  & 9.174 & 9.187 & 0.013 & 13.828 & 14.000 & 0.172 \\
        \bottomrule
    \end{tabular}
\end{table}

\section{Conclusion}
\label{sec:conclusion}

In this work, we present an efficient $G^0W^0$ implementation within a combined NAO and PP framework by integrating the LibRPA library with the DFT software package ABACUS. 
The methodological improvements presented in this work lie in three key aspects: First, for the treatment of the $\Gamma$-point singularity, we incorporated analytical head and wing corrections for the averaged inverse dielectric function, enabling significantly faster $\bfk$-point convergence. Second, we introduced a compression scheme within the LRI-based workflow, which substantially reduces memory usage and computational cost while improving numerical stability in matrix operations. This effectively removes one of the major bottlenecks in applying $G^0W^0$ to large-scale systems. Third, we addressed the long-standing issue of using a pseudopotential originally constructed for KS-DFT in $G^0W^0$ calculations. Specifically, a practical and easily applicable criterion was applied based on the frequency-dependent macroscopic dielectric constant to assess pseudopotential reliability prior to performing $G^0W^0$, thereby avoiding unnecessary computational expenditure caused by a suboptimal pseudopotential. Together, these developments form the basis of a highly efficient $G^0W^0$ computational scheme in ABACUS+LibRPA. We have validated the algorithmic improvements through systematic benchmarks. Further cross-code comparisons with established packages such as VASP and FHI-aims demonstrate that our $G^0W^0$ band gaps for typical semiconductors agree within 0.1--0.3 eV, confirming the accuracy and reliability of our implementation.

This work sets a solid foundation for performing reliable $G^0W^0$ calculations within the pseudopotential-based NAO basis set framework. The low-scaling $G^0W^0$ algorithm developed within LibRPA, together with a compact NAO-PP basis set representation, offers a promising route to large-scale $G^0W^0$ calculations for systems with thousands of atoms. Furthermore, the local NAO representation of the full $GW$ self-energy matrix is perfectly suitable for graph-neural-network-based machine learning models. Thus, we expect that the present implementation will provide valuable training data for developing machine-learning $GW$ approaches.

\begin{acknowledgments}
  We acknowledge the funding support from “Advanced Materials-National Science and Technology Major Project (Grant Number 2025ZD0618401)” and from National Natural Science Foundation of China (Grant Nos 12374067, 12134012, 12188101 and 12204332) and Guangdong Basic and Applied Basic Research Foundation (Project Numbers 2021A1515110603).
  This work was also supported by the Strategic Priority Research Program of Chinese Academy of Sciences under Grant No. XDB0500201, the National Key Research and Development Program of China (Grant Nos. 2022YFA1403800 and 2023YFA1507004), and the
robotic AI-Scientist platform of the Chinese Academy of Sciences. M.-Y. Z. acknowledges the support by the IOP-Humboldt Postdoctoral Fellowship in Physics of Institute of Physics, Chinese Academy of Sciences (Grant No. 202402). We thank the electronic structure team (from AI for Science Institute, Beijing) for improving the ABACUS package from various aspects. The numerical calculations in this study were partly carried out on the ORISE Supercomputer.  
\end{acknowledgments}

\appendix
\section{Compression transformation of auxiliary basis set in periodic systems}

\label{ap:periodic compression}
In periodic systems, the compression transformation follows the same idea as presented in Sec.~\ref{imp:compression}, but the lattice translations carried by the auxiliary basis functions must be kept explicitly. We use $\rho,\lambda$ to label the standard OBS-generated ABF set and $\mu,\nu$ to label the enhanced OBS+-generated ABF set. At a given imaginary-time point $i\tau_j$, the same real-space response function can then be written in either representation as

\begin{equation}
    \begin{aligned}
    \chi^0(\mathbf{r},\mathbf{r}',i\tau_j)
    &= \sum_{\rho\lambda}\sum_{\mathbf{R}_1\mathbf{R}_2}
    P_\rho^{\mathrm{std}}(\mathbf{r}-\boldsymbol{\tau}_{\rho}-\mathbf{R}_1)\,
    \chi_{\rho\lambda}^{0,\mathrm{RI}}(\mathbf{R}_2-\mathbf{R}_1,i\tau_j)\,
    P_\lambda^{\mathrm{std}}(\mathbf{r}'-\boldsymbol{\tau}_{\lambda}-\mathbf{R}_2) \\
    &= \sum_{\mu\nu}\sum_{\mathbf{R}_1\mathbf{R}_2}
    P_\mu^{\mathrm{enh}}(\mathbf{r}-\boldsymbol{\tau}_{\mu}-\mathbf{R}_1)\,
    \chi_{\mu\nu}^{0,\mathrm{LRI}}(\mathbf{R}_2-\mathbf{R}_1,i\tau_j)\,
    P_\nu^{\mathrm{enh}}(\mathbf{r}'-\boldsymbol{\tau}_{\nu}-\mathbf{R}_2).
    \end{aligned}
\end{equation}

To make the projection step explicit, we first define the standard-basis projected real-space response matrix
\begin{equation}
    \chi_{\rho\lambda}^{0,\mathrm{proj}}(\mathbf{R},i\tau_j)
    \equiv
    \langle P_{\rho\mathbf{0}}^{\mathrm{std}}|\chi^0(i\tau_j)|P_{\lambda\mathbf{R}}^{\mathrm{std}}\rangle .
\end{equation}
Substituting the two expansions above into this definition gives, still entirely in real space,
\begin{equation}
    \begin{aligned}
    \chi_{\rho\lambda}^{0,\mathrm{proj}}(\mathbf{R},i\tau_j)
    &= \sum_{\mathbf{R}_1\mathbf{R}_2}\sum_{\rho'\lambda'}
    \langle P_{\rho\mathbf{0}}^{\mathrm{std}}|P_{\rho'\mathbf{R}_1}^{\mathrm{std}}\rangle\,
    \chi_{\rho'\lambda'}^{0,\mathrm{RI}}(\mathbf{R}_2-\mathbf{R}_1,i\tau_j)\,
    \langle P_{\lambda'\mathbf{R}_2}^{\mathrm{std}}|P_{\lambda\mathbf{R}}^{\mathrm{std}}\rangle \\
    &= \sum_{\mathbf{R}_1\mathbf{R}_2}\sum_{\mu\nu}
    \langle P_{\rho\mathbf{0}}^{\mathrm{std}}|P_{\mu\mathbf{R}_1}^{\mathrm{enh}}\rangle\,
    \chi_{\mu\nu}^{0,\mathrm{LRI}}(\mathbf{R}_2-\mathbf{R}_1,i\tau_j)\,
    \langle P_{\nu\mathbf{R}_2}^{\mathrm{enh}}|P_{\lambda\mathbf{R}}^{\mathrm{std}}\rangle .
    \end{aligned}
\end{equation}
It is therefore convenient to introduce the corresponding real-space overlap matrices,
\begin{equation}
    ({\cal S}^{\mathrm{std,std}})_{\rho\lambda}(\mathbf{R})
    = \langle P_{\rho\mathbf{0}}^{\mathrm{std}}|P_{\lambda\mathbf{R}}^{\mathrm{std}}\rangle ,
\end{equation}
\begin{equation}
    ({\cal S}^{\mathrm{std,enh}})_{\rho\mu}(\mathbf{R})
    = \langle P_{\rho\mathbf{0}}^{\mathrm{std}}|P_{\mu\mathbf{R}}^{\mathrm{enh}}\rangle ,
\end{equation}
\begin{equation}
    ({\cal S}^{\mathrm{enh,std}})_{\nu\lambda}(\mathbf{R})
    = \langle P_{\nu\mathbf{0}}^{\mathrm{enh}}|P_{\lambda\mathbf{R}}^{\mathrm{std}}\rangle .
\end{equation}
With these definitions, the projected response takes the form of a lattice convolution,
\begin{equation}
    \begin{aligned}
    \chi_{\rho\lambda}^{0,\mathrm{proj}}(\mathbf{R},i\tau_j)
    &= \sum_{\mathbf{R}_1\mathbf{R}_2}\sum_{\rho'\lambda'}
    ({\cal S}^{\mathrm{std,std}})_{\rho\rho'}(\mathbf{R}_1)\,
    \chi_{\rho'\lambda'}^{0,\mathrm{RI}}(\mathbf{R}_2-\mathbf{R}_1,i\tau_j)\,
    ({\cal S}^{\mathrm{std,std}})_{\lambda'\lambda}(\mathbf{R}-\mathbf{R}_2) \\
    &= \sum_{\mathbf{R}_1\mathbf{R}_2}\sum_{\mu\nu}
    ({\cal S}^{\mathrm{std,enh}})_{\rho\mu}(\mathbf{R}_1)\,
    \chi_{\mu\nu}^{0,\mathrm{LRI}}(\mathbf{R}_2-\mathbf{R}_1,i\tau_j)\,
    ({\cal S}^{\mathrm{enh,std}})_{\nu\lambda}(\mathbf{R}-\mathbf{R}_2).
    \end{aligned}
\end{equation}
Only after the sums over $\mathbf{R}_1$ and $\mathbf{R}_2$ are written out in this way do we apply the lattice Fourier transform, i.e., the spatial part of Eq.~\eqref{eq:chi0_qw_from_Rtau}. This converts the real-space convolution into ordinary matrix multiplication in $\mathbf{q}$ space. Defining
\begin{equation}
    \chi_{\rho\lambda}^{0,\mathrm{proj}}(\mathbf{q},i\tau_j)
    \equiv \sum_{\mathbf{R}} e^{i\mathbf{q}\cdot\mathbf{R}}
    \chi_{\rho\lambda}^{0,\mathrm{proj}}(\mathbf{R},i\tau_j),
\end{equation}
one obtains

\begin{equation}
    \begin{aligned}
    \chi_{\rho\lambda}^{0,\mathrm{proj}}(\mathbf{q},i\tau_j)
    &= \sum_{\rho'\lambda'}
    ({\cal S}^{\mathrm{std,std}})_{\rho\rho'}(\mathbf{q})\,
    \chi_{\rho'\lambda'}^{0,\mathrm{RI}}(\mathbf{q},i\tau_j)\,
    ({\cal S}^{\mathrm{std,std}})_{\lambda'\lambda}(\mathbf{q}) \\
    &= \sum_{\mu\nu}
    ({\cal S}^{\mathrm{std,enh}})_{\rho\mu}(\mathbf{q})\,
    \chi_{\mu\nu}^{0,\mathrm{LRI}}(\mathbf{q},i\tau_j)\,
    ({\cal S}^{\mathrm{enh,std}})_{\nu\lambda}(\mathbf{q}),
    \end{aligned}
    \label{eq:chi0_periodic_compression}
\end{equation}

Here, the $\mathbf{q}$-space overlaps are defined as
\begin{equation}
    ({\cal S}^{\mathrm{std,std}})_{\rho\lambda}(\mathbf{q})
    = \sum_{\mathbf{R}}
    \langle P_{\rho\mathbf{0}}^{\mathrm{std}}|P_{\lambda\mathbf{R}}^{\mathrm{std}}\rangle
    e^{i\mathbf{q}\cdot\mathbf{R}},
\end{equation}
\begin{equation}
    ({\cal S}^{\mathrm{std,enh}})_{\rho\mu}(\mathbf{q})
    = \sum_{\mathbf{R}}
    \langle P_{\rho\mathbf{0}}^{\mathrm{std}}|P_{\mu\mathbf{R}}^{\mathrm{enh}}\rangle
    e^{i\mathbf{q}\cdot\mathbf{R}},
\end{equation}
with $({\cal S}^{\mathrm{enh,std}})_{\nu\lambda}(\mathbf{q}) = [({\cal S}^{\mathrm{std,enh}})^\dagger]_{\nu\lambda}(\mathbf{q})$. Applying $\left({\cal S}^{\mathrm{std,std}}(\mathbf{q})\right)^{-1}$ from both sides yields the compressed response matrix in the standard ABF representation,
\begin{equation}
    \chi^{0,\mathrm{RI}}(\mathbf{q},i\tau_j)
    = \left({\cal S}^{\mathrm{std,std}}(\mathbf{q})\right)^{-1}
    {\cal S}^{\mathrm{std,enh}}(\mathbf{q})\,
    \chi^{0,\mathrm{LRI}}(\mathbf{q},i\tau_j)\,
    {\cal S}^{\mathrm{enh,std}}(\mathbf{q})
    \left({\cal S}^{\mathrm{std,std}}(\mathbf{q})\right)^{-1}.
    \label{eq:chi0_compression_periodic}
\end{equation}
When the dielectric matrix is built, the remaining minimax cosine transform from $i\tau_j$ to $i\omega_k$ is then performed exactly as in Eq.~\eqref{eq:chi0_qw_from_Rtau}. For the actual implementation, it is convenient to introduce the compression matrix
\begin{equation}
    U(\mathbf{q})
    = \left({\cal S}^{\mathrm{std,std}}(\mathbf{q})\right)^{-1}
    {\cal S}^{\mathrm{std,enh}}(\mathbf{q}).
\end{equation}
The response function is compressed at each imaginary-time point as
\begin{equation}
    \chi^{0,\mathrm{RI}}(\mathbf{q},i\tau_j)
    = U(\mathbf{q})\,
    \chi^{0,\mathrm{LRI}}(\mathbf{q},i\tau_j)\,
    U^\dagger(\mathbf{q}),
\end{equation}
After the dielectric matrix has been inverted and the compact screened Coulomb matrix $W^{c,\mathrm{RI}}(\mathbf{q},i\omega_k)$ has been obtained, the reverse path is followed. First, Eq.~\eqref{eq:Wc_Rtau_from_qw} is used to transform it back to $W^{c,\mathrm{RI}}(\mathbf{q},i\tau_j)$. The latter is then upfolded to the enhanced ABF representation as
\begin{equation}
    W^{c,\mathrm{LRI}}(\mathbf{q},i\tau_j)
    = U^\dagger(\mathbf{q})\,
    W^{c,\mathrm{RI}}(\mathbf{q},i\tau_j)\,
    U(\mathbf{q}).
\end{equation}
Finally, an inverse lattice Fourier transform brings the screened Coulomb matrix back to real space,
\begin{equation}
    W_{\mu\nu}^{c,\mathrm{LRI}}(\mathbf{R},i\tau_j)
    = \frac{1}{N_\mathbf{q}}\sum_\mathbf{q}
    e^{-i\mathbf{q}\cdot\mathbf{R}}
    W_{\mu\nu}^{c,\mathrm{LRI}}(\mathbf{q},i\tau_j),
\end{equation}
after which it can be inserted directly into Eq.~\eqref{eq:sigmac-R} for the evaluation of the correlation self-energy. In this way, the large enhanced ABF set is used only where it is needed for accuracy, while the expensive dielectric inversion and storage remain in the compact standard ABF representation.

\section{Detailed analysis of the LRI error for $\chi^0$ and the self-energy}
\label{ap:chi0-compression}

In this appendix, we provide a more detailed analysis of the observation made in Sec.~\ref{res:compression} regarding the different sensitivities of $\chi^0$ and the self-energy to the LRI approximation.

\begin{figure}[htbp]
    \centering
    \includegraphics[width=0.9\textwidth]{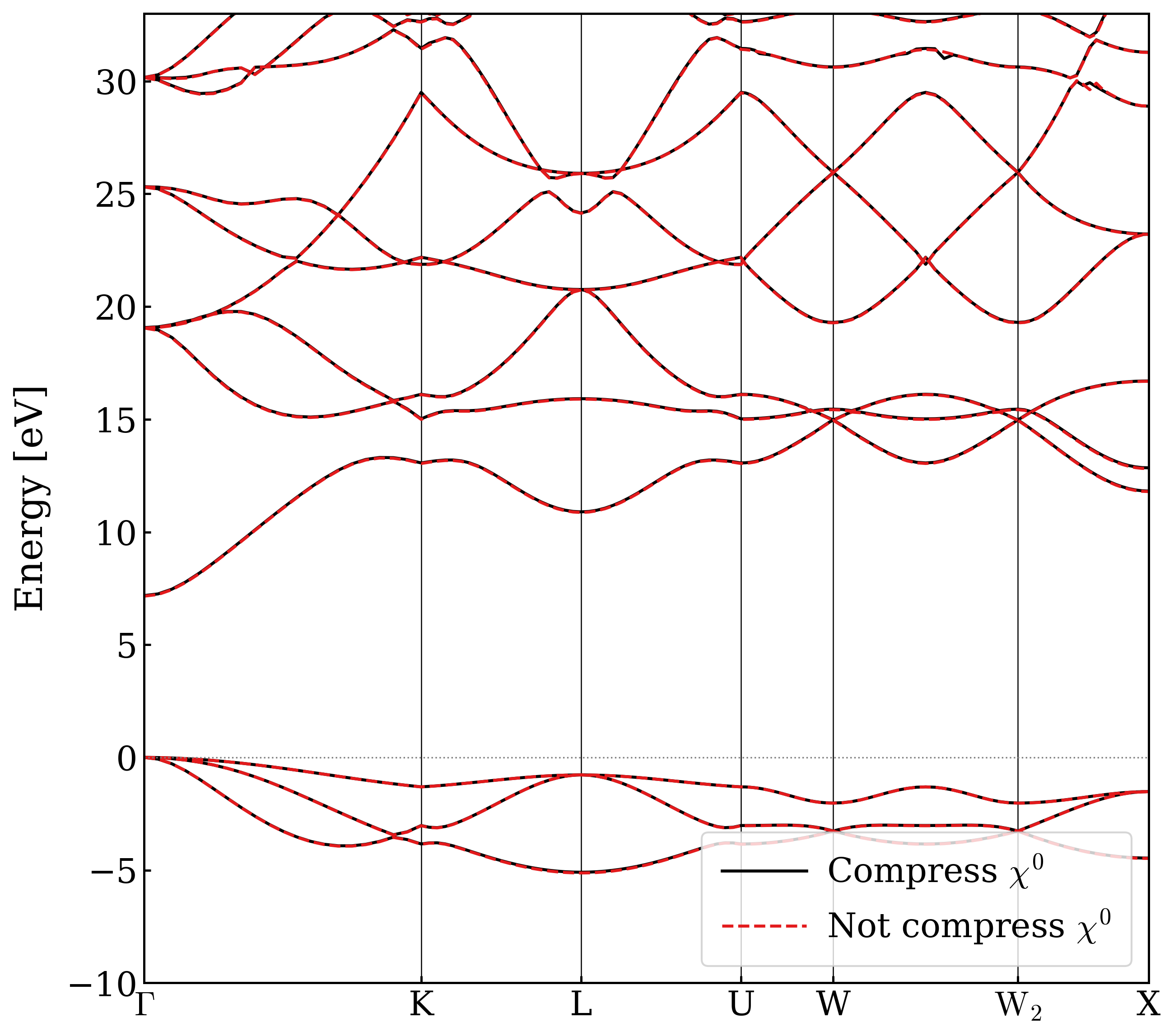}
    \caption{Comparison of $G^0W^0$ band structures for MgO computed with and without $\chi^0$ compression.}
    \label{fig:MgO_chi0_comparison}
\end{figure}

As shown in Fig.~\ref{fig:MgO_chi0_comparison}, for MgO the band structures computed with and without $\chi^0$ compression are nearly indistinguishable, with band gap differences below 0.03~eV. Computing $\chi^0$ with a small ABF set (387 functions) instead of a large ABF set (896 functions) accelerates this specific step by approximately a factor of 5, and the speedup can reach up to an order of magnitude for certain systems. These results suggest that for certain materials, the primary source of error in the LRI approximation lies not in the calculation of the response function $\chi^0$, but rather in the subsequent steps of the self-energy calculation. Specifically, we find that using exclusively the small ABF set for representing $W^{0(c)}$ and calculating the self-energy $\Sigma^c$ introduces significant errors, indicating that the LRI error predominantly affects the self-energy calculation rather than the initial response function calculation.

However, in systems where the LRI error is more pronounced, employing a large ABF set for computing $\chi^0$ becomes essential to obtain accurate results, making the compression of $\chi^0$ particularly valuable in such cases. Since it is generally difficult to determine \textit{a priori} whether starting with a large ABF set and then compressing $\chi^0$ is necessary for a given system, this compression strategy is especially useful in practical $G^0W^0$ calculations with SOC included. In practice, one usually first performs a $G^0W^0$ calculation without SOC, which can then be used to judge whether $\chi^0$ compression is required before carrying out the more expensive SOC-$G^0W^0$ calculation.

\section{\texorpdfstring{Comparison of $G^0W^0$ band structures obtained with different pseudopotentials for GaAs}{Comparison of G0W0 band structures obtained with different pseudopotentials for GaAs}}
\label{ap:pp_comparison}

In this appendix, we provide a direct comparison of the $G^0W^0$ quasiparticle band structures of GaAs obtained with Dojo pseudopotentials containing different numbers of valence electrons. As discussed in Sec.~\ref{sec:result}, the dielectric function criterion already indicates that pseudopotentials with fewer valence electrons (Ga~$3d^{10} 4s^2 4p^1$ and As~$4s^2 4p^3$) yield less reliable screening. Here we show that this deficiency propagates directly into the quasiparticle band structure.

\begin{figure}[htbp]
    \centering
    \includegraphics[width=0.9\textwidth]{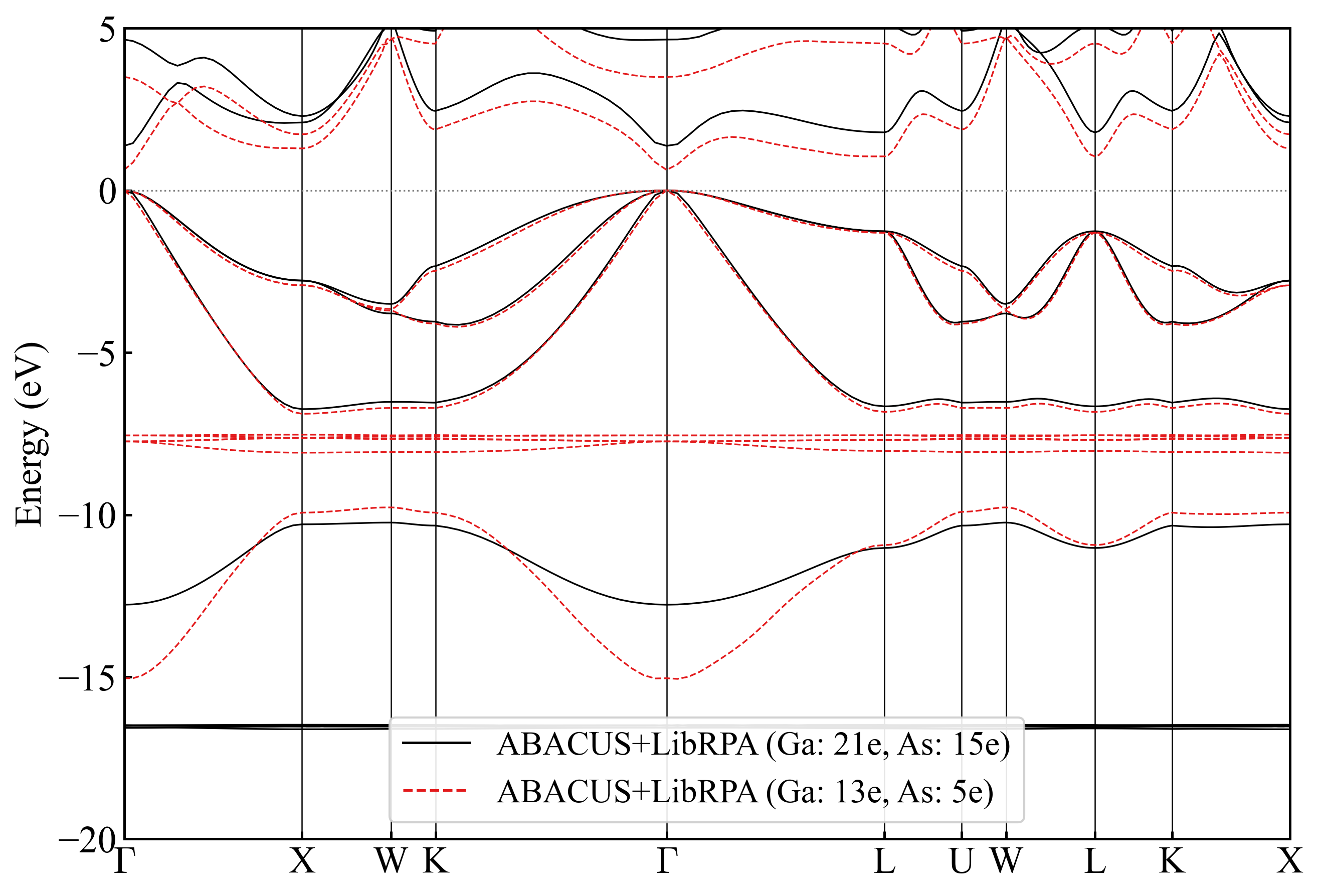}
    \caption{Comparison of $G^0W^0$@PBE quasiparticle band structures of GaAs obtained with Dojo pseudopotentials containing different numbers of valence electrons. Black solid lines: Ga~($3s^2 3p^6 3d^{10} 4s^2 4p^1$) / As~($3d^{10} 4s^2 4p^3$); red dashed lines: Ga~($3d^{10} 4s^2 4p^1$) / As~($4s^2 4p^3$). \allowbreak The valence band maximum is aligned to 0~eV for each calculation separately.}
    \label{fig:band_GaAs_appendix}
\end{figure}

Fig.~\ref{fig:band_GaAs_appendix} compares the $G^0W^0$@PBE band structures computed with two sets of pseudopotentials: (i)~Ga~($3s^2 3p^6 3d^{10} 4s^2 4p^1$) / As~($3d^{10} 4s^2 4p^3$) and (ii)~Ga~($3d^{10} 4s^2 4p^1$) / As~($4s^2 4p^3$). The valence bands obtained with set~(ii) exhibit significant deviations from set~(i), with errors reaching several eV for the deeper valence states. The conduction bands are also noticeably affected, although to a lesser extent. These results confirm that the pseudopotential quality assessed through the dielectric function (Fig.~\ref{fig:head_GaAs}) is a reliable predictor of the accuracy achievable in the subsequent $G^0W^0$ calculation.

Notably, the most striking discrepancy between the two pseudopotential sets lies in the treatment of the Ga-3$d$ semicore states. With set~(ii), the Ga-3$d$ states located at approximately $-15$~eV in the PBE band structure receive a large positive $G^0W^0$ correction of approximately $+7$~eV, which pushes them upward toward the main valence band region. This is qualitatively incorrect, as semicore $d$-states should generally be shifted to deeper binding energies by the $G^0W^0$ self-energy correction.

This erroneous behavior originates from the frozen-core approximation inherent in the $G^0W^0$ formalism. The Ga-13e pseudopotential includes the $3d^{10}$ semicore states in the valence space but freezes the same-shell $3s^2$ and $3p^6$ states in the core. Since the exchange self-energy $\Sigma_x$ for a semicore $d$-state receives dominant contributions from states within the same principal shell, the omission of the $3s$ and $3p$ exchange contributions results in an insufficiently attractive $\Sigma_x$ for the $3d$ states. Although the correlation self-energy $\Sigma_c$ partially compensates by pushing the $3d$ states toward the Fermi level, the net effect is dominated by the missing exchange, leading to an overall upward shift of the $3d$ quasiparticle energies.

This phenomenon was analyzed in detail by Li~\textit{et~al.}\cite{Li_gwpp_2012} for GaN and ZnS, where pseudopotentials with analogous frozen-core treatments were also found to produce qualitatively incorrect $G^0W^0$ corrections for semicore $d$-states. Their analysis demonstrated that including the same-shell $sp$ states in the valence space restores the proper exchange contribution, which pushes the $d$-states to deeper binding energies. The correlation contribution acts in the opposite direction but with smaller magnitude, resulting in a net downward shift that correctly describes the quasiparticle renormalization of the $d$-states. Our observations for GaAs are fully consistent with this picture and further validate the dielectric-function-based pseudopotential selection criterion proposed in Sec.~\ref{imp:select pp}.

\bibliography{main}

@article{ZhuT21,
  title = {All-electron {Gaussian}-based {{G0W0}} for valence and core excitation energies of periodic systems},
  author = {Zhu, Tianyu and Chan, Garnet Kin-Lic},
  date = {2021-01-04},
  year = 2021,
  journal = {J. Chem. Theory Comput.},
  volume = {17},
  number = {2},
  pages = {727-741},
  publisher = {American Chemical Society},
  issn = {1549-9618},
  doi = {10.1021/acs.jctc.0c00704},
  url = {https://doi.org/10.1021/acs.jctc.0c00704},
}

@misc{Zhang2026arXiv,
  title={Low-scaling $GW$ calculation of quasi-particle energies within numerical atomic orbital framework},
  author={Zhang, Min-Ye and Lin, Peize and Shi, Rong and Ren, Xinguo},
  note={arXiv:2603.27292},
  year={2026}
}

@article{lin_force_2025,
  title    = {Efficient hybrid-functional-based force and stress calculations for periodic systems with thousands of atoms},
  journal  = {J. Chem. Theory Comput.},
  volume   = {},
  author   = {Lin, Peize  and Ji, Yuyang and He, Lixin and Ren, Xinguo and He},
  month    = {},
  year     = {2025},
  note     = {Publisher: American Chemical Society},
  pages    = {}
}

@article{soler_siesta_2002,
  title    = {The {{SIESTA}} method for ab initio order-{{N}} materials simulation},
  volume   = {14},
  issn     = {0953-8984},
  url      = {https://dx.doi.org/10.1088/0953-8984/14/11/302},
  doi      = {10.1088/0953-8984/14/11/302},
  number   = {11},
  urldate  = {2024-04-24},
  journal  = {Journal of Physics: Condensed Matter},
  author   = {Soler, José M. and Artacho, Emilio and Gale, Julian D. and García, Alberto and Junquera, Javier and Ordejón, Pablo and Sánchez-Portal, Daniel},
  month    = mar,
  year     = {2002},
  pages    = {2745},
}

@article{SG15,
  title     = {Optimized norm-conserving Vanderbilt pseudopotentials},
  author    = {Hamann, D. R.},
  journal   = {Phys. Rev. B},
  volume    = {88},
  issue     = {8},
  pages     = {085117},
  numpages  = {10},
  year      = {2013},
  month     = {Aug},
  publisher = {American Physical Society},
  doi       = {10.1103/PhysRevB.88.085117},
  url       = {https://link.aps.org/doi/10.1103/PhysRevB.88.085117}
}

@article{chen_electronic_2011,
  title    = {Electronic structure interpolation via atomic orbitals},
  volume   = {23},
  issn     = {0953-8984, 1361-648X},
  url      = {https://iopscience.iop.org/article/10.1088/0953-8984/23/32/325501},
  doi      = {10.1088/0953-8984/23/32/325501},
  number   = {32},
  urldate  = {2024-04-15},
  journal  = {Journal of Physics: Condensed Matter},
  author   = {Chen, Mohan and Guo, G-C and He, Lixin},
  month    = aug,
  year     = {2011},
  pages    = {325501}
}

@article{li_large-scale_2016,
  title    = {Large-scale ab initio simulations based on systematically improvable atomic basis},
  volume   = {112},
  issn     = {09270256},
  url      = {https://linkinghub.elsevier.com/retrieve/pii/S0927025615004140},
  doi      = {10.1016/j.commatsci.2015.07.004},
  urldate  = {2024-04-15},
  journal  = {Computational Materials Science},
  author   = {Li, Pengfei and Liu, Xiaohui and Chen, Mohan and Lin, Peize and Ren, Xinguo and Lin, Lin and Yang, Chao and He, Lixin},
  month    = feb,
  year     = {2016},
  pages    = {503--517}
}

@article{cao_sym_2025,
  author  = {Cao, Yu and Zhang, Min-Ye and Lin, Peize and Chen, Mohan and Ren, Xinguo},
  title   = {Applying Space-Group Symmetry to Speed Up Hybrid-Functional Calculations within the Framework of Numerical Atomic Orbitals},
  journal = {Journal of Chemical Theory and Computation},
  volume  = {21},
  number  = {16},
  pages   = {8086-8105},
  year    = {2025},
  doi     = {10.1021/acs.jctc.5c00537},
  note    = {PMID: 40789044},
  url     = {https://doi.org/10.1021/acs.jctc.5c00537},
  eprint  = {https://doi.org/10.1021/acs.jctc.5c00537
             }
}

@article{lin_ab_2024,
  title      = {Ab initio electronic structure calculations based on numerical atomic orbitals: {Basic} fomalisms and recent progresses},
  volume     = {14},
  issn       = {1759-0876, 1759-0884},
  shorttitle = {Ab initio electronic structure calculations based on numerical atomic orbitals},
  url        = {https://wires.onlinelibrary.wiley.com/doi/10.1002/wcms.1687},
  doi        = {10.1002/wcms.1687},
  number     = {1},
  urldate    = {2024-04-15},
  journal    = {WIREs Computational Molecular Science},
  author     = {Lin, Peize and Ren, Xinguo and Liu, Xiaohui and He, Lixin},
  month      = jan,
  year       = {2024},
  pages      = {e1687}
}

@article{lin_efficient_2021,
  title     = {Efficient {Hybrid} {Density} {Functional} {Calculations} for {Large} {Periodic} {Systems} {Using} {Numerical} {Atomic} {Orbitals}},
  volume    = {17},
  copyright = {https://doi.org/10.15223/policy-029},
  issn      = {1549-9618, 1549-9626},
  url       = {https://pubs.acs.org/doi/10.1021/acs.jctc.0c00960},
  doi       = {10.1021/acs.jctc.0c00960},
  number    = {1},
  urldate   = {2024-04-15},
  journal   = {Journal of Chemical Theory and Computation},
  author    = {Lin, Peize and Ren, Xinguo and He, Lixin},
  month     = jan,
  year      = {2021},
  pages     = {222--239}
}

@article{hohenberg_inhomogeneous_1964,
  title    = {Inhomogeneous {Electron} {Gas}},
  volume   = {136},
  url      = {https://link.aps.org/doi/10.1103/PhysRev.136.B864},
  doi      = {10.1103/PhysRev.136.B864},
  number   = {3B},
  urldate  = {2024-04-24},
  journal  = {Physical Review},
  author   = {Hohenberg, P. and Kohn, W.},
  month    = nov,
  year     = {1964},
  note     = {Publisher: American Physical Society},
  pages    = {B864--B871}
}

@article{kohn_self-consistent_1965,
  title    = {Self-{Consistent} {Equations} {Including} {Exchange} and {Correlation} {Effects}},
  volume   = {140},
  url      = {https://link.aps.org/doi/10.1103/PhysRev.140.A1133},
  doi      = {10.1103/PhysRev.140.A1133},
  number   = {4A},
  urldate  = {2024-04-24},
  journal  = {Physical Review},
  author   = {Kohn, W. and Sham, L. J.},
  month    = nov,
  year     = {1965},
  note     = {Publisher: American Physical Society},
  pages    = {A1133--A1138}
}

@article{Ren_gw_2021,
  title     = {All-electron periodic ${G}_{0}{W}_{0}$ implementation with numerical atomic orbital basis functions: Algorithm and benchmarks},
  author    = {Ren, Xinguo and Merz, Florian and Jiang, Hong and Yao, Yi and Rampp, Markus and Lederer, Hermann and Blum, Volker and Scheffler, Matthias},
  journal   = {Phys. Rev. Mater.},
  volume    = {5},
  issue     = {1},
  pages     = {013807},
  numpages  = {20},
  year      = {2021},
  month     = {Jan},
  publisher = {American Physical Society},
  doi       = {10.1103/PhysRevMaterials.5.013807},
  url       = {https://link.aps.org/doi/10.1103/PhysRevMaterials.5.013807}
}

@article{Spencer_coulomb_2008,
  title     = {Efficient calculation of the exact exchange energy in periodic systems using a truncated Coulomb potential},
  author    = {Spencer, James and Alavi, Ali},
  journal   = {Phys. Rev. B},
  volume    = {77},
  issue     = {19},
  pages     = {193110},
  numpages  = {4},
  year      = {2008},
  month     = {May},
  publisher = {American Physical Society},
  doi       = {10.1103/PhysRevB.77.193110},
  url       = {https://link.aps.org/doi/10.1103/PhysRevB.77.193110}
}

@article{Friedrich_hw_2009,
  title   = {Efficient calculation of the Coulomb matrix and its expansion around k=0 within the FLAPW method},
  journal = {Computer Physics Communications},
  volume  = {180},
  number  = {3},
  pages   = {347-359},
  year    = {2009},
  issn    = {0010-4655},
  doi     = {https://doi.org/10.1016/j.cpc.2008.10.009},
  url     = {https://www.sciencedirect.com/science/article/pii/S0010465508003664},
  author  = {Christoph Friedrich and Arno Schindlmayr and Stefan Blügel}
}

@article{Friedrich_hw_2010,
  title     = {Efficient implementation of the $GW$ approximation within the all-electron FLAPW method},
  author    = {Friedrich, Christoph and Bl\"ugel, Stefan and Schindlmayr, Arno},
  journal   = {Phys. Rev. B},
  volume    = {81},
  issue     = {12},
  pages     = {125102},
  numpages  = {16},
  year      = {2010},
  month     = {Mar},
  publisher = {American Physical Society},
  doi       = {10.1103/PhysRevB.81.125102},
  url       = {https://link.aps.org/doi/10.1103/PhysRevB.81.125102}
}

@article{Rangel/etal:2020,
Author = {Tonatiuh Rangel and Mauro {Del Ben} and Daniele Varsano and Gabriel Antonius and
Fabien Bruneval and Felipe H. {da Jornada} and Michiel J.~{van Setten} and Okan K. Orhan and
David D. {O'Regan} and Andrew Canning and Andrea Ferretti and Andrea Marini and
Gian-Marco Rignanese and Jack Deslippe and Steven G. Louie and Jeffrey B. Neaton},
Title = {Reproducibility in $G_0W_0$ calculations for solids},
Journal = {Comput. Phys. Commun.},
Year = {2020},
Volume = {255},
pages = {107242},
}

@article{PhysRevB.93.115203,
  title = {$GW$ with linearized augmented plane waves extended by high-energy local orbitals},
  author = {Jiang, Hong and Blaha, Peter},
  journal = {Phys. Rev. B},
  volume = {93},
  issue = {11},
  pages = {115203},
  numpages = {11},
  year = {2016},
  month = {Mar},
  publisher = {American Physical Society},
  doi = {10.1103/PhysRevB.93.115203},
  url = {https://link.aps.org/doi/10.1103/PhysRevB.93.115203}
}

@article{jiang_fhi-gap_2013,
  title   = {FHI-gap: A GW code based on the all-electron augmented plane wave method},
  journal = {Computer Physics Communications},
  volume  = {184},
  number  = {2},
  pages   = {348-366},
  year    = {2013},
  issn    = {0010-4655},
  doi     = {https://doi.org/10.1016/j.cpc.2012.09.018},
  url     = {https://www.sciencedirect.com/science/article/pii/S0010465512003049},
  author  = {Hong Jiang and Ricardo I. Gómez-Abal and Xin-Zheng Li and Christian Meisenbichler and Claudia Ambrosch-Draxl and Matthias Scheffler}
}

@article{Jin_pyatb_2023,
  title   = {PYATB: An efficient Python package for electronic structure calculations using ab initio tight-binding model},
  journal = {Computer Physics Communications},
  volume  = {291},
  pages   = {108844},
  year    = {2023},
  issn    = {0010-4655},
  doi     = {https://doi.org/10.1016/j.cpc.2023.108844},
  url     = {https://www.sciencedirect.com/science/article/pii/S0010465523001893},
  author  = {Gan Jin and Hongsheng Pang and Yuyang Ji and Zujian Dai and Lixin He}
}

@article{Lee_velocity_2018,
  title     = {Tight-binding calculations of optical matrix elements for conductivity using nonorthogonal atomic orbitals: Anomalous Hall conductivity in bcc Fe},
  author    = {Lee, Chi-Cheng and Lee, Yung-Ting and Fukuda, Masahiro and Ozaki, Taisuke},
  journal   = {Phys. Rev. B},
  volume    = {98},
  issue     = {11},
  pages     = {115115},
  numpages  = {8},
  year      = {2018},
  month     = {Sep},
  publisher = {American Physical Society},
  doi       = {10.1103/PhysRevB.98.115115},
  url       = {https://link.aps.org/doi/10.1103/PhysRevB.98.115115}
}

@article{Freysoldt_dielectric_2007,
  title   = {Dielectric anisotropy in the GW space–time method},
  journal = {Computer Physics Communications},
  volume  = {176},
  number  = {1},
  pages   = {1-13},
  year    = {2007},
  issn    = {0010-4655},
  doi     = {https://doi.org/10.1016/j.cpc.2006.07.018},
  url     = {https://www.sciencedirect.com/science/article/pii/S0010465506003134},
  author  = {Christoph Freysoldt and Philipp Eggert and Patrick Rinke and Arno Schindlmayr and R.W. Godby and Matthias Scheffler}
}

@article{ren_resolution--identity_2012,
  title    = {Resolution-of-identity approach to {Hartree}–{Fock}, hybrid density functionals, {RPA}, {MP2} and {GW} with numeric atom-centered orbital basis functions},
  volume   = {14},
  issn     = {1367-2630},
  url      = {https://dx.doi.org/10.1088/1367-2630/14/5/053020},
  doi      = {10.1088/1367-2630/14/5/053020},
  number   = {5},
  urldate  = {2024-04-24},
  journal  = {New Journal of Physics},
  author   = {Ren, Xinguo and Rinke, Patrick and Blum, Volker and Wieferink, Jürgen and Tkatchenko, Alexandre and Sanfilippo, Andrea and Reuter, Karsten and Scheffler, Matthias},
  month    = may,
  year     = {2012},
  note     = {Publisher: IOP Publishing},
  pages    = {053020}
}

@article{ihrig_accurate_2015,
  title    = {Accurate localized resolution of identity approach for linear-scaling hybrid density functionals and for many-body perturbation theory},
  volume   = {17},
  issn     = {1367-2630},
  url      = {https://dx.doi.org/10.1088/1367-2630/17/9/093020},
  doi      = {10.1088/1367-2630/17/9/093020},
  number   = {9},
  urldate  = {2024-04-24},
  journal  = {New Journal of Physics},
  author   = {Ihrig, Arvid Conrad and Wieferink, J\"urgen and Zhang, Igor Ying and Ropo, Matti and Ren, Xinguo and Rinke, Patrick and Scheffler, Matthias and Blum, Volker},
  month    = sep,
  year     = {2015},
  note     = {Publisher: IOP Publishing},
  pages    = {093020}
}

@article{Li_gwpp_2012,
  title     = {Impact of widely used approximations to the G0W0 method: an all-electron perspective},
  author    = {Li, Xin-Zheng and G{\'o}mez-Abal, Ricardo and Jiang, Hong and Ambrosch-Draxl, Claudia and Scheffler, Matthias},
  journal   = {New Journal of Physics},
  volume    = {14},
  number    = {2},
  pages     = {023006},
  year      = {2012},
  publisher = {IOP Publishing},
  doi       = {10.1088/1367-2630/14/2/023006},
  url       = {https://doi.org/10.1088/1367-2630/14/2/023006}
}

@article{Gomez_gwpp_2008,
  title     = {Influence of the Core-Valence Interaction and of the Pseudopotential Approximation on the Electron Self-Energy in Semiconductors},
  author    = {G\'omez-Abal, Ricardo and Li, Xinzheng and Scheffler, Matthias and Ambrosch-Draxl, Claudia},
  journal   = {Phys. Rev. Lett.},
  volume    = {101},
  issue     = {10},
  pages     = {106404},
  numpages  = {4},
  year      = {2008},
  month     = {Sep},
  publisher = {American Physical Society},
  doi       = {10.1103/PhysRevLett.101.106404},
  url       = {https://link.aps.org/doi/10.1103/PhysRevLett.101.106404}
}

@article{van_setten_pseudodojo_2018,
  title   = {The PseudoDojo: Training and grading a 85 element optimized norm-conserving pseudopotential table},
  journal = {Computer Physics Communications},
  volume  = {226},
  pages   = {39-54},
  year    = {2018},
  issn    = {0010-4655},
  doi     = {https://doi.org/10.1016/j.cpc.2018.01.012},
  url     = {https://www.sciencedirect.com/science/article/pii/S0010465518300250},
  author  = {M.J. {van Setten} and M. Giantomassi and E. Bousquet and M.J. Verstraete and D.R. Hamann and X. Gonze and G.-M. Rignanese}
}

@article{kaltak_minimax_2014,
  title     = {Cubic scaling algorithm for the random phase approximation: Self-interstitials and vacancies in Si},
  author    = {Kaltak, Merzuk and Klime{\v{s}}, Ji{\v{r}}{\'\i} and Kresse, Georg},
  journal   = {Physical Review B},
  volume    = {90},
  number    = {5},
  pages     = {054115},
  year      = {2014},
  publisher = {American Physical Society}
}

@article{kaltak_minimax2_2014,
  title     = {Low scaling algorithms for the random phase approximation: Imaginary time and Laplace transformations},
  author    = {Kaltak, Merzuk and Klimes, Jiri and Kresse, Georg},
  journal   = {Journal of chemical theory and computation},
  volume    = {10},
  number    = {6},
  pages     = {2498--2507},
  year      = {2014},
  publisher = {ACS Publications}
}

@article{Liu_minimax_2016,
  title     = {Cubic scaling $GW$: Towards fast quasiparticle calculations},
  author    = {Liu, Peitao and Kaltak, Merzuk and Klime\ifmmode \check{s}\else \v{s}\fi{}, Ji\ifmmode \check{r}\else \v{r}\fi{}\'{\i} and Kresse, Georg},
  journal   = {Phys. Rev. B},
  volume    = {94},
  issue     = {16},
  pages     = {165109},
  numpages  = {13},
  year      = {2016},
  month     = {Oct},
  publisher = {American Physical Society},
  doi       = {10.1103/PhysRevB.94.165109},
  url       = {https://link.aps.org/doi/10.1103/PhysRevB.94.165109}
}

@article{Azizi_greenx_2023,
  doi       = {10.21105/joss.05570},
  url       = {https://doi.org/10.21105/joss.05570},
  year      = {2023},
  publisher = {The Open Journal},
  volume    = {8},
  number    = {90},
  pages     = {5570},
  author    = {Azizi, Maryam and Wilhelm, Jan and Golze, Dorothea and Giantomassi, Matteo and Panadés-Barrueta, Ramón L. and Delesma, Francisco A. and Buccheri, Alexander and Gulans, Andris and Rinke, Patrick and Draxl, Claudia and Gonze, Xavier},
  title     = {Time-frequency component of the GreenX library: minimax grids for efficient RPA and GW calculations},
  journal   = {Journal of Open Source Software}
}

@article{Azizi_minimax_2024,
  title     = {Validation of the GreenX library time-frequency component for efficient $GW$ and RPA calculations},
  author    = {Azizi, Maryam and Wilhelm, Jan and Golze, Dorothea and Delesma, Francisco A. and Panad\'es-Barrueta, Ram\'on L. and Rinke, Patrick and Giantomassi, Matteo and Gonze, Xavier},
  journal   = {Phys. Rev. B},
  volume    = {109},
  issue     = {24},
  pages     = {245101},
  numpages  = {15},
  year      = {2024},
  month     = {Jun},
  publisher = {American Physical Society},
  doi       = {10.1103/PhysRevB.109.245101},
  url       = {https://link.aps.org/doi/10.1103/PhysRevB.109.245101}
}

@article{chen_systematically_2010,
  title    = {Systematically improvable optimized atomic basis sets for \textit{ab initio} calculations},
  volume   = {22},
  issn     = {0953-8984, 1361-648X},
  url      = {https://iopscience.iop.org/article/10.1088/0953-8984/22/44/445501},
  doi      = {10.1088/0953-8984/22/44/445501},
  number   = {44},
  urldate  = {2024-04-15},
  journal  = {Journal of Physics: Condensed Matter},
  author   = {Chen, Mohan and Guo, G-C and He, Lixin},
  month    = oct,
  year     = {2010},
  pages    = {445501},
}

@article{lin_strategy_2021,
  title    = {Strategy for constructing compact numerical atomic orbital basis sets by incorporating the gradients of reference wavefunctions},
  volume   = {103},
  issn     = {2469-9950, 2469-9969},
  url      = {https://link.aps.org/doi/10.1103/PhysRevB.103.235131},
  doi      = {10.1103/PhysRevB.103.235131},
  number   = {23},
  urldate  = {2024-04-15},
  journal  = {Physical Review B},
  author   = {Lin, Peize and Ren, Xinguo and He, Lixin},
  month    = jun,
  year     = {2021},
  pages    = {235131},
}

@article{blum_ab_2009,
  title    = {\textit{{Ab} initio} molecular simulations with numeric atom-centered orbitals},
  volume   = {180},
  issn     = {0010-4655},
  url      = {https://www.sciencedirect.com/science/article/pii/S0010465509002033},
  doi      = {10.1016/j.cpc.2009.06.022},
  number   = {11},
  urldate  = {2024-05-07},
  journal  = {Computer Physics Communications},
  author   = {Blum, Volker and Gehrke, Ralf and Hanke, Felix and Havu, Paula and Havu, Ville and Ren, Xinguo and Reuter, Karsten and Scheffler, Matthias},
  month    = nov,
  year     = {2009},
  pages    = {2175--2196}
}

@article{abinit_Gonze_2005,
  author    = {Gonze, X. and Rignanese, G.-M. and Verstraete, M. and Beuken, J.-M. and Pouillon, Y. and Caracas, R. and Jollet, F. and Torrent, M. and Zerah, G. and Mikami, M. and Ghosez, Ph. and Veithen, M. and Raty, J.-Y. and Olevano, V. and Bruneval, F. and Reining, L. and Godby, R. and Onida, G. and D.C. Allan, D.R. Hamann},
  doi       = {10.1524/zkri.220.5.558.65066},
  issn      = {2196-7105, 2194-4946},
  journal   = {Zeitschrift für Kristallographie - Crystalline Materials},
  month     = {January},
  number    = {5/6},
  pages     = {558-562},
  publisher = {Walter de Gruyter GmbH},
  source    = {Crossref},
  title     = {A brief introduction to the {ABINIT} software package},
  volume    = {220},
  year      = {2005}
}

@article{cp2k_Graml_2024,
  author  = {Graml, Maximilian and Zollner, Klaus and Hernangómez-Pérez, Daniel and Faria Junior, Paulo and Wilhelm, Jan},
  year    = {2024},
  month   = {02},
  pages   = {},
  title   = {Low-Scaling GW Algorithm Applied to Twisted Transition-Metal Dichalcogenide Heterobilayers},
  volume  = {20},
  journal = {Journal of chemical theory and computation},
  doi     = {10.1021/acs.jctc.3c01230}
}

@article{exciting_Gulans_2014,
  title     = {Accurate all-electron ${G}_{0}{W}_{0}$ quasiparticle energies employing the full-potential augmented plane-wave method},
  author    = {Nabok, Dmitrii and Gulans, Andris and Draxl, Claudia},
  journal   = {Phys. Rev. B},
  volume    = {94},
  issue     = {3},
  pages     = {035118},
  numpages  = {9},
  year      = {2016},
  month     = {Jul},
  publisher = {American Physical Society},
  doi       = {10.1103/PhysRevB.94.035118},
  url       = {https://link.aps.org/doi/10.1103/PhysRevB.94.035118}
}

@article{GPAW_Huser_2013,
  author  = {H\"user, Falco and Olsen, Thomas and Thygesen, Kristian S.},
  title   = {Quasiparticle GW calculations for solids, molecules, and two-dimensional materials},
  journal = {Phys. Rev. B},
  volume  = {87},
  number  = {23},
  pages   = {235132},
  year    = {2013},
  doi     = {10.1103/PhysRevB.87.235132}
}

@article{BerkeleyGW_Deslippe_2012,
  title    = {BerkeleyGW: A massively parallel computer package for the calculation of the quasiparticle and optical properties of materials and nanostructures},
  journal  = {Computer Physics Communications},
  volume   = {183},
  number   = {6},
  pages    = {1269-1289},
  year     = {2012},
  issn     = {0010-4655},
  doi      = {https://doi.org/10.1016/j.cpc.2011.12.006},
  url      = {https://www.sciencedirect.com/science/article/pii/S0010465511003912},
  author   = {Jack Deslippe and Georgy Samsonidze and David A. Strubbe and Manish Jain and Marvin L. Cohen and Steven G. Louie}
}

@article{yambo_Sangalli_2019,
  author  = {Sangalli, Davide and Ferretti, Andrea and Miranda, Henrique and Attaccalite, Claudio and Marri, I and Cannuccia, E and Melo, P and Marsili, Margherita and Paleari, F and Marrazzo, A and Prandini, G and Bonfa, Pietro and Atambo, M and Affinito, Fabio and Palummo, Maurizia and Molina-Sánchez, Alejandro and Hogan, Conor and Grüning, Myrta and Varsano, Daniele and Marini, A},
  year    = {2019},
  month   = {05},
  pages   = {},
  title   = {Many-body perturbation theory calculations using the yambo code},
  volume  = {31},
  journal = {Journal of Physics: Condensed Matter},
  doi     = {10.1088/1361-648X/ab15d0}
}

@article{vasp_gw_2006,
  title = {Implementation and performance of the frequency-dependent $GW$ method within the PAW framework},
  author = {Shishkin, M. and Kresse, G.},
  journal = {Phys. Rev. B},
  volume = {74},
  issue = {3},
  pages = {035101},
  numpages = {13},
  year = {2006},
  month = {Jul},
  publisher = {American Physical Society},
  doi = {10.1103/PhysRevB.74.035101},
  url = {https://link.aps.org/doi/10.1103/PhysRevB.74.035101}
}

@article{vasp_gw_2014,
  title     = {Predictive $GW$ calculations using plane waves and pseudopotentials},
  author    = {Klime\ifmmode \check{s}\else \v{s}\fi{}, Ji\ifmmode \check{r}\else \v{r}\fi{}\'{\i} and Kaltak, Merzuk and Kresse, Georg},
  journal   = {Phys. Rev. B},
  volume    = {90},
  issue     = {7},
  pages     = {075125},
  numpages  = {15},
  year      = {2014},
  month     = {Aug},
  publisher = {American Physical Society},
  doi       = {10.1103/PhysRevB.90.075125},
  url       = {https://link.aps.org/doi/10.1103/PhysRevB.90.075125}
}

@article{rpa_shi_2024,
  title     = {Subquadratic-scaling real-space random phase approximation correlation energy calculations for periodic systems with numerical atomic orbitals},
  author    = {Shi, Rong and Lin, Peize and Zhang, Min-Ye and He, Lixin and Ren, Xinguo},
  journal   = {Phys. Rev. B},
  volume    = {109},
  issue     = {3},
  pages     = {035103},
  numpages  = {15},
  year      = {2024},
  month     = {Jan},
  publisher = {American Physical Society},
  doi       = {10.1103/PhysRevB.109.035103},
  url       = {https://link.aps.org/doi/10.1103/PhysRevB.109.035103}
}

@article{librpa_shi_2025,
  title    = {LibRPA: A software package for low-scaling first-principles calculations of random phase approximation electron correlation energy based on numerical atomic orbitals},
  journal  = {Computer Physics Communications},
  volume   = {309},
  pages    = {109496},
  year     = {2025},
  issn     = {0010-4655},
  doi      = {https://doi.org/10.1016/j.cpc.2024.109496},
  url      = {https://www.sciencedirect.com/science/article/pii/S0010465524004193},
  author   = {Rong Shi and Min-Ye Zhang and Peize Lin and Lixin He and Xinguo Ren}
}

@article{hafner_ab-initio_2008,
  title      = {Ab-initio simulations of materials using {VASP}: {Density}-functional theory and beyond},
  volume     = {29},
  copyright  = {Copyright © 2008 Wiley Periodicals, Inc.},
  issn       = {1096-987X},
  shorttitle = {Ab-initio simulations of materials using {VASP}},
  url        = {https://onlinelibrary.wiley.com/doi/abs/10.1002/jcc.21057},
  doi        = {10.1002/jcc.21057},
  number     = {13},
  urldate    = {2024-05-11},
  journal    = {Journal of Computational Chemistry},
  author     = {Hafner, J\"urgen},
  year       = {2008},
  pages      = {2044--2078},
}

@article{hedin_gw_1965,
  title     = {New Method for Calculating the One-Particle Green's Function with Application to the Electron-Gas Problem},
  author    = {Hedin, Lars},
  journal   = {Phys. Rev.},
  volume    = {139},
  issue     = {3A},
  pages     = {A796--A823},
  numpages  = {0},
  year      = {1965},
  month     = {Aug},
  publisher = {American Physical Society},
  doi       = {10.1103/PhysRev.139.A796},
  url       = {https://link.aps.org/doi/10.1103/PhysRev.139.A796}
}

@article{gloze_compendium_2019,
  author  = {Golze, Dorothea  and Dvorak, Marc  and Rinke, Patrick },
  title   = {The GW Compendium: A Practical Guide to Theoretical Photoemission Spectroscopy},
  journal = {Frontiers in Chemistry},
  volume  = {Volume 7 - 2019},
  year    = {2019},
  url     = {https://www.frontiersin.org/journals/chemistry/articles/10.3389/fchem.2019.00377},
  doi     = {10.3389/fchem.2019.00377},
  issn    = {2296-2646}
}

@article{lin_accuracy_2020,
  title    = {Accuracy of {Localized} {Resolution} of the {Identity} in {Periodic} {Hybrid} {Functional} {Calculations} with {Numerical} {Atomic} {Orbitals}},
  volume   = {11},
  url      = {https://doi.org/10.1021/acs.jpclett.0c00481},
  doi      = {10.1021/acs.jpclett.0c00481},
  number   = {8},
  urldate  = {2024-04-24},
  journal  = {The Journal of Physical Chemistry Letters},
  author   = {Lin, Peize and Ren, Xinguo and He, Lixin},
  month    = apr,
  year     = {2020},
  note     = {Publisher: American Chemical Society},
  pages    = {3082--3088}
}

@article{gw100_2015,
  author  = {van Setten, Michiel J. and Caruso, Fabio and Sharifzadeh, Sahar and Ren, Xinguo and Scheffler, Matthias and Liu, Fang and Lischner, Johannes and Lin, Lin and Deslippe, Jack R. and Louie, Steven G. and Yang, Chao and Weigend, Florian and Neaton, Jeffrey B. and Evers, Ferdinand and Rinke, Patrick},
  title   = {GW100: Benchmarking G0W0 for Molecular Systems},
  journal = {Journal of Chemical Theory and Computation},
  volume  = {11},
  number  = {12},
  pages   = {5665-5687},
  year    = {2015},
  doi     = {10.1021/acs.jctc.5b00453},
  note    = {PMID: 26642984},
  url     = { 
             https://doi.org/10.1021/acs.jctc.5b00453
             },
  eprint  = { 
             https://doi.org/10.1021/acs.jctc.5b00453
             }
}

@article{2d_Rasmussen_2016,
  title     = {Efficient many-body calculations for two-dimensional materials using exact limits for the screened potential: Band gaps of ${\mathbf{MoS}}_{2}, h$-BN, and phosphorene},
  author    = {Rasmussen, Filip A. and Schmidt, Per S. and Winther, Kirsten T. and Thygesen, Kristian S.},
  journal   = {Phys. Rev. B},
  volume    = {94},
  issue     = {15},
  pages     = {155406},
  numpages  = {9},
  year      = {2016},
  month     = {Oct},
  publisher = {American Physical Society},
  doi       = {10.1103/PhysRevB.94.155406},
  url       = {https://link.aps.org/doi/10.1103/PhysRevB.94.155406}
}

@article{gw_Reining_2018,
  author   = {Reining, Lucia},
  title    = {The GW approximation: content, successes and limitations},
  journal  = {WIREs Computational Molecular Science},
  volume   = {8},
  number   = {3},
  pages    = {e1344},
  doi      = {https://doi.org/10.1002/wcms.1344},
  url      = {https://wires.onlinelibrary.wiley.com/doi/abs/10.1002/wcms.1344},
  year     = {2018}
}

@article{exx_Levchenko_2015,
  title    = {Hybrid functionals for large periodic systems in an all-electron, numeric atom-centered basis framework},
  journal  = {Computer Physics Communications},
  volume   = {192},
  pages    = {60-69},
  year     = {2015},
  issn     = {0010-4655},
  doi      = {https://doi.org/10.1016/j.cpc.2015.02.021},
  url      = {https://www.sciencedirect.com/science/article/pii/S001046551500079X},
  author   = {Sergey V. Levchenko and Xinguo Ren and Jürgen Wieferink and Rainer Johanni and Patrick Rinke and Volker Blum and Matthias Scheffler}
}

@article{exx_kokott_2024,
  author   = {Kokott, Sebastian and Merz, Florian and Yao, Yi and Carbogno, Christian and Rossi, Mariana and Havu, Ville and Rampp, Markus and Scheffler, Matthias and Blum, Volker},
  title    = {Efficient all-electron hybrid density functionals for atomistic simulations beyond 10 000 atoms},
  journal  = {The Journal of Chemical Physics},
  volume   = {161},
  number   = {2},
  pages    = {024112},
  year     = {2024},
  month    = {07},
  issn     = {0021-9606},
  doi      = {10.1063/5.0208103},
  url      = {https://doi.org/10.1063/5.0208103},
}

@article{li2022deep,
  title     = {Deep-learning density functional theory Hamiltonian for efficient ab initio electronic-structure calculation},
  author    = {Li, He and Wang, Zun and Zou, Nianlong and Ye, Meng and Xu, Runzhang and Gong, Xiaoxun and Duan, Wenhui and Xu, Yong},
  journal   = {Nature Computational Science},
  volume    = {2},
  number    = {6},
  pages     = {367--377},
  year      = {2022},
  publisher = {Nature Publishing Group US New York}
}

@article{wang2024deeph,
  title   = {Deeph-2: Enhancing deep-learning electronic structure via an equivariant local-coordinate transformer},
  author  = {Wang, Yuxiang and Li, He and Tang, Zechen and Tao, Honggeng and Wang, Yanzhen and Yuan, Zilong and Chen, Zezhou and Duan, Wenhui and Xu, Yong},
  journal = {arXiv preprint arXiv:2401.17015},
  year    = {2024}
}

@article{wang2024universal,
  title     = {Universal materials model of deep-learning density functional theory Hamiltonian},
  author    = {Wang, Yuxiang and Li, Yang and Tang, Zechen and Li, He and Yuan, Zilong and Tao, Honggeng and Zou, Nianlong and Bao, Ting and Liang, Xinghao and Chen, Zezhou and others},
  journal   = {Science Bulletin},
  volume    = {69},
  number    = {16},
  pages     = {2514--2521},
  year      = {2024},
  publisher = {Elsevier}
}

@article{cheng2025efficient,
  title   = {Efficient construction of effective Hamiltonians with a hybrid machine learning method},
  author  = {Cheng, Yang and Zhang, Binhua and Li, Xueyang and Yu, Hongyu and Xu, Changsong and Xiang, Hongjun},
  journal = {arXiv preprint arXiv:2505.04925},
  year    = {2025}
}

@article{zhang2025advancing,
  title     = {Advancing nonadiabatic molecular dynamics simulations in solids with E (3) equivariant deep neural hamiltonians},
  author    = {Zhang, Changwei and Zhong, Yang and Tao, Zhi-Guo and Qin, Xinming and Shang, Honghui and Lan, Zhenggang and Prezhdo, Oleg V and Gong, Xin-Gao and Chu, Weibin and Xiang, Hongjun},
  journal   = {Nature Communications},
  volume    = {16},
  number    = {1},
  pages     = {2033},
  year      = {2025},
  publisher = {Nature Publishing Group UK London}
}

@article{su2023efficient,
  title     = {Efficient determination of the Hamiltonian and electronic properties using graph neural network with complete local coordinates},
  author    = {Su, Mao and Yang, Ji-Hui and Xiang, Hong-Jun and Gong, Xin-Gao},
  journal   = {Machine Learning: Science and Technology},
  volume    = {4},
  number    = {3},
  pages     = {035010},
  year      = {2023},
  publisher = {IOP Publishing}
}

@article{zhong2023transferable,
  title     = {Transferable equivariant graph neural networks for the Hamiltonians of molecules and solids},
  author    = {Zhong, Yang and Yu, Hongyu and Su, Mao and Gong, Xingao and Xiang, Hongjun},
  journal   = {npj Computational Materials},
  volume    = {9},
  number    = {1},
  pages     = {182},
  year      = {2023},
  publisher = {Nature Publishing Group UK London}
}

@article{spacetime_rieger_1999,
  title    = {The GW space-time method for the self-energy of large systems},
  journal  = {Computer Physics Communications},
  volume   = {117},
  number   = {3},
  pages    = {211-228},
  year     = {1999},
  issn     = {0010-4655},
  doi      = {https://doi.org/10.1016/S0010-4655(98)00174-X},
  url      = {https://www.sciencedirect.com/science/article/pii/S001046559800174X},
  author   = {Martin M. Rieger and L. Steinbeck and I.D. White and H.N. Rojas and R.W. Godby}
}

@article{spacetime_rojas_1995,
  title     = {Space-Time Method for Ab Initio Calculations of Self-Energies and Dielectric Response Functions of Solids},
  author    = {Rojas, H. N. and Godby, R. W. and Needs, R. J.},
  journal   = {Phys. Rev. Lett.},
  volume    = {74},
  issue     = {10},
  pages     = {1827--1830},
  numpages  = {0},
  year      = {1995},
  month     = {Mar},
  publisher = {American Physical Society},
  doi       = {10.1103/PhysRevLett.74.1827},
  url       = {https://link.aps.org/doi/10.1103/PhysRevLett.74.1827}
}

@article{abacus_2025_jcp,
    author = {Zhou, Weiqing and Zheng, Daye and Liu, Qianrui and Lu, Denghui and Liu, Yu and Lin, Peize and Huang, Yike and Peng, Xingliang and Bao, Jie J. and Cai, Chun and Jin, Zuxin and Wu, Jing and Zhang, Haochong and Jin, Gan and Ji, Yuyang and Shen, Zhenxiong and Liu, Xiaohui and Sun, Liang and Cao, Yu and Sun, Menglin and Liu, Jianchuan and Chen, Tao and Liu, Renxi and Li, Yuanbo and Han, Haozhi and Liang, Xinyuan and Bao, Taoni and Deng, Zichao and Liu, Tao and Chen, Nuo and Ren, Hongxu and Zhang, Xiaoyang and Liu, Zhaoqing and Fu, Yiwei and Liu, Maochang and Li, Zhuoyuan and Wen, Tongqi and Tang, Zechen and Xu, Yong and Duan, Wenhui and Wang, Xiaoyang and Gu, Qiangqiang and Dai, Fu-Zhi and Zheng, Qijing and Zhong, Yang and Xiang, Hongjun and Gong, Xingao and Zhao, Jin and Zhang, Yuzhi and Ou, Qi and Jiang, Hong and Liu, Shi and Xu, Ben and Xu, Shenzhen and Ren, Xinguo and He, Lixin and Zhang, Linfeng and Chen, Mohan},
    title = {ABACUS: An electronic structure analysis package for the AI era},
    journal = {The Journal of Chemical Physics},
    volume = {163},
    number = {19},
    pages = {192501},
    year = {2025},
    month = {11},
    issn = {0021-9606},
    doi = {10.1063/5.0297563},
    url = {https://doi.org/10.1063/5.0297563},
}

@article{openmx_2003,
  title = {Variationally optimized atomic orbitals for large-scale electronic structures},
  author = {Ozaki, T.},
  journal = {Phys. Rev. B},
  volume = {67},
  issue = {15},
  pages = {155108},
  numpages = {5},
  year = {2003},
  month = {Apr},
  publisher = {American Physical Society},
  doi = {10.1103/PhysRevB.67.155108},
  url = {https://link.aps.org/doi/10.1103/PhysRevB.67.155108}
}

@article{openmx_2004,
  title = {Numerical atomic basis orbitals from H to Kr},
  author = {Ozaki, T. and Kino, H.},
  journal = {Phys. Rev. B},
  volume = {69},
  issue = {19},
  pages = {195113},
  numpages = {19},
  year = {2004},
  month = {May},
  publisher = {American Physical Society},
  doi = {10.1103/PhysRevB.69.195113},
  url = {https://link.aps.org/doi/10.1103/PhysRevB.69.195113}
}

@article{aims_soc_2017,
  title = {One-hundred-three compound band-structure benchmark of post-self-consistent spin-orbit coupling treatments in density functional theory},
  author = {Huhn, William P. and Blum, Volker},
  journal = {Phys. Rev. Mater.},
  volume = {1},
  issue = {3},
  pages = {033803},
  numpages = {18},
  year = {2017},
  month = {Aug},
  publisher = {American Physical Society},
  doi = {10.1103/PhysRevMaterials.1.033803},
  url = {https://link.aps.org/doi/10.1103/PhysRevMaterials.1.033803}
}

@article{Vurgaftman/etal:2001,
  title = {Band parameters for III--V compound semiconductors and their alloys},
  author = {Vurgaftman, I. and Meyer, J. R. and Ram-Mohan, L. R.},
  journal = {J. Appl. Phys.},
  volume = {89},
  pages = {5815--5875},
  year = {2001},
  publisher = {AIP Publishing}
}

@book{tikhonov_ill_posed_1995,
  title = {Numerical Methods for the Solution of Ill-Posed Problems},
  author = {Goncharsky, A. and Stepanov, V. and Tikhonov, A. and Yagola, A.},
  year = {1995},
  publisher = {Kluwer Academic Publishers},
  address = {Dordrecht}
}

\end{document}